\input harvmac.tex
\input epsf.tex
\input amssym
\input ulem.sty


\let\includefigures=\iftrue
\let\useblackboard=\iftrue
\newfam\black

\def\figin{\epsfcheck\figin}\def\figins{\epsfcheck\figins}
\def\epsfcheck{\ifx\epsfbox\UnDeFiNeD
\message{(NO epsf.tex, FIGURES WILL BE IGNORED)}
\gdef\figin##1{\vskip2in}\gdef\figins##1{\hskip.5in}
\else\message{(FIGURES WILL BE INCLUDED)}%
\gdef\figin##1{##1}\gdef\figins##1{##1}\fi}
\def\DefWarn#1{}
\def\figinsert{\goodbreak\midinsert}
\def\ifig#1#2#3{\DefWarn#1\xdef#1{fig.~\the\figno}
\writedef{#1\leftbracket fig.\noexpand~\the\figno} %
\figinsert\figin{\centerline{#3}}\medskip\centerline{\vbox{\baselineskip12pt
\advance\hsize by -1truein\noindent\footnotefont{\bf
Fig.~\the\figno:} #2}}
\bigskip\endinsert\global\advance\figno by1}


\includefigures
\message{If you do not have epsf.tex (to include figures),}
\message{change the option at the top of the tex file.}
\input epsf
\def\figin{\epsfcheck\figin}\def\figins{\epsfcheck\figins}
\def\epsfcheck{\ifx\epsfbox\UnDeFiNeD
\message{(NO epsf.tex, FIGURES WILL BE IGNORED)}
\gdef\figin##1{\vskip2in}\gdef\figins##1{\hskip.5in}
\else\message{(FIGURES WILL BE INCLUDED)}%
\gdef\figin##1{##1}\gdef\figins##1{##1}\fi}
\def\DefWarn#1{}
\def\figinsert{\goodbreak\midinsert}
\def\ifig#1#2#3{\DefWarn#1\xdef#1{fig.~\the\figno}
\writedef{#1\leftbracket fig.\noexpand~\the\figno}%
\figinsert\figin{\centerline{#3}}\medskip\centerline{\vbox{
\baselineskip12pt\advance\hsize by -1truein
\noindent\footnotefont{\bf Fig.~\the\figno:} #2}}
\endinsert\global\advance\figno by1}
\else
\def\ifig#1#2#3{\xdef#1{fig.~\the\figno}
\writedef{#1\leftbracket fig.\noexpand~\the\figno}%
\global\advance\figno by1} \fi

\def\figin{\epsfcheck\figin}\def\figins{\epsfcheck\figins}
\def\epsfcheck{\ifx\epsfbox\UnDeFiNeD
\message{(NO epsf.tex, FIGURES WILL BE IGNORED)}
\gdef\figin##1{\vskip2in}\gdef\figins##1{\hskip.5in}
\else\message{(FIGURES WILL BE INCLUDED)}%
\gdef\figin##1{##1}\gdef\figins##1{##1}\fi}
\def\DefWarn#1{}
\def\figinsert{\goodbreak\midinsert}
\def\ifig#1#2#3{\DefWarn#1\xdef#1{fig.~\the\figno}
\writedef{#1\leftbracket fig.\noexpand~\the\figno} %
\figinsert\figin{\centerline{#3}}\medskip\centerline{\vbox{\baselineskip12pt
\advance\hsize by -1truein\noindent\footnotefont{\bf
Fig.~\the\figno:} #2}}
\bigskip\endinsert\global\advance\figno by1}

\def \la {\langle}
\def \ra {\rangle}

\def \pa {\partial}

\def\OO{{\cal OO}}

\catcode`\@=11
\def\slash#1{\mathord{\mathpalette\c@ncel{#1}}}
\overfullrule=0pt

\def\GG{{\cal G}}

\def\NN{{\cal N}}
\def\OO{{\cal O}}

\def\underrel#1\over#2{\mathrel{\mathop{\kern\z@#1}\limits_{#2}}}

\catcode`\@=12



\def\zbar{{\bar z}}

\def\bT{{\bf T}}

\def\p{{\partial}}

\def\bt{{\bar t}}

\def\DH{{\Delta_{H}}}

\def\OH{{\OO_{H}}}

\def \vac {|0\rangle}


\def\unlockat{\catcode`\@=11}
\def\lockat{\catcode`\@=12}

\unlockat

\def\newsec#1{\global\advance\secno by1\message{(\the\secno. #1)}
\global\subsecno=0\global\subsubsecno=0\eqnres@t\noindent
{\bf\the\secno. #1}
\writetoca{{\secsym} {#1}}\par\nobreak\medskip\nobreak}
\global\newcount\subsecno \global\subsecno=0
\def\subsec#1{\global\advance\subsecno
by1\message{(\secsym\the\subsecno. #1)}
\ifnum\lastpenalty>9000\else\bigbreak\fi\global\subsubsecno=0
\noindent{\it\secsym\the\subsecno. #1}
\writetoca{\string\quad {\secsym\the\subsecno.} {#1}}
\par\nobreak\medskip\nobreak}
\global\newcount\subsubsecno \global\subsubsecno=0
\def\subsubsec#1{\global\advance\subsubsecno by1
\message{(\secsym\the\subsecno.\the\subsubsecno. #1)}
\ifnum\lastpenalty>9000\else\bigbreak\fi
\noindent\quad{\secsym\the\subsecno.\the\subsubsecno.}{#1}
\writetoca{\string\qquad{\secsym\the\subsecno.\the\subsubsecno.}{#1}}
\par\nobreak\medskip\nobreak}

\def\subsubseclab#1{\DefWarn#1\xdef
#1{\noexpand\hyperref{}{subsubsection}%
{\secsym\the\subsecno.\the\subsubsecno}%
{\secsym\the\subsecno.\the\subsubsecno}}%
\writedef{#1\leftbracket#1}\wrlabeL{#1=#1}}
\lockat


\lref\DickensRE{
  T.~A.~Dickens, U.~J.~Lindqwister, W.~R.~Somsky and L.~G.~Yaffe,
  ``The Coherent State Variational Algorithm. 2. Implementation and Testing,''
Nucl.\ Phys.\ B {\bf 309}, 1 (1988)..
}

\lref\BrownYN{
  F.~R.~Brown and L.~G.~Yaffe,
  ``The Coherent State Variational Algorithm: A Numerical Method for Solving Large $N$ Gauge Theories,''
Nucl.\ Phys.\ B {\bf 271}, 267 (1986)..
}

\lref\YaffeVF{
  L.~G.~Yaffe,
  ``Large n Limits as Classical Mechanics,''
Rev.\ Mod.\ Phys.\  {\bf 54}, 407 (1982)..
}

\lref\OsbornCR{
  H.~Osborn and A.~C.~Petkou,
  ``Implications of conformal invariance in field theories for general dimensions,''
Annals Phys.\  {\bf 231}, 311 (1994).
[hep-th/9307010].
}

\lref\DolanHV{
  F.~A.~Dolan and H.~Osborn,
  ``Conformal partial waves and the operator product expansion,''
Nucl.\ Phys.\ B {\bf 678}, 491 (2004).
[hep-th/0309180].
}

\lref\FitzpatrickDM{
  A.~L.~Fitzpatrick and J.~Kaplan,
  ``Unitarity and the Holographic S-Matrix,''
JHEP {\bf 1210}, 032 (2012).
[arXiv:1112.4845 [hep-th]].
}

\lref\FitzpatrickYX{
  A.~L.~Fitzpatrick, J.~Kaplan, D.~Poland and D.~Simmons-Duffin,
  ``The Analytic Bootstrap and AdS Superhorizon Locality,''
JHEP {\bf 1312}, 004 (2013).
[arXiv:1212.3616 [hep-th]].
}

\lref\KomargodskiEK{
  Z.~Komargodski and A.~Zhiboedov,
  ``Convexity and Liberation at Large Spin,''
JHEP {\bf 1311}, 140 (2013).
[arXiv:1212.4103 [hep-th]].
}

\lref\BanerjeeKJH{
  S.~Banerjee, K.~Papadodimas, S.~Raju, P.~Samantray and P.~Shrivastava,
  ``A Bound on Thermal Relativistic Correlators at Large Spacelike Momenta,''
SciPost Phys.\  {\bf 8}, no. 4, 064 (2020).
[arXiv:1902.07203 [hep-th]].
}

\lref\PappadopuloJK{
  D.~Pappadopulo, S.~Rychkov, J.~Espin and R.~Rattazzi,
  ``OPE Convergence in Conformal Field Theory,''
Phys.\ Rev.\ D {\bf 86}, 105043 (2012).
[arXiv:1208.6449 [hep-th]].
}

\lref\KarlssonDBD{
  R.~Karlsson, M.~Kulaxizi, A.~Parnachev and P.~Tadi\' c,
  ``Leading Multi-Stress Tensors and Conformal Bootstrap,''
JHEP {\bf 2001}, 076 (2020).
[arXiv:1909.05775 [hep-th]].
}
\lref\IliesiuFAO{
  L.~Iliesiu, M.~Kologlu, R.~Mahajan, E.~Perlmutter and D.~Simmons-Duffin,
  ``The Conformal Bootstrap at Finite Temperature,''
JHEP {\bf 1810}, 070 (2018).
[arXiv:1802.10266 [hep-th]].
}
\lref\RattazziGJ{
  R.~Rattazzi, S.~Rychkov and A.~Vichi,
  ``Central Charge Bounds in 4D Conformal Field Theory,''
Phys.\ Rev.\ D {\bf 83}, 046011 (2011).
[arXiv:1009.2725 [hep-th]].
}
\lref\DattaJEO{
  S.~Datta, P.~Kraus and B.~Michel,
  ``Typicality and thermality in 2d CFT,''
JHEP {\bf 1907}, 143 (2019).
[arXiv:1904.00668 [hep-th]].
}
\lref\FitzpatrickYJB{
  A.~L.~Fitzpatrick, K.~W.~Huang, D.~Meltzer, E.~Perlmutter and D.~Simmons-Duffin,
  ``Model-Dependence of Minimal-Twist OPEs in $d>2$ Holographic CFTs,''
[arXiv:2007.07382 [hep-th]].
}
\lref\AmadoKGR{
  I.~Amado, B.~Sundborg, L.~Thorlacius and N.~Wintergerst,
  ``Black holes from large N singlet models,''
JHEP {\bf 1803}, 075 (2018).
[arXiv:1712.06963 [hep-th]].
}
\lref\AmadoPGY{
  I.~Amado, B.~Sundborg, L.~Thorlacius and N.~Wintergerst,
  ``Probing emergent geometry through phase transitions in free vector and matrix models,''
JHEP {\bf 1702}, 005 (2017).
[arXiv:1612.03009 [hep-th]].
}
\lref\HaggiManiRU{
  P.~Haggi-Mani and B.~Sundborg,
  ``Free large N supersymmetric Yang-Mills theory as a string theory,''
JHEP {\bf 0004}, 031 (2000).
[hep-th/0002189].
}
\lref\SundborgUE{
  B.~Sundborg,
  ``The Hagedorn transition, deconfinement and N=4 SYM theory,''
Nucl.\ Phys.\ B {\bf 573}, 349 (2000).
[hep-th/9908001].
}
\lref\AharonySX{
  O.~Aharony, J.~Marsano, S.~Minwalla, K.~Papadodimas and M.~Van Raamsdonk,
  ``The Hagedorn - deconfinement phase transition in weakly coupled large N gauge theories,''
Adv.\ Theor.\ Math.\ Phys.\  {\bf 8}, 603 (2004).
[hep-th/0310285].
}
\lref\KulaxiziTKD{
  M.~Kulaxizi, G.~S.~Ng and A.~Parnachev,
  ``Subleading Eikonal, AdS/CFT and Double Stress Tensors,''
JHEP {\bf 1910}, 107 (2019).
[arXiv:1907.00867 [hep-th]].
}
\lref\KarlssonGHX{
  R.~Karlsson, M.~Kulaxizi, A.~Parnachev and P.~Tadi\' c,
  ``Stress tensor sector of conformal correlators operators in the Regge limit,''
JHEP {\bf 2007}, 019 (2020).
[arXiv:2002.12254 [hep-th]].
}
\lref\LiDQM{
  Y.~Z.~Li and H.~Y.~Zhang,
  ``More on Heavy-Light Bootstrap up to Double-Stress-Tensor,''
[arXiv:2004.04758 [hep-th]].
}
\lref\LiZBA{
  Y.~Z.~Li,
  ``Heavy-light Bootstrap from Lorentzian Inversion Formula,''
JHEP {\bf 2007}, 046 (2020).
[arXiv:1910.06357 [hep-th]].
}
\lref\DolanUT{
  F.~A.~Dolan and H.~Osborn,
  ``Conformal four point functions and the operator product expansion,''
Nucl.\ Phys.\ B {\bf 599}, 459 (2001).
[hep-th/0011040].
}
\lref\GiustoMUP{
  S.~Giusto, M.~R.~R.~Hughes and R.~Russo,
  ``The Regge limit of AdS$_3$ holographic correlators,''
[arXiv:2007.12118 [hep-th]].
}
\lref\FitzpatrickZQZ{
  A.~L.~Fitzpatrick and K.~W.~Huang,
  ``Universal Lowest-Twist in CFTs from Holography,''
JHEP {\bf 1908}, 138 (2019).
[arXiv:1903.05306 [hep-th]].
}
\lref\PerlmutterIYA{
  E.~Perlmutter,
  ``Virasoro conformal blocks in closed form,''
JHEP {\bf 1508}, 088 (2015).
[arXiv:1502.07742 [hep-th]].
}
\lref\PenedonesUE{
  J.~Penedones,
  ``Writing CFT correlation functions as AdS scattering amplitudes,''
JHEP {\bf 1103}, 025 (2011).
[arXiv:1011.1485 [hep-th]].
}
\lref\FitzpatrickEFK{
  A.~L.~Fitzpatrick, K.~W.~Huang and D.~Li,
  ``Probing universalities in d > 2 CFTs: from black holes to shockwaves,''
JHEP {\bf 1911}, 139 (2019).
[arXiv:1907.10810 [hep-th]].
}
\lref\LiDQM{
  Y.~Z.~Li and H.~Y.~Zhang,
  ``More on Heavy-Light Bootstrap up to Double-Stress-Tensor,''
[arXiv:2004.04758 [hep-th]].
}
\lref\LiZBA{
  Y.~Z.~Li,
  ``Heavy-light Bootstrap from Lorentzian Inversion Formula,''
JHEP {\bf 2007}, 046 (2020).
[arXiv:1910.06357 [hep-th]].
}
\lref\DelacretazNIT{
  L.~V.~Delacr\'etaz,
  ``Heavy Operators and Hydrodynamic Tails,''
SciPost Phys.\  {\bf 9}, no. 3, 034 (2020).
[arXiv:2006.01139 [hep-th]].
}
\lref\LiuJHS{
  J.~Liu, E.~Perlmutter, V.~Rosenhaus and D.~Simmons-Duffin,
  ``$d$-dimensional SYK, AdS Loops, and $6j$ Symbols,''
JHEP {\bf 1903}, 052 (2019).
[arXiv:1808.00612 [hep-th]].
}
\lref\HijanoZSA{
  E.~Hijano, P.~Kraus, E.~Perlmutter and R.~Snively,
  ``Witten Diagrams Revisited: The AdS Geometry of Conformal Blocks,''
JHEP {\bf 1601}, 146 (2016).
[arXiv:1508.00501 [hep-th]].
}
\lref\GobeilFZY{
  Y.~Gobeil, A.~Maloney, G.~S.~Ng and J.~q.~Wu,
  ``Thermal Conformal Blocks,''
SciPost Phys.\  {\bf 7}, no. 2, 015 (2019).
[arXiv:1802.10537 [hep-th]].
}
\lref\KarlssonTXU{
  R.~Karlsson,
  ``Multi-stress tensors and next-to-leading singularities in the Regge limit,''
JHEP {\bf 2008}, 037 (2020).
[arXiv:1912.01577 [hep-th]].
}
\lref\KarlssonQFI{
  R.~Karlsson, M.~Kulaxizi, A.~Parnachev and P.~Tadi\' c,
  ``Black Holes and Conformal Regge Bootstrap,''
JHEP {\bf 1910}, 046 (2019).
[arXiv:1904.00060 [hep-th]].
}
\lref\KulaxiziDXO{
  M.~Kulaxizi, G.~S.~Ng and A.~Parnachev,
  ``Black Holes, Heavy States, Phase Shift and Anomalous Dimensions,''
SciPost Phys.\  {\bf 6}, no. 6, 065 (2019).
[arXiv:1812.03120 [hep-th]].
}
\lref\LiTPF{
  Y.~Z.~Li, Z.~F.~Mai and H.~Lü,
  ``Holographic OPE Coefficients from AdS Black Holes with Matters,''
JHEP {\bf 1909}, 001 (2019).
[arXiv:1905.09302 [hep-th]].
}
\lref\MaldacenaRE{
  J.~M.~Maldacena,
  ``The Large N limit of superconformal field theories and supergravity,''
Int.\ J.\ Theor.\ Phys.\  {\bf 38}, 1113 (1999), [Adv.\ Theor.\ Math.\ Phys.\  {\bf 2}, 231 (1998)].
[hep-th/9711200].
}
\lref\WittenQJ{
  E.~Witten,
  ``Anti-de Sitter space and holography,''
Adv.\ Theor.\ Math.\ Phys.\  {\bf 2}, 253 (1998).
[hep-th/9802150].
}
\lref\GubserBC{
  S.~S.~Gubser, I.~R.~Klebanov and A.~M.~Polyakov,
  ``Gauge theory correlators from noncritical string theory,''
Phys.\ Lett.\ B {\bf 428}, 105 (1998).
[hep-th/9802109].
}
\lref\SimmonsDuffinGJK{
  D.~Simmons-Duffin,
  ``The Conformal Bootstrap,''
[arXiv:1602.07982 [hep-th]].
}
\lref\RychkovIQZ{
  S.~Rychkov,
  ``EPFL Lectures on Conformal Field Theory in D>= 3 Dimensions,''
[arXiv:1601.05000 [hep-th]].
}
\lref\PolandEPD{
  D.~Poland, S.~Rychkov and A.~Vichi,
  ``The Conformal Bootstrap: Theory, Numerical Techniques, and Applications,''
Rev.\ Mod.\ Phys.\  {\bf 91}, 015002 (2019).
[arXiv:1805.04405 [hep-th]].
}
\lref\AldayEUA{
  L.~F.~Alday, M.~Kologlu and A.~Zhiboedov,
  ``Holographic Correlators at Finite Temperature,''
[arXiv:2009.10062 [hep-th]].
}
\lref\ElShowkAG{
  S.~El-Showk and K.~Papadodimas,
  ``Emergent Spacetime and Holographic CFTs,''
JHEP {\bf 1210}, 106 (2012).
[arXiv:1101.4163 [hep-th]].
}
\lref\IliesiuZLZ{
  L.~Iliesiu, M.~Kologlu and D.~Simmons-Duffin,
  ``Bootstrapping the 3d Ising model at finite temperature,''
JHEP {\bf 1912}, 072 (2019).
[arXiv:1811.05451 [hep-th]].
}
\lref\KatzRLA{
  E.~Katz, S.~Sachdev, E.~S.~Sorensen and W.~Witczak-Krempa,
  ``Conformal field theories at nonzero temperature: Operator product expansions, Monte Carlo, and holography,''
Phys.\ Rev.\ B {\bf 90}, no. 24, 245109 (2014).
[arXiv:1409.3841 [cond-mat.str-el]].
}
\lref\PetkouYNM{
  A.~C.~Petkou and A.~Stergiou,
  ``Dynamics of Finite-Temperature Conformal Field Theories from Operator Product Expansion Inversion Formulas,''
Phys.\ Rev.\ Lett.\  {\bf 121}, no. 7, 071602 (2018).
[arXiv:1806.02340 [hep-th]].
}
\lref\ManentiWXS{
  A.~Manenti,
  ``Thermal CFTs in momentum space,''
JHEP {\bf 2001}, 009 (2020).
[arXiv:1905.01355 [hep-th]].
}
\lref\AharonySX{
  O.~Aharony, J.~Marsano, S.~Minwalla, K.~Papadodimas and M.~Van Raamsdonk,
  ``The Hagedorn - deconfinement phase transition in weakly coupled large N gauge theories,''
Adv.\ Theor.\ Math.\ Phys.\  {\bf 8}, 603 (2004).
[hep-th/0310285].
}
\lref\LashkariVGJ{
  N.~Lashkari, A.~Dymarsky and H.~Liu,
  ``Eigenstate Thermalization Hypothesis in Conformal Field Theory,''
J.\ Stat.\ Mech.\  {\bf 1803}, no. 3, 033101 (2018).
[arXiv:1610.00302 [hep-th]].
}
\lref\LashkariHWQ{
  N.~Lashkari, A.~Dymarsky and H.~Liu,
  ``Universality of Quantum Information in Chaotic CFTs,''
JHEP {\bf 1803}, 070 (2018).
[arXiv:1710.10458 [hep-th]].
}
\lref\SredneckiETH{
  M.~Srednicki,
  ``Chaos and quantum thermalization,''
Phys. Rev. E, {\bf 50} (1994), 888.
}

\lref\DeutschETH{
  J.~M.~Deutsch,
  ``Quantum statistical mechanics in a closed system,''
Phys. Rev. A, {\bf 43} (1991), 2046.
}
\lref\DeutschETHRev{
  J.~M.~Deutsch,
  ``Eigenstate Thermalization Hypothesis,''
 Rept.Prog.Phys., {\bf 81} (2018), 082001.
}

\lref\DunjkoETH{
  M.~Rigol, V.~Dunjko, M.~Olshanii, 
  ``Thermalization and its mechanism for generic isolated quantum systems,''
	Nature, {\bf 452} (2008), 854.
}
\lref\DAlessioRWT{
  L.~D'Alessio, Y.~Kafri, A.~Polkovnikov and M.~Rigol,
  ``From quantum chaos and eigenstate thermalization to statistical mechanics and thermodynamics,''
Adv.\ Phys.\  {\bf 65}, no. 3, 239 (2016).
[arXiv:1509.06411 [cond-mat.stat-mech]].
}

\lref\DelacretazCFK{
  L.~V.~Delacr\'etaz, T.~Hartman, S.~A.~Hartnoll and A.~Lewkowycz,
  ``Thermalization, Viscosity and the Averaged Null Energy Condition,''
JHEP {\bf 1810}, 028 (2018).
[arXiv:1805.04194 [hep-th]].
}

\lref\BasuKZO{
  P.~Basu, D.~Das, S.~Datta and S.~Pal,
  ``Thermality of eigenstates in conformal field theories,''
Phys.\ Rev.\ E {\bf 96}, no. 2, 022149 (2017).
[arXiv:1705.03001 [hep-th]].
}

\lref\FitzpatrickQMA{
  A.~L.~Fitzpatrick, J.~Kaplan, M.~T.~Walters and J.~Wang,
  ``Eikonalization of Conformal Blocks,''
JHEP {\bf 1509}, 019 (2015).
[arXiv:1504.01737 [hep-th]].
}
\lref\PappadopuloJK{
  D.~Pappadopulo, S.~Rychkov, J.~Espin and R.~Rattazzi,
  ``OPE Convergence in Conformal Field Theory,''
Phys.\ Rev.\ D {\bf 86}, 105043 (2012).
[arXiv:1208.6449 [hep-th]].
}
\lref\ChaiZGQ{
  N.~Chai, S.~Chaudhuri, C.~Choi, Z.~Komargodski, E.~Rabinovici and M.~Smolkin,
  ``Thermal Order in Conformal Theories,''
Phys.\ Rev.\ D {\bf 102}, no. 6, 065014 (2020).
[arXiv:2005.03676 [hep-th]].
}
\lref\MaxfieldRKN{
  H.~Maxfield,
  ``A view of the bulk from the worldline,''
[arXiv:1712.00885 [hep-th]].
}
\lref\ParnachevFNA{
  A.~Parnachev,
  ``Near Lightcone Thermal Conformal Correlators and Holography,''
[arXiv:2005.06877 [hep-th]].
}
\lref\CollierWEQ{
  S.~Collier, A.~Maloney, H.~Maxfield and I.~Tsiares,
  ``Universal dynamics of heavy operators in CFT$_{2}$,''
JHEP {\bf 2007}, 074 (2020).
[arXiv:1912.00222 [hep-th]].
}
\lref\AsplundCOA{
  C.~T.~Asplund, A.~Bernamonti, F.~Galli and T.~Hartman,
  ``Holographic Entanglement Entropy from 2d CFT: Heavy States and Local Quenches,''
JHEP {\bf 1502}, 171 (2015).
[arXiv:1410.1392 [hep-th]].
}
\lref\FitzpatrickZHA{
  A.~L.~Fitzpatrick, J.~Kaplan and M.~T.~Walters,
  ``Virasoro Conformal Blocks and Thermality from Classical Background Fields,''
JHEP {\bf 1511}, 200 (2015).
[arXiv:1501.05315 [hep-th]].
}
\lref\FitzpatrickFOA{
  A.~L.~Fitzpatrick, J.~Kaplan, M.~T.~Walters and J.~Wang,
  ``Hawking from Catalan,''
JHEP {\bf 1605}, 069 (2016).
[arXiv:1510.00014 [hep-th]].
}
\lref\FitzpatrickDLT{
  A.~L.~Fitzpatrick and J.~Kaplan,
  ``Conformal Blocks Beyond the Semi-Classical Limit,''
JHEP {\bf 1605}, 075 (2016).
[arXiv:1512.03052 [hep-th]].
}
\lref\AnousKSS{
  T.~Anous, T.~Hartman, A.~Rovai and J.~Sonner,
  ``Black Hole Collapse in the 1/c Expansion,''
JHEP {\bf 1607}, 123 (2016).
[arXiv:1603.04856 [hep-th]].
}
\lref\FitzpatrickIVE{
  A.~L.~Fitzpatrick, J.~Kaplan, D.~Li and J.~Wang,
  ``On information loss in AdS$_{3}$/CFT$_{2}$,''
JHEP {\bf 1605}, 109 (2016).
[arXiv:1603.08925 [hep-th]].
}
\lref\KrausNWO{
  P.~Kraus and A.~Maloney,
  ``A cardy formula for three-point coefficients or how the black hole got its spots,''
JHEP {\bf 1705}, 160 (2017).
[arXiv:1608.03284 [hep-th]].
}
\lref\FitzpatrickMJQ{
  A.~L.~Fitzpatrick and J.~Kaplan,
  ``On the Late-Time Behavior of Virasoro Blocks and a Classification of Semiclassical Saddles,''
JHEP {\bf 1704}, 072 (2017).
[arXiv:1609.07153 [hep-th]].
}
\lref\FaulknerHLL{
  T.~Faulkner and H.~Wang,
  ``Probing beyond ETH at large $c$,''
JHEP {\bf 1806}, 123 (2018).
[arXiv:1712.03464 [hep-th]].
}
\lref\BeskenBSU{
  M.~Besken, S.~Datta and P.~Kraus,
  ``Quantum thermalization and Virasoro symmetry,''
J.\ Stat.\ Mech.\  {\bf 2006}, 063104 (2020).
[arXiv:1907.06661 [hep-th]].
}
\lref\CollierWEQ{
  S.~Collier, A.~Maloney, H.~Maxfield and I.~Tsiares,
  ``Universal dynamics of heavy operators in CFT$_{2}$,''
JHEP {\bf 2007}, 074 (2020).
[arXiv:1912.00222 [hep-th]].
}
\lref\BriganteBQ{
  M.~Brigante, G.~Festuccia and H.~Liu,
  ``Inheritance principle and non-renormalization theorems at finite temperature,''
Phys.\ Lett.\ B {\bf 638}, 538 (2006).
[hep-th/0509117].
}
\lref\BelinHEA{
  A.~Belin and J.~de Boer,
  ``Random Statistics of OPE Coefficients and Euclidean Wormholes,''
[arXiv:2006.05499 [hep-th]].
}
\lref\PerlmutterIYA{
  E.~Perlmutter,
  ``Virasoro conformal blocks in closed form,''
JHEP {\bf 1508}, 088 (2015).
[arXiv:1502.07742 [hep-th]].
}
\lref\deBoerBOV{
  J.~de Boer and D.~Engelhardt,
  ``Remarks on thermalization in 2D CFT,''
Phys.\ Rev.\ D {\bf 94}, no. 12, 126019 (2016).
[arXiv:1604.05327 [hep-th]].
}
\lref\HeTXY{
  S.~He, F.~L.~Lin and J.~j.~Zhang,
  ``Dissimilarities of reduced density matrices and eigenstate thermalization hypothesis,''
JHEP {\bf 1712}, 073 (2017).
[arXiv:1708.05090 [hep-th]].
}
\lref\HeVYF{
  S.~He, F.~L.~Lin and J.~j.~Zhang,
  ``Subsystem eigenstate thermalization hypothesis for entanglement entropy in CFT,''
JHEP {\bf 1708}, 126 (2017).
[arXiv:1703.08724 [hep-th]].
}
\lref\ChenYZE{
  H.~Chen, C.~Hussong, J.~Kaplan and D.~Li,
  ``A Numerical Approach to Virasoro Blocks and the Information Paradox,''
JHEP {\bf 1709}, 102 (2017).
[arXiv:1703.09727 [hep-th]].
}
\lref\BrehmIPF{
  E.~M.~Brehm, D.~Das and S.~Datta,
  ``Probing thermality beyond the diagonal,''
Phys.\ Rev.\ D {\bf 98}, no. 12, 126015 (2018).
[arXiv:1804.07924 [hep-th]].
}
\lref\RomeroBermudezDIM{
  A.~Romero-Berm\'udez, P.~Sabella-Garnier and K.~Schalm,
  ``A Cardy formula for off-diagonal three-point coefficients; or, how the geometry behind the horizon gets disentangled,''
JHEP {\bf 1809}, 005 (2018).
[arXiv:1804.08899 [hep-th]].
}
\lref\HikidaKHG{
  Y.~Hikida, Y.~Kusuki and T.~Takayanagi,
  ``Eigenstate thermalization hypothesis and modular invariance of two-dimensional conformal field theories,''
Phys.\ Rev.\ D {\bf 98}, no. 2, 026003 (2018).
[arXiv:1804.09658 [hep-th]].
}
\lref\AnousYKU{
  T.~Anous and J.~Sonner,
  ``Phases of scrambling in eigenstates,''
SciPost Phys.\  {\bf 7}, 003 (2019).
[arXiv:1903.03143 [hep-th]].
}
\lref\LaiE{
  H.~H.~Lai and K.~Yang,
  ``Entanglement entropy scaling laws and eigenstate typicality in
free fermion systems,''
Phys. Rev. B {\bf 91}, no.\ 8, 081110 (2015).
[arXiv:1409.1224 [cond-mat]].
}
\lref\GuoPVI{
  W.~Z.~Guo, F.~L.~Lin and J.~Zhang,
  ``Note on ETH of descendant states in 2D CFT,''
JHEP {\bf 1901}, 152 (2019).
[arXiv:1810.01258 [hep-th]].
}
\lref\MaloneyHDG{
  A.~Maloney, G.~S.~Ng, S.~F.~Ross and I.~Tsiares,
  ``Thermal Correlation Functions of KdV Charges in 2D CFT,''
JHEP {\bf 1902}, 044 (2019).
[arXiv:1810.11053 [hep-th]].
}
\lref\MaloneyYRZ{
  A.~Maloney, G.~S.~Ng, S.~F.~Ross and I.~Tsiares,
  ``Generalized Gibbs Ensemble and the Statistics of KdV Charges in 2D CFT,''
JHEP {\bf 1903}, 075 (2019).
[arXiv:1810.11054 [hep-th]].
}
\lref\DymarskyLHF{
  A.~Dymarsky and K.~Pavlenko,
  ``Generalized Gibbs Ensemble of 2d CFTs at large central charge in the thermodynamic limit,''
JHEP {\bf 1901}, 098 (2019).
[arXiv:1810.11025 [hep-th]].
}
\lref\DymarskyETQ{
  A.~Dymarsky and K.~Pavlenko,
  ``Generalized Eigenstate Thermalization Hypothesis in 2D Conformal Field Theories,''
Phys.\ Rev.\ Lett.\  {\bf 123}, no. 11, 111602 (2019).
[arXiv:1903.03559 [hep-th]].
}
\lref\DymarskyIWX{
  A.~Dymarsky and K.~Pavlenko,
  ``Exact generalized partition function of 2D CFTs at large central charge,''
JHEP {\bf 1905}, 077 (2019), [JHEP {\bf 2019}, 077 (2020)].
[arXiv:1812.05108 [hep-th]].
}
\lref\BrehmFYY{
  E.~M.~Brehm and D.~Das,
  ``Korteweg–de Vries characters in large central charge CFTs,''
Phys.\ Rev.\ D {\bf 101}, no. 8, 086025 (2020).
[arXiv:1901.10354 [hep-th]].
}
\lref\DiFrancescoNK{
  P.~Di Francesco, P.~Mathieu and D.~Senechal,
  ``Conformal Field Theory,''
}
\lref\CaputaETA{
  P.~Caputa, J.~Simón, A.~\v Stikonas and T.~Takayanagi,
  ``Quantum Entanglement of Localized Excited States at Finite Temperature,''
JHEP {\bf 1501}, 102 (2015).
[arXiv:1410.2287 [hep-th]].
}
\lref\DasUAX{
  D.~Das, Y.~Kusuki and S.~Pal,
  ``Universality in asymptotic bounds and its saturation in $2$D CFT,''
JHEP {\bf 2104}, 288 (2021).
[arXiv:2011.02482 [hep-th]].
}
\lref\EngelsoyTSP{
  J.~Engels\"oy, J.~Larana-Aragon, B.~Sundborg and N.~Wintergerst,
  ``Operator thermalisation in d $>$ 2: Huygens or resurgence,''
JHEP {\bf 2009}, 103 (2020).
[arXiv:2007.00589 [hep-th]].
}
\lref\SabellaGarnierTSI{
  P.~Sabella-Garnier, K.~Schalm, T.~Vakhtel and J.~Zaanen,
  ``Thermalization/Relaxation in integrable and free field theories: an Operator Thermalization Hypothesis,''
[arXiv:1906.02597 [cond-mat.stat-mech]].
}
\lref\BukvaOLY{
  A.~Bukva, P.~Sabella-Garnier and K.~Schalm,
  ``Operator thermalization vs eigenstate thermalization,''
[arXiv:1911.06292 [cond-mat.stat-mech]].
}
\lref\DymarskyNTG{
  A.~Dymarsky, N.~Lashkari and H.~Liu,
  ``Subsystem ETH,''
Phys.\ Rev.\ E {\bf 97}, 012140 (2018).
[arXiv:1611.08764 [cond-mat.stat-mech]].
}

\Title{
\vbox{\baselineskip8pt
}}
{\vbox{
\centerline{Thermalization in Large-$N$ CFTs}
}}

\vskip.1in
 \centerline{
Robin Karlsson, Andrei Parnachev and Petar Tadi\' c \footnote{}{karlsson, parnachev, tadicp $@$ maths.tcd.ie  } } \vskip.1in
\centerline{\it 
School of Mathematics, Trinity College Dublin, Dublin 2, Ireland}

\vskip.4in \centerline{\bf Abstract}{ 
\noindent In $d$-dimensional CFTs with a large number of degrees of freedom an important set of  operators   consists
 of the stress tensor and its products, multi stress tensors.
Thermalization of such operators,  the equality between their expectation values  in heavy states and at finite temperature,
is equivalent to a universal  behavior of their OPE coefficients with a pair of identical heavy operators.
We verify this behavior in a number of examples which include holographic and 
free CFTs and provide a  bootstrap argument for the  general case.
In a free CFT we  check the thermalization of multi stress tensor operators directly
and also confirm the equality between the contributions of multi stress tensors to heavy-heavy-light-light correlators and 
 to the corresponding thermal light-light two-point functions
 by disentangling the contributions of  other light
operators.
Unlike  multi stress tensors, these light operators  violate the Eigenstate Thermalization Hypothesis
and do not thermalize.

}
 
\Date{February 2021}

\listtoc\writetoc
\vskip 1.0in \noindent

\newsec{Introduction and summary}

Holography  \refs{\MaldacenaRE\WittenQJ-\GubserBC} provides us with a  useful tool to study 
$d$-dimensional CFTs at large central charge $C_T$, especially when combined with  modern CFT techniques 
(see e.g.\ \refs{\RychkovIQZ\SimmonsDuffinGJK-\PolandEPD} for reviews).
One of the basic  objects in this setup is a Witten diagram with a single graviton exchange
which contributes to  four-point functions.
It can be decomposed into the conformal
blocks of the stress-tensor and of the double-trace operators made out of external fields  \HijanoZSA.

When a pair of the external operators denoted by $\OO_H$ is taken to be heavy, with the conformal dimension $\DH \sim C_T$,
and the other pair denoted by $\OO_L$ stays light, the resulting heavy-heavy-light-light (HHLL) correlator
describes a light probe interacting with a heavy state. 
In this case, operators which are comprised out of many stress tensors (multi stress tensor operators) contribute,
 together with the multi-trace operators involving $\OO_L$.
As we review below, the OPE coefficients of the scalar operators with a (unit-normalized) multi stress tensor operator $T^k_{\tau,s}$,
which contains $k$ stress tensors and has  twist $\tau$ and spin $s$,
scale like $\lambda_{\OO_\Delta \OO_\Delta T^k_{\tau,s}} \sim \Delta^k/ C_T^{k/2}$ for large $\Delta$.

The contribution of a given multi stress tensor operator to the HHLL four-point function $\la \OO_H \OO_L \OO_L \OO_H \ra$ can be compared to the
contribution of the same operator to the corresponding two-point function at finite temperature\foot{See \refs{\ElShowkAG\PappadopuloJK\KatzRLA\LashkariVGJ\LashkariHWQ\IliesiuFAO\GobeilFZY\DelacretazCFK\PetkouYNM\IliesiuZLZ\BanerjeeKJH\ManentiWXS\SabellaGarnierTSI\BukvaOLY\ChaiZGQ\DelacretazNIT\EngelsoyTSP-\AldayEUA} for 
some previous work on finite temperature conformal field theories in $d>2$.} $\beta^{-1}$, $\la \OO_L \OO_L \ra_\beta$.
In this paper we argue that they are the same in generic large-$C_T$ CFTs.
As we explain later, this means that  OPE coefficients of $T^k_{\tau,s}$ 
with the two  heavy operators $\OO_H$,
$\la \OO_H T^k_{\tau,s} \OO_H \ra$, are equal to their finite temperature expectation values, $\la T^k_{\tau,s} \ra_\beta$.
The relation between the inverse temperature $\beta$ and the conformal dimension $\DH$ is set by considering the stress tensor
($k=1, \tau=d-2, s=2$), but the equality between the 
thermal expectation values and the OPE coefficients for all other multi stress tensor operators is a nontrivial statement.
We call it ``the thermalization of the stress tensor sector'' \foot{We show this explicitly
for certain primary heavy operators $\OH$ in free CFTs. We also observe that other light operators do not satisfy the thermalization property 
that the stress tensor sector enjoys.
 }.
It is directly related to the Eigenstate Thermalization Hypothesis (ETH) \refs{\DeutschETH\SredneckiETH\DunjkoETH\DAlessioRWT-\DeutschETHRev}, as we review below.
Hence, we argue that all multi stress tensor operators in the large-$C_T$ CFTs satisfy the ETH.
In $d=2$ the ETH and thermalization have been studied in e.g.\  
\refs{\AsplundCOA\CaputaETA\FitzpatrickZHA\FitzpatrickFOA\FitzpatrickDLT\FitzpatrickIVE\AnousKSS\deBoerBOV\KrausNWO\FitzpatrickMJQ\LaiE\BasuKZO\HeVYF\ChenYZE\HeTXY\BrehmIPF\FaulknerHLL\RomeroBermudezDIM\HikidaKHG\GuoPVI\MaloneyHDG\MaloneyYRZ\DymarskyLHF\DymarskyIWX\BrehmFYY\DymarskyETQ\AnousYKU\DattaJEO\BeskenBSU\CollierWEQ-\DasUAX}.

Here we want to address the $d>2$ case.
In holographic theories CFT and bootstrap techniques provide  a lot of  data which indicates that the
thermalization of the stress tensor sector happens
\refs{\KulaxiziDXO\FitzpatrickZQZ\KarlssonQFI\LiTPF\KulaxiziTKD\FitzpatrickEFK\KarlssonDBD\LiZBA\KarlssonTXU\KarlssonGHX\LiDQM-\FitzpatrickYJB}.
Some of the OPE coefficients in holographic CFTs were computed using  two-point functions in a
black hole background \FitzpatrickZQZ\ -- these are thermal correlators according to the standard holographic dictionary.
It is also worth noting that the leading $\Delta$ behavior of the OPE
coefficients in holographic models  does not depend on the coefficients of the higher derivative terms in the bulk lagrangian \KarlssonGHX\
(this should not be confused with the universality of the OPE coefficients of the minimal-twist multi stress tensors \FitzpatrickZQZ).
Such a universality follows from the thermalization of the stress tensor sector as we discuss below.

A natural question is whether the thermalization of the stress tensor sector is just a property of holographic CFTs
or if it holds more generally.
In this paper we argue for the latter scenario.
We compute the OPE coefficients (and the thermal expectation values) for a number of multi stress tensor
operators in a free CFT and observe thermalization as well as universality of OPE coefficients. 
 We also provide a bootstrap argument for all CFTs with a large central charge.

The rest of the paper is organized as follows.
In Section 2, we begin by considering the thermalization of multi stress tensor operators $T^k_{\tau,s}$. 
The heavy state we consider is created by a scalar operator $\OH$ with dimension $\Delta_H\sim C_T$ and by thermalization of a multi stress tensor operator we mean\foot{Here we are suppressing the tensor structure.  Note that all terms scale like $C_T^{k/2}$ which
is consistent with $T^k_{\tau,s}$ being unit-normalized.}
\eqn\thermIntro{ \langle \OO_H |   T^k_{\tau,s}  | \OO_H\rangle  \Big|_{\Delta_H^{k}\over C_T^{k/2}}  
    = \lambda_{ \OH \OH T^k_{\tau,s}  }\Big|_{\Delta_H^{k}\over C_T^{k/2}} = \la T^k_{\tau,s} \ra_\beta,    
}
where the heavy state $| \OO_H\rangle$ on the sphere of unit radius is created by the operator $\OO_H$,
$\lambda_{\OH\OH T^k_{\tau,s}}$ are the OPE coefficients of $T^k_{\tau,s}$ in the $\OH\times\OH$ OPE and  $|_{\Delta_H^{k}/ C_T^{k/2}}$ 
means we keep only leading  terms that scale like 
$\Delta_H^{k}/ C_T^{k/2} \sim C_T^{k/2}$.
 In \thermIntro\ $\la T^k_{\tau,s} \ra_\beta$  is the one-point function on the sphere at finite temperature $\beta^{-1}$.
Note that the OPE coefficients involving the stress tensor  are fixed by the Ward identity, 
and hence eq.~\thermIntro\ for the stress tensor establishes a relation between the temperature $\beta^{-1}$ and $\DH$.
By the large-$C_T$ factorization\foot{See \YaffeVF\  for a general discussion of large-$N$ factorization and
\refs{\BrownYN,\DickensRE} and \ElShowkAG\ for the discussion
in the context of gauge theories and CFTs respectively. 
The factorization  holds in  adjoint models in the 't Hooft limit at finite temperature,
but there are counterexamples, like e.g. a direct product 
of low-$C_T$ CFTs. However the factorization of multi stress tensors would still apply in these models.},
 the thermal one-point functions of multi stress tensors can be related to the thermal one-point function of the stress tensor itself. Explicitly,
\eqn\thermFactIntro{
  \langle T^k_{\tau,s}\rangle_{\beta} = c_{\tau,s}^k(\langle T_{d-2,2}^1\rangle_{\beta})^k =  c_{\tau,s}^k( \lambda_{\OH\OH T^1_{d-2,2}} )^k,
}
where $c_{\tau,s}^k$ are theory-independent coefficients that appear because of the index structure  in $\langle T^k_{\tau,s}\rangle_{\beta}$.
In the second equality in \thermFactIntro\ we used \thermIntro\ for the stress tensor.
Note that \thermIntro\ and \thermFactIntro\ imply that the leading $\DH$ behavior of the multi stress tensor OPE coefficients
is universal, i.e.\ it does not depend on the theory\foot{This amounts to the large-$C_T$ factorization of 
correlators $\langle \OO_H | T_{\mu\nu}\ldots  T_{\alpha\beta} |\OO_H \rangle $ in heavy states. }.
We provide a bootstrap argument for this universality in all large-$C_T$ theories.
Also note that \thermFactIntro\ is written for multi-trace operators $T^k_{\tau,s} $ which do not contain derivatives, but
 the presence of derivatives does not affect the statement of universality.

In Section 3, we check the universality by computing a number of the multi stress tensor OPE coefficients
in a free $SU(N)$ adjoint scalar theory in $d=4$  dimensions. 
We compare the leading $\DH$ behavior  in the free
theory with  results from holography/bootstrap and find  perfect agreement in all cases listed below.
After fixing the coefficients for the stress tensor case in Section 3.1, we look at the first
nontrivial case, $T^2_{4,4}$ in Section 3.2.
Section 3.3 is devoted to the  double stress tensor with two derivatives,
$T^2_{4,6}$.
This is an operator whose finite temperature expectation value vanishes in the large 
volume limit (on the plane), but is finite on the sphere.
In Section 3.4 we consider minimal twist multi stress tensors of the type $T^k_{2 k,2 k}$.
Section 3.5 is devoted to multi stress tensors with non-minimal twist, $T^2_{6,2}$ and $T^2_{8,0}$.

In Section 4, we verify that \thermIntro\ holds in the free adjoint scalar theory for a variety of operators.
In this section we again consider $d=4$, but in addition, take the infinite volume limit.
This is for technical reasons -- it is easier to compute a finite temperature expectation value
on the plane than on the sphere.
We spell out the index structure in \thermIntro\ in detail and go over all the examples discussed in the previous section.
In addition, we discuss some triple stress tensor operators.

We continue in Section 5 by studying thermal two-point functions in the free adjoint scalar model in $d=4$. 
By decomposing the correlator into thermal blocks we read off the product of thermal one-point functions and the OPE coefficients for several operators of low dimension and observe agreement with the results of
Sections 3 and 4.
Due to the presence of multiple operators with the same dimension and spin, we have to solve a mixing problem to find which operators contribute to the thermal two-point function.
 
In Section 6 we  explain the relation between our results and the Eigenstate Thermalization Hypothesis. 
We observe that unlike multi stress tensors, other light operators explicitly violate the Eigenstate Thermalization Hypothesis
and do not thermalize. We end with a discussion in Section 7.
 
Appendices A, B, and C contain explicit calculations of OPE coefficients while in Appendices D and E thermal one-point functions are calculated. In Appendix F we review the statement that the thermal one-point functions of multi-trace operators with derivatives vanish on $S^1\times {\bf R}^{d-1}$. In appendix G we study a free scalar in two dimensions and calculate
thermal two-point functions of  certain quasi-primary operators. In Appendix H we  consider a free scalar vector model in four dimensions. 
Appendix I discusses the factorization of multi-trace operators in the large volume limit.

%
%
%
%
%
%
\newsec{Thermalization and universality}
In the following we consider large-$C_T$ CFTs on a $(d-1)$-dimensional sphere of radius $R$, which
we set to unity for most of this section.
As reviewed in \KarlssonGHX, the stress tensor sector of conformal four-point functions
consists of the contributions of the stress tensor and all its composites (multi stress tensors).
The HHLL correlators we consider involve two heavy operators inserted at $x^{0}_{E}=\pm\infty$
and two light operators inserted on the Euclidean cylinder, with angular separation $\varphi$ 
and time separation $x_{E}^{0}$. 
The correlator in a heavy state (the HHLL correlator on the cylinder) is related to the correlator on the plane by a conformal transformation
\eqn\corr{    \la \OO_H|\OO(x_{E}^{0},\varphi) \OO(0)|\OO_H\ra = \lim_{x_4\to\infty}x_4^{2\DH}(z \zbar)^{-\Delta/2}\la \OO_H(x_4) \OO(1) \OO(z,\zbar) \OO_H(0) \ra,    }  
where the cross-ratios $(z, \zbar)$ on the plane are related to the  coordinates $(x_{E}^{0},\varphi)$ via
\eqn\zzbar{   z= e^{-{x_{E}^{0}} -i\varphi}, \qquad  \zbar= e^{ -{x_{E}^{0}}+i\varphi}.   }

The stress tensor sector of the HHLL correlator is given by
\eqn\defStressTensorSector{
  \GG(z,\zbar) = \lim_{x_4\to\infty} x_4^{2\DH}\langle \OH(x_4)\OO(1)\OO(z,\zbar)\OH(0)\rangle \Big|_{\rm multi\ stress \ tensors}
}
and can be expanded in conformal blocks 
\eqn\tChExp{
  \GG(z,\zbar) = {1\over [(1-z)(1-\zbar)]^\Delta}\sum_{T^{k}_{\tau,s}} P^{(HH,LL)}_{T^k_{\tau,s}} g_{\tau,s}^{(0,0)}(1-z,1-\zbar),
}
where $\tau,s,k$ label the twist, spin, and multiplicity of multi stress tensors.
We are interested in the double scaling limit where the central charge and the dimension of $\OH$ are large,
$C_T,\DH \rightarrow \infty$ with their ratio $\mu \propto \DH/C_T$ fixed.
In this limit the products of the OPE coefficients which appear in \tChExp\ are given by
\eqn\prodope{  P^{(HH,LL)}_{T^k_{\tau,s}} = \left(-{1\over2}\right)^s\lambda_{ \OO \OO T^{k}_{\tau,s}  }\lambda_{ \OH \OH T^k_{\tau,s}  }\Bigg|_{\left({\DH\over C_T}\right)^{k}},   }
where we only keep the leading, $\left({\Delta_H\over \sqrt{C_T}   }\right)^{k}$  term in the OPE coefficients  $\lambda_{ \OH \OH T^k_{\tau,s}  }$, but retain
 all terms in the OPE coefficients of the light operators $\lambda_{ \OO \OO T^k_{\tau,s}  }$.
The contribution of the conformal family of a  multi stress operator $T^k_{\tau,s}$ to the HHLL correlator is therefore
\eqn\hhlltk{   \la \OO_H| \OO(x^0_E,\varphi) \OO(0) |\OO_H\ra |_{T^k_{\tau,s}}  =  { P^{(HH,LL)}_{T^k_{\tau,s}}g_{\tau,s}^{(0,0)}(1-z,1-\zbar)\over  [\sqrt{z\zbar}(1-z)(1-\zbar)]^{\Delta}}.
}

We now consider these CFTs at finite temperature $\beta^{-1}$.
To isolate the contribution of the conformal family associated with $T^k_{\tau,s}$, we can write the thermal correlator as
\eqn\ftcorr{\eqalign{
     \la \OO(x^0_E,\varphi) \OO(0)&\ra_\beta  =  {1\over Z(\beta)} \sum_i e^{-\beta \Delta_i}   \la \OO_i| \OO(x^0_E,\varphi) \OO(0) |\OO_i\ra\cr
     =&{1\over {[\sqrt{z \zbar}(1-z)(1-\zbar)]^\Delta}}\sum_{T^{k}_{\tau,s}}  \left(-{1\over 2}\right)^s\lambda_{ \OO \OO  T^k_{\tau,s}  }  g_{\tau,s}^{(0,0)}(1-z,1-\zbar) \  \la T^k_{\tau,s} \ra_\beta\cr
     &+\ldots,
}}
where 
\eqn\tkbeta{  \la T^k_{\tau,s} \ra_\beta = {1\over Z(\beta)} \sum_i e^{-\beta \Delta_i}      \lambda_{\OO_i \OO_i   T^k_{\tau,s} }
                          }
is the finite temperature one-point function on the sphere of the $T^k_{\tau,s} $ operator and the dots denote contributions from other operators. 
In \tkbeta\ $Z(\beta)$ is the partition function and the sum runs over all operators, including descendants\foot{The corresponding conformal blocks can be obtained in the usual way by applying the quadratic conformal Casimir and solving the resulting
differential equation \DolanHV.}.
Note that 
\eqn\deff{    \la T^k_{\tau,s} \ra_\beta = \beta^{-(\tau +s)} f^k_{\tau,s}(\beta).  }
Here and below the indices are suppressed 
(see e.g. \IliesiuZLZ\ for the explicit form) and $f^k_{\tau,s}(\beta) \sim C_T^{k/2}$ is a theory-dependent nontrivial function of $\beta$ which approaches a constant $f^k_{\tau,s}(0)$ in the large volume ($\beta\rightarrow0$) limit.

Consider the thermalization of the stress tensor sector:
\eqn\thermalization{   \langle \OO_H |   T^k_{\tau,s}  | \OO_H\rangle\Big|_{\Delta_H^{k}\over C_T^{k/2}} 
= \lambda_{ \OH \OH T^k_{\tau,s}  }\Big|_{\Delta_H^{k}\over C_T^{k/2}}   = \la T^k_{\tau,s} \ra_\beta.    }
Note that  $T^k_{\tau,s} $ is unit-normalized, so all terms in \thermalization\ scale like $C_T^{{k/2}}$.
Eq.\  \thermalization\ implies the equality between \hhlltk\ and the corresponding term in \ftcorr.
Note that the left-hand side of \thermalization\ is a function of the energy density while the right-hand side is a function of temperature.
The relationship is fixed by considering   the stress tensor case: the corresponding 
function $f^1_{d-2,2}(\beta)$ is determined by the free energy on the sphere (see Section 6).

In the following, we will first discuss the case where the multi stress operators $T^k_{\tau,s}$ do not have any derivatives inserted,
and then show that the derivatives do not change the conclusions.
Assuming large-$C_T$ factorization, the leading $C_T$ behavior of $\la T^k_{\tau,s} \ra_\beta$
on the sphere  is determined by that of the stress tensor.
Schematically, 
\eqn\tkbetaf{  \la T^k_{\tau,s} \ra_\beta =  c^k_{\tau,s} \ ( \la T^1_{d-2,2} \ra_\beta )^k +\ldots,
}
where $c^k_{\tau,s}$ are numerical coefficients, which depend on $k, \tau,s$, but are independent
of the details of the theory and the dots stand for terms subleading in $C_T^{-1}$.
By combining  \tkbetaf\ and \thermalization , one can formulate a universality condition
\eqn\universality{    \lambda_{ \OH \OH T^k_{\tau,s}}\Big|_{\Delta_H^{k}\over C_T^{k/2}}  =  c^k_{\tau,s} \  (  \lambda_{ \OH \OH T^1_{d-2,2}}  )^k=    c^k_{\tau,s} \left({{d}\over{1-d}}\right)^{k} {\Delta_H^k \over C_T^{k\over2} },
          }
where the last equality follows from the stress tensor Ward identity for the three-point function which fixes $\lambda_{ \OH \OH T^1_{d-2,2}} $ ($T^1_{d-2,2}$ here is unit-normalized). In other words, thermalization and large-$C_T$ factorization imply that
the leading $\Delta^k/C_T^{k/2}$ behavior of the multi stress tensor OPE coefficients is completely fixed
and given by \universality\ in all large-$C_T$ CFTs.

In the paragraph above we considered multi stress tensor operators that did not contain any derivatives in them.
However, the story largely remains the same when the derivatives are included, as long as their number does not
scale with $C_T$.
Indeed, the three-point function  involving the stress-tensor with added derivatives, $\p_\alpha\ldots \p_\beta T_{\mu\nu}$
still behaves like $ \lambda_{ \OH \OH \p_\alpha\ldots \p_\beta T_{\mu\nu} }\simeq \DH/ \sqrt{C_T} $ up to a theory-independent coefficient.
Hence,  \universality\ still  holds, provided thermalization and  large-$C_T$ factorization
hold on the sphere. 

Note that due to conformal invariance, correlators on the sphere depend on $R$ only through the ratio $\beta/R$.
 Moreover, in the large volume limit, factors of $R$ need to drop out of \hhlltk\ and \ftcorr\ to have a well defined limit. 
 To see this we use that $(1-z)\to 0$ and $(1-\zbar)\to 0$ when $R\to \infty$ and the conformal blocks behave as (see e.g.\ \PolandEPD)
\eqn\blockRlimit{\eqalign{
	g_{\tau,s}^{(0,0)}(1-z,1-\zbar) &\sim \NN_{d,s}[(1-z)(1-\zbar)]^{\tau+s\over 2}C^{(d/2-1)}_s\Big({(1-z)+(1-\zbar)\over 2\sqrt{(1-z)(1-\zbar)}}\Big) \cr
	&\sim \NN_{d,s}{|x|^{\tau+s}\over R^{\tau+s}}C^{(d/2-1)}_s\Big({x_E^0\over|x|}\Big),
}
}
where $|x|=\sqrt{(x_E^0)^2+{\bf x}^2}$, $C^{(d/2-1)}_s({x_E^0\over|x|})$ is a Gegenbauer polynomial and $\NN_{d,s}={s!\over (d/2-1)_s}$. Including the factor $[(1-z)(1-\zbar)]^{-\Delta}$ from \hhlltk\ in \blockRlimit\ this agrees with the thermal block on $S^1\times {\bf R}^{d-1}$ in \IliesiuFAO. Now from the thermalization of the stress tensor we will find in the large volume limit that 
\eqn\opeScalLargeR{
	{\DH\over C_T} \propto \Big({R\over \beta}\Big)^d,
}
and from \universality\ and \blockRlimit\ it follows that
\eqn\stressLarge{
	g_{\tau,s}^{(0,0)}(1-z,1-\zbar)\lambda_{ \OO \OO T^k_{\tau,s}}\lambda_{ \OH \OH T^k_{\tau,s}}\Big|_{\Delta_H^{k}\over C_T^{k}} \propto R^{d k-(\tau+s)}\beta^{-d k}.
}
The dimension of multi stress tensors $T^{k}_{\tau,s}$ is given by $\tau+s=d k+n$ where $n=0,2,\ldots$. Therefore, the only multi stress tensors that contribute in the large volume limit have dimensions $d k$. Restoring $R$ in \hhlltk-\ftcorr\ and inserting \stressLarge\ one finds that $R$ drops out in the large volume limit. The correct dependence $\beta^{-(\tau+s)}$ from \deff\ in the $R\to \infty$ limit is also recovered in \hhlltk\ using \stressLarge. The multi stress tensor operators that contribute in the large volume limit are therefore of the schematic form $T_{\mu_1\nu_1}T_{\mu_2\nu_2}\cdots T_{\mu_k\nu_k}$ with arbitrarily many contractions and no derivatives. 

In holographic theories thermalization and the Wilson line prescription for the correlator allows one to compute
the universal part of the OPE coefficients (see \refs{\KulaxiziTKD,\ParnachevFNA} for  explicit computations in the $d=4$ case).
It is also easy to check explicitly that the universality \universality\ holds for holographic theories with a Gauss-Bonnet gravitational coupling added.
While the statement was shown to be true for the leading twist OPE coefficients in \FitzpatrickZQZ, it was not immediately obvious for multi stress tensors
of non-minimal twist. 
Some such OPE coefficients were computed in \refs{\FitzpatrickZQZ,\KarlssonGHX}.  (See e.g. eqs.\ (5.48), (5.51), (5.52), (5.57) and (D.1)-(D.5)  in \KarlssonGHX).
Indeed, the leading $\Delta^k/C_T^{k/2}$ behavior of these OPE coefficients is independent of the Gauss-Bonnet coupling.

What about a general large-$C_T$ theory? We first consider the OPE coefficients of double-stress tensors. To this end, consider the four point function\foot{This correlator for finite $\Delta$ was recently considered in holographic CFTs with $\Delta_{\rm gap}\gg1$ and $\Delta\ll\Delta_{\rm gap}$ in \FitzpatrickYJB .} $\langle \OO T_{\mu\nu}T_{\rho\sigma}\OO\rangle$ where $\OO$ is a scalar operator with scaling dimension $\Delta$. In the direct channel $\OO\times\OO\to \OO'\to T_{\mu\nu}\times T_{\rho\sigma}$ for finite $\Delta$ and large $C_T$, the leading contribution in the large-$C_T$ limit comes from the identity operator $\OO\times\OO\to {\bf 1}\to T_{\mu\nu}\times T_{\rho\sigma}$. The subleading contributions in the direct-channel are due to single trace operators as well as double trace operators made out of the external operators of the schematic form $T^{2}_{\tau ,s}$ and $[\OO \OO]_{n,l} = :\OO \pa^{2n}\pa_{1}\ldots\pa_{l}\OO:$. The exchange of the identity operator is reproduced in the cross-channel $\OO\times T_{\mu\nu}\to [\OO T_{\alpha\beta}]_{n,l} \to \OO \times T_{\rho\sigma}$ by mixed double-trace operators $[\OO T_{\alpha\beta}]_{n,l}$ with OPE coefficients fixed by the MFT \refs{\FitzpatrickDM\FitzpatrickYX-\KomargodskiEK}. 
The subleading contributions in $1/C_T$ are  then due to corrections to the anomalous dimension and OPE coefficients of $[\OO T_{\alpha\beta}]_{n,l}$ and single trace operators  in the $\OO\times T_{\mu\nu}$ OPE. An important example of the latter is the exchange of the single trace operator $\OO$, whose contribution is universally fixed by the stress tensor Ward identity to be $(\lambda_{\OO T^{1}_{d-2,2}\OO})^2\propto \Delta^2/ C_T$ times the conformal block. This gives a universal contribution to $\lambda_{\OO\OO T^2_{\tau,s}}$ as was also noted  in \FitzpatrickYJB. 

We now want to consider the case where $\Delta\sim C_T$ and study the OPE coefficients of the double-stress tensor operators in the $\OO\times\OO$ OPE. Firstly, note that the contribution from $T^2_{\tau,s}$ to the four-point function expanded in the direct channel is proportional to $\lambda_{\OO\OO T^2_{\tau,s}}\lambda_{TT T^2_{\tau,s}}$. The OPE coefficients $\lambda_{TTT^2_{\tau,s}}$ are fixed by the MFT and are independent of $\Delta$ and therefore the dependence on the scaling dimension comes solely from the OPE coefficients $\lambda_{\OO\OO T^2_{\tau,s}}$. In the cross-channel, we analyze two kinds of contributions: from the exchanged operator $\OO$ and from all other operators $\OO'\neq \OO$. From the operator $\OO$ we get a universal contribution to the OPE coefficients in the direct channel $\lambda_{\OO\OO T^{2}_{\tau,s}}$, that we denote by $\lambda^{(1)}_{\OO\OO T^{2}_{\tau,s}}$. This contribution is universal since it only depends on $(\lambda_{\OO T^{1}_{d-2,2}\OO})^2\propto \Delta^{2}/C_T$ in the cross-channel, which is fixed by the Ward identity. The contributions from other operators $\OO'$ to the same OPE coefficient will be denoted by $\lambda^{(2)}_{\OO\OO T^{2}_{\tau,s}}$, such that $\lambda_{\OO\OO T^{2}_{\tau,s}}=\lambda^{(1)}_{\OO\OO T^{2}_{\tau,s}}+ \lambda^{(2)}_{\OO\OO T^{2}_{\tau,s}}$. Note that it also follows from the stress tensor Ward identity that the only scalar primary that appears in the cross-channel is $\OO$. The operator $\OO'$ therefore necessarily has spin $s\neq0$. 

To prove universality we need to show that $\lambda^{(2)}_{\OO\OO T^2_{\tau,s}}\ll \Delta^2/C_T$ in limit $1\ll \Delta \propto C_T$ by studying the $\Delta$ dependence of the OPE coefficients $\lambda_{\OO T^{1}_{d-2,2}\OO'}$ in the cross-channel. 
For operators $\OO'$, such that $\Delta_{\OO'}\ll\Delta$, we expect that these OPE coefficients are heavily suppressed.
 It would be interesting to understand if one could put a general bound on the contribution of these operators in the cross-channel in any large-$C_T$ theory.
 On the other hand,  assuming  thermalization, the OPE coefficients due to operators $\OO'$ such that $\Delta_{\OO'}\sim \Delta$ have been calculated in \DelacretazNIT . The obtained results are in agreement with our expectation, namely, these OPE coefficients are suppressed in $1\ll \Delta \propto C_T$ limit.
Additionally, in the cross-channel we have double-trace operators $[\OO T_{\alpha\beta}]_{n,l}$, whose OPE is fixed by the MFT and it does not get $\Delta$-enhanced. Note that in holographic theories with a large gap, $1\ll\Delta_{\rm gap}\ll C_T$, in the regime $\Delta\ll\Delta_{\rm gap}$ there is a coupling $\lambda_{\OO T^{1}_{d-2,2}T^{1}_{d-2,2}}$ which scales like  ${1\over \Delta_{\rm gap}^2\sqrt{C_T}}$ and its contribution to multi-stress tensor OPE coefficients was studied in \FitzpatrickYJB. This is different from the regime considered in this paper where $\Delta\gg \Delta_{\rm gap}$.

One can iteratively extend the argument given here to multi stress tensors operators (with $k>2$) by considering multi stress tensors as external operators. For example, to argue
the universality of $\lambda_{\OO\OO T^{3}_{\tau,s}}$ one may consider $\la\OO T^{1}_{d-2,2} T^{2}_{\tau,s}\OO \ra$. 
The bootstrap argument above can be applied again by using the fact that OPE coefficients $\lambda_{\OO\OO T^{2}_{\tau ,s}}$ are universal, and the
OPE coefficients $\lambda_{\OO T^{2}_{\tau,s}\OO'}$ are again expected to be subleading.


\newsec{OPE coefficients in the free  adjoint scalar model}
In this section we consider a four-dimensional theory of a free scalar in the adjoint  representation of $SU(N)$, see  \refs{\SundborgUE\HaggiManiRU\AharonySX\BriganteBQ\AmadoPGY-\AmadoKGR} for related work. 
 The relation between $N$ and the central charge $C_T$ in this theory is \OsbornCR\
\eqn\ct{C_T={4\over 3}(N^2-1),
}
and we consider the large-$N$ (large-$C_T$) limit.
The propagator for the scalar field ${\phi^i}_j$  is given by
\eqn\fundprop{
  \langle {\phi^i}_j(x){\phi^k}_l(y)\rangle = \left({\delta^{i}}_l{\delta^{k}}_j-{1\over N}{\delta^{i}}_j{\delta^{k}}_l\right){1\over|x-y|^2}.
}
A single trace scalar operator with dimension $\Delta$ is given by 
\eqn\singlop{
  \OO_{\Delta}(x) = {1\over\sqrt{\Delta}N^{\Delta\over 2}}:Tr(\phi^\Delta):(x), 
}
where $:\ldots:$ denotes the oscillator normal ordering 
 and the normalization is fixed by
\eqn\normaliz{\la\OO_{\Delta}(x)\OO_{\Delta}(y) \ra = {1\over{|x-y|^{2\Delta}}}.
}

The CFT data that we compute in this section are the OPE coefficients of multi stress tensors in the $\OO_\Delta\times\OO_\Delta$ OPE. 
Assuming we can take $\Delta\to \DH \sim C_T$, the large-$\Delta$ limit of these OPE coefficients is shown to be universal. 
One may worry  that for $\DH\sim C_T$ we can no longer trust the planar expansion, but, as we show in Appendix C,
 the large-$\Delta$ limit of the planar result yields the correct expression even for $\DH\sim C_T$.
 
\subsec{Stress tensor}
The stress tensor operator is given by 
\eqn\stress{
  T_{\mu\nu}(x) = {1\over3\sqrt{C_T}}:Tr\left(\pa_\mu\phi\pa_\nu\phi-{1\over 2}\phi\pa_\mu\pa_\nu\phi-({\rm trace})\right):(x),
} 
where the normalization 
\eqn\stressnorm{
  \langle T^{\mu\nu}(x)T_{\rho\sigma}(0)\rangle = {1\over |x|^8}\left({I^{(\mu}}_\rho(x) {I^{\nu)}}_\sigma (x)-({\rm traces})\right),
}
with ${I^\mu}_\nu(x):={\delta^\mu}_\nu-{2x^\mu x_\nu\over |x|^2}$.
The OPE coefficient is fixed by the stress tensor Ward identity to be 
\eqn\TOPE{
  \lambda_{\OO_\Delta\OO_{\Delta}T^1_{2,2}} =-{{4\Delta}\over{3\sqrt{C_{T}}}}
  .
}
\noindent It is also useful to find \TOPE\ using Wick contractions since an analogous calculation will be necessary for multi stress tensors.
We do this explicitly in Appendix A.

\subsec{Double-stress tensor with minimal twist}
In this section we study the minimal-twist composite operator made out of two stress tensors
\eqn\doublestress{
  (T^2)_{\mu\nu\rho\sigma}(x) = {1\over\sqrt{2}}:T_{(\mu\nu}T_{\rho\sigma)}:(x)-({\rm traces}),
}
with the normalization  
\eqn\doublestressnorm{
  \langle(T^2)^{\mu\nu\rho\sigma}(x)(T^2)_{\kappa\lambda\delta\omega}(0)\rangle = {1\over |x|^{16}}\left({I^{(\mu}}_\kappa{I^{\nu}}_\lambda{I^{\rho}}_\delta{I^{\sigma)}}_\omega-({\rm traces})\right).
}

Consider the following three-point function 
\eqn\tsqthreept{
  \langle\OO_\Delta(x_1)\OO_{\Delta}(x_2)(T^{2})_{\mu\nu\rho\sigma}(x_3)\rangle = {\lambda_{\OO_\Delta\OO_\Delta T^2_{4,4}}\over |x_{12}|^{2\Delta-4}|x_{13}|^{4}|x_{23}|^{4}}\left(Z_\mu Z_\nu Z_\rho Z_\sigma-({\rm traces})\right),
}
where $Z^\mu={x_{13}^\mu\over |x_{13}|^2}-{x_{12}^\mu\over |x_{12}|^2}$. It is shown in Appendix A that the OPE coefficient $\lambda_{\OO_\Delta\OO_\Delta T^2_{4,4}}$ is given at leading order in the large-$C_T$ limit by
\eqn\opeTSq{
  \lambda_{\OO_\Delta\OO_\Delta T^2_{4,4}} = {8\sqrt{2}\Delta(\Delta-1)\over 9C_T}.
}
\noindent Evaluating $P^{(HH,LL)}_{T^2_{4,4}}$ defined by \prodope\ in the large-$\Delta$ limit\foot{By the large-$\Delta$ limit, we strictly speaking mean 
$1\ll \Delta\ll    C_T$.
However in this paper we often extrapolate this to the $\Delta\sim C_T$ regime.}, we obtain
\eqn\holdt{\eqalign{P^{(HH,LL)}_{T^2_{4,4}}&=\left(-{1\over 2}\right)^{4}\lambda_{\OO_{H}\OO_{H} T^2_{4,4}}\lambda_{\OO_{\Delta}\OO_{\Delta} T^2_{4,4}}\Bigg|_{\left({\DH\over C_T}\right)^{2}}\cr
&={8\over 81}{{\Delta_{H}^2}\over{C_T^2}} \left(\Delta^2 +\OO\left(\Delta\right)\right)= \mu^2\left({{\Delta^2}\over{28800}}+\OO(\Delta)\right),
}}
\noindent where we use the following relation 
\eqn\defmu{
\mu = {160\over 3}{\DH \over C_T}.}
\noindent The  result \holdt\ agrees with the leading behavior 
of the corresponding OPE coefficients computed using holography in \FitzpatrickZQZ\ and bootstrap in \refs{\KulaxiziTKD,\KarlssonDBD}.


\subsec{Double-stress tensor with minimal twist and spin $s=6$}
We consider double-stess tensor operator with two (uncontracted) derivatives inserted
\eqn\tdertt{(T^2)_{\mu\nu\rho\sigma\eta\kappa}(x)={1\over{2\sqrt{182}}}:\left(T_{(\mu\nu}\pa_{\rho}\pa_{\sigma}T_{\eta\kappa)}(x)-{7\over 6}\left(\pa_{(\rho}T_{\mu\nu}\right)\left(\pa_{\sigma}T_{\eta\kappa)}\right)(x)-({\rm traces})\right):.
}
\noindent Using the conformal algebra eq.~(C.2), it is straightforward to check that this operator is primary. It is unit-normalized such that
\eqn\doublestressnormtd{
  \langle(T^2)^{\mu\nu\rho\sigma\eta\kappa}(x)(T^2)_{\alpha\beta\gamma\delta\xi\epsilon}(0)\rangle = {1\over |x|^{20}}\left({I^{(\mu}}_\alpha {I^{\nu}}_\beta{I^{\rho}}_\gamma{I^{\sigma}}_\delta {I^{\eta}}_\xi {I^{\kappa)}}_\epsilon  -({\rm traces})\right).
}
By a calculation similar to those summarized in Appendix A, we observe that the OPE coefficient of $(T^2)_{\mu\nu\rho\sigma\eta\kappa}$ in the $\OO_\Delta\times\OO_\Delta$ OPE is given at leading order in the large-$C_T$ limit by
\eqn\lambdatdertt{\lambda_{\OO_\Delta\OO_\Delta T^2_{4,6}} ={8\over 3}\sqrt{{2\over 91}}{{\Delta(\Delta -1)}\over{C_T}}.
}
\noindent Evaluating $P^{(HH,LL)}_{T^2_{4,6}}$, defined by \prodope , in the large-$\Delta$ limit, we obtain
\eqn\holdttd{\eqalign{P^{(HH,LL)}_{T^2_{4,6}}&=\left(-{1\over 2}\right)^{6}\lambda_{\OO_{H}\OO_{H} T^2_{4,6}}\lambda_{\OO_{\Delta}\OO_{\Delta} T^2_{4,6}}\Bigg|_{\left({\DH\over C_T}\right)^{2}}\cr
&={2\over 819}{{\Delta_{H}^2}\over{C_T^2}} \left(\Delta^2 +\OO\left(\Delta\right)\right)= \mu^2\left({{\Delta^2}\over{1164800}}+\OO(\Delta)\right).
}}
\noindent The  result \holdttd\ agrees with the leading behavior 
of the corresponding OPE coefficients computed using holography in \FitzpatrickZQZ\ and bootstrap in \refs{\KulaxiziTKD,\KarlssonDBD}.

\subsec{Minimal-twist multi stress tensors}
We now consider multi stress tensors  $T^k_{2 k, 2 k}$. Just like the double stress tensor ($k=2$), we show that these have universal OPE coefficients in the large-$\Delta$ limit for any $k$. 

Consider the unit-normalized minimal-twist multi stress tensor operator given by 
\eqn\multistress{
  (T^k)_{\mu_1\mu_2\ldots\mu_{2k}}(x) = {1\over \sqrt{k!}}:T_{(\mu_1\mu_2}T_{\mu_3\mu_4}\cdots T_{\mu_{2k-1}\mu_{2k})}:(x)-({\rm traces}).
}
The OPE coefficient of $(T^k)_{\mu_1\mu_2\ldots\mu_{2k}}$ in the $\OO_\Delta\times\OO_\Delta$ OPE, in the large-$C_T$ limit is given by\foot{See
Appendix A for detailed computations of similar OPE coefficients.}
\eqn\OPEMultiStress{
  \lambda_{\OO_\Delta\OO_\Delta T^k_{2k,2k}} = \left(-{4\over 3}\right)^k{1\over\sqrt{k!}C_T^{k/2}}{\Gamma(\Delta+1)\over\Gamma(\Delta-k+1)}.
}
\noindent First, we write $P^{(HH,LL)}_{T^3_{6,6}}$, defined by \prodope , in the large-$\Delta$ limit. We obtain this OPE coefficient from \OPEMultiStress\ for $k=3$,
\eqn\pholttt{\eqalign{P^{(HH,LL)}_{T^3_{6,6}}&=\left(-{1\over 2}\right)^{6}\lambda_{\OO_{H}\OO_{H} T^3_{6,6}}\lambda_{\OO_{\Delta}\OO_{\Delta} T^3_{6,6}}\Bigg|_{\left({\DH\over C_T}\right)^{3}}\cr
&={32\over 2187}{{\Delta_{H}^3}\over{C_T^3}} \left(\Delta^3 +\OO\left(\Delta^2\right)\right)= \mu^3\left({{\Delta^3}\over{10368000}}+\OO(\Delta^2)\right).
}}
\noindent 
The  result \pholttt\ agrees with the leading behavior 
of the corresponding OPE coefficients computed using holography in \FitzpatrickZQZ\ and bootstrap in \KarlssonDBD.


Additionally, we consider the OPE coefficient $P^{(HH,LL)}_{T^k_{2k,2k}}$ in the large-$\Delta$ limit for general $k$,
\eqn\pkgen{\eqalign{P^{(HH,LL)}_{T^k_{2k,2k}}&=\left(-{1\over 2}\right)^{2k}\lambda_{\OO_{H}\OO_{H} T^k_{2k,2k}}\lambda_{\OO_{\Delta}\OO_{\Delta} T^k_{2k,2k}}\Bigg|_{\left({\DH\over C_T}\right)^{k}}\cr
&={1\over{k!}}\left({2\over 3}\right)^{2k}{{\Delta_{H}^k}\over{C_T^k}} \left(\Delta^k +\OO\left(\Delta^{k-1}\right)\right)= \mu^k\left({{\Delta^k}\over{120^{k} k!}}+\OO(\Delta^{k-1})\right).
}}
\noindent If we consider the limit $1-\zbar \ll 1-z \ll 1$, such that $\mu(1-\zbar)(1-z)^3$ is held fixed, only operators $T^k_{2k,2k}$ contribute to the heavy-heavy-light-light four-point function given by eq.~\defStressTensorSector. The conformal blocks of $T^k_{2k,2k}$ in this limit are given by
\eqn\conblocks{g^{(0,0)}_{2k,2k}(1-z,1-\zbar)\approx(1-\zbar)^{k}(1-z)^{3k},
}
\noindent and we can sum all contributions in eq.~\tChExp\ explicitly to obtain
\eqn\summing{\GG(z,\zbar)\approx{1\over{\left((1-z)(1-\zbar)\right)^{\Delta}}}e^{{{\mu \Delta}\over 120}(1-\zbar)(1-z)^{3}}.
}
\noindent Notice that the term in the exponential is precisely the stress-tensor conformal block in the limit $1-\zbar \ll 1-z \ll 1$ times its OPE coefficient. Therefore, the OPE coefficients \OPEMultiStress\ imply the exponentiation of stress-tensor conformal block.
We conclude that these OPE coefficients are the same as the ones computed using holography and bootstrap in the limit of large $\Delta$.

\subsec{Double-stress tensors with non-minimal twist}
So far we have shown that the minimal-twist multi stress tensor OPE coefficients are universal in the limit of large $\Delta$. In this subsection, we extend this to show that the simplest non-minimal twist double-stress tensors also have universal OPE coefficients at large $\Delta$. 

The subleading twist double-stress tensor with twist $\tau=6$ is of the schematic form $:{T^{\mu}}_\alpha T^{\alpha\nu}:$ and has dimension $\Delta=8$ and spin $s=2$. It is given by 
\eqn\defSub{
  (T^{2})_{\mu\nu}(x) = {1\over \sqrt{2}}:{T_{\mu\alpha}} {T^{\alpha}}_{\nu}:(x)-({\rm trace}).
}
The normalization in \defSub\ is again chosen such that $(T^{2})_{\mu\nu}$ is unit-normalized, see Appendix B for details. 

The OPE coefficient of $(T^{2})_{\mu\nu}$ in the $\OO_\Delta\times\OO_\Delta$ OPE is found from the three-point function in the large-$C_T$ limit, for details see Appendix B,
\eqn\threeptSub{
  \langle \OO_\Delta(x_1)\OO_\Delta(x_2)(T^{2})^{\mu\nu}(x_3)\rangle = {4\sqrt{2}\Delta(\Delta-1)\over {9C_T}}{Z^\mu Z^\nu-({\rm trace})\over |x_{12}|^{2\Delta-6}|x_{13}|^{6} |x_{23}|^{6}},
}
from which we read off the OPE coefficient
\eqn\opeCoeffSub{
  \lambda_{\OO_\Delta\OO_\Delta T^{2}_{6,2}} = {4\sqrt{2}\Delta(\Delta-1)\over9C_T}.
}
\noindent Evaluating $P^{(HH,LL)}_{T^2_{6,2}}$, defined by \prodope , in the large-$\Delta$ limit, we obtain
\eqn\pholst{\eqalign{P^{(HH,LL)}_{T^2_{6,2}}&=\left(-{1\over 2}\right)^2\lambda_{\OO_H \OO_H T^{2}_{6,2}}\lambda_{\OO_{\Delta}\OO_{\Delta}T^{2}_{6,2}}\Bigg|_{\left({\DH\over C_T}\right)^{2}}\cr
&={8\over 81}{\DH^2 \over{C_T^2}}\left(\Delta^2 +\OO\left({\Delta}\right)\right)=\mu ^2\left({{\Delta ^2}\over{28800}}+\OO(\Delta)\right).
}}
\noindent The result \pholst\ agrees with the leading behavior 
of the corresponding OPE coefficients computed using holography in \FitzpatrickZQZ\ and bootstrap in \KarlssonGHX.

We further consider the scalar double-stress tensor with $\Delta=8$ and spin $s=0$ which is given by 
\eqn\scalarDouble{
	(T^{2})(x) = {1\over 3\sqrt{2}} :T_{\mu\nu}T^{\mu\nu}:(x). 
}
\noindent The three point function $\langle \OO_\Delta(x_1)\OO_\Delta(x_2)(T^{2})(x_3)\rangle$ is found in Appendix B to be 
\eqn\threeptScalarDouble{
	\langle \OO_\Delta(x_1)\OO_\Delta(x_2)(T^{2})(x_3)\rangle = {2\sqrt{2}\Delta(\Delta-1)\over 9C_T}{1\over |x_{12}|^{2\Delta-8}|x_{13}|^8|x_{23}|^8},
}
from which we read off the OPE coefficient
\eqn\opeCoeffScalarDouble{
	\lambda_{\OO_\Delta\OO_\Delta T^2_{8,0}} = {2\sqrt{2}\Delta(\Delta-1)\over 9C_T}.
}
\noindent We write $P^{(HH,LL)}_{T^{2}_{8,0}}$ in the large-$\Delta$ limit
\eqn\pholsc{\eqalign{P^{(HH,LL)}_{T^{2}_{8,0}}&=\lambda_{\OO_H\OO_H T^2_{8,0}}\lambda_{\OO_{\Delta}\OO_{\Delta}T^2_{8,0}}\Bigg|_{\left({\DH\over C_T}\right)^{2}} \cr
&= {8\over 81}{\DH^2\over C_T^2}\left(\Delta^2+\OO(\Delta)\right)=\mu ^2\left({{\Delta ^2}\over{28800}}+\OO(\Delta)\right).
}}
\noindent The  result \pholsc\ agrees with the leading behavior 
of the corresponding OPE coefficients computed using holography in \FitzpatrickZQZ\ and bootstrap in \KarlssonGHX.

%
%
%
%

\newsec{Thermal one-point functions in the free adjoint scalar model}
In this section we explicitly show that multi stress tensor operators thermalize in the free theory by
 calculating the thermal one-point function of some of these operators on $S^1\times {\bf{R}}^{3}$. 
One-point functions of primary symmetric traceless operators  at finite temperature are fixed by symmetry 
up to a dimensionless coefficient $b_\OO$  (see e.g.\ \refs{\ElShowkAG,\IliesiuFAO})
\eqn\onept{
  \langle \OO_{\mu_1\cdots\mu_{s_\OO}}\rangle_\beta = {b_\OO\over \beta^{\Delta_\OO}}\left(e_{\mu_1}\cdots e_{\mu_{s_\OO}}-({\rm traces})\right).
}
Here $e_{\mu}$ is a unit vector along the thermal circle. 

To compare the thermal one-point functions and OPE coefficients from the previous section, we need to derive a relation between ${\Delta_H\over C_T}$ and the  temperature\foot{See also Section 6 and Appendix D for alternative derivations.} $\beta^{-1}$. Here $\Delta_H\sim N^2$ refers to the scaling dimension of a heavy operator $\OO_H$ with OPE coefficients given by the large-$\Delta$ limit of those obtained in Section 3. One can relate the inverse temperature $\beta$ to the parameter $\mu={160\over3}{\DH\over C_T}$ using the
Stefan-Boltzmann's law $E/{\rm vol}(S^{3}) = N^2\pi^{2}/30\beta^4$. 
The energy of the state $E$ is related to its conformal dimension $\Delta$ via 
$E=\Delta/R$.
One can then use ${\rm vol}(S^{3})=2\pi^2 R^3$ and the relation between $N$ and $C_T$ given by \ct, to find
\eqn\muu{\eqalign{\mu ={160\over 3}{\Delta_H \over C_T}={160 \over 3}  E\  {R\over C_T}={8\over3}\Big({\pi R\over \beta}\Big)^4.
}}

\subsec{Stress tensor}
The thermal one-point function for the stress tensor $T^{1}_{2,2}=T_{\mu\nu}$ is calculated in Appendix D
 where we find that $b_{T^1_{2,2}}$ is given by 
\eqn\bT{
  b_{T^1_{2,2}} = -{2\pi^4 N\over 15\sqrt{3}}. 
}
Using  \muu\ and \bT\ one arrives at
\eqn\stressTensorTherm{
  b_{T^{1}_{2,2}}\beta^{-4}=\lambda_{\OH\OH T^1_{2,2}}.
}

\subsec{Double-stress tensor with minimal twist}
In this section we calculate the thermal one-point function of the double-stress tensor operator with $\tau=4$ and spin $s=4$. 
The operator  is written explicitly in \doublestress . 
The leading  contribution to the thermal one-point function of $(T^2)_{\mu\nu\rho\sigma}$ follows from the large-$N$ factorization and is given by 
\eqn\tsqonept{\eqalign{
  \langle (T^2)_{\mu\nu\rho\sigma} \rangle_\beta &= {1\over \sqrt{2}}\langle T_{(\mu\nu}\rangle_\beta\langle T_{\rho\sigma)}\rangle_\beta -({\rm traces})\cr
  &={2\sqrt{2}\pi^8N^2\over675\beta^8}\left(e_\mu e_\nu e_\rho e_\sigma-({\rm traces})\right).
}}
Using the relation \muu\ and the OPE coefficient \opeTSq, we observe the thermalization of this operator, 
\eqn\thermTT{
  b_{T^2_{4,4}}\beta^{-8}= \lambda_{\OO_H \OO_H T^2_{4,4}}\Big|_{\Delta_H^2\over C_T}.
}

\subsec{Minimal-twist multi stress tensors}
Consider now multi stress tensors $T^k_{2k,2k}$ with twist $\tau=2k$ and spin $s=2k$. 
We show that these operators thermalize for any $k$ by calculating their thermal one-point functions:
\eqn\thermalOneMulti{
  \langle (T^k)_{\mu_1\mu_2\ldots\mu_{2k}}\rangle_{\beta} = {b_{T^k_{2k,2k}}\over\beta^{4k}}(e_{\mu_1}e_{\mu_2}\cdots e_{\mu_{2k}}-({\rm traces})),
}
where the leading behavior of $b_{T^k_{2k,2k}}$ follows from the large-$N$ factorization:
\eqn\bmulti{ \eqalign{
  b_{T^k_{2k,2k}}&= {1\over \sqrt{k!}}(b_{T^{1}_{2,2}})^k = {(-{2\over 5})^kN^k \pi^{4k}\over 3^{3k\over 2}\sqrt{k!}}.
}}
Eqs. \muu\ and \bmulti\ may be combined to yield
\eqn\thermmulti{
  b_{T^k_{2k,2k}}\beta^{-4k} = \lambda_{\OO_H \OO_H T^k_{2k,2k}}\Big|_{\Delta_H^{k}\over C_T^{k/2}}.
}

\subsec{Double-stress tensors with non-minimal twist}
The subleading twist double-stress tensor is of the schematic form $:{T^{\mu}}_\alpha T^{\alpha\nu}:$ and has twist $\tau=6$ and spin $s=2$. 
The explicit form can be found in \defSub.
The  leading term in the thermal one-point function is given by
\eqn\thermaloneSub{\eqalign{
  \langle (T^{2})^{\mu\nu}\rangle_\beta &= {1\over\sqrt{2}}\langle T^{\mu\alpha}\rangle_\beta\langle {T^{\nu}}_\alpha\rangle_\beta -({\rm trace})\cr
  &={b^2_{T^1_{2,2}}\over 2\sqrt{2}\beta^8}(e^{\mu}e^{\nu}-{1\over 4}\delta^{\mu\nu})\cr
  &={\sqrt{2}N^2\pi^8\over {675\beta^8}}(e^{\mu}e^{\nu}-{1\over 4}\delta^{\mu\nu}),
}}
therefore,
\eqn\thermaloneSubExpl{
  b_{T^2_{6,2}} = {\sqrt{2}N^2\pi^8\over 675}.
}
Taking the large-$\Delta$ limit of the OPE coefficient in \opeCoeffSub\ and substituting \muu, we observe thermalization, 
\eqn\thermSub{
  b_{T^{2}_{6,2}}\beta^{-8}= \lambda_{\OO_H \OO_H T^{2}_{6,2}}\Big|_{\Delta_H^{2}\over C_T}.
}

We further consider the scalar double-stress tensor with $\tau=8$ and $s=0$ which is given by \scalarDouble . The thermal one-point function for this operator is
\eqn\oneptScalarDouble{\eqalign{
  \langle (T^{2})\rangle_\beta &={1\over 3\sqrt{2}} \la T_{\mu\nu}\ra_{\beta} \la T^{\mu\nu}\ra_{\beta}\cr
  &= {1\over 3\sqrt{2}}{3\over 4} b_{T^1_{2,2}}^2\beta^{-8} = {\pi^8N^2\over 675\sqrt{2}\beta^{8}},
}}
where the factor of ${3\over4}$ in the first line comes from the index contractions. Hence, 
\eqn\ttscb{
b_{T^2_{8,0}}={\pi^8N^2\over 675\sqrt{2}}.
}
Using  \ttscb ,  \opeCoeffScalarDouble\ and \muu, we again observe thermalization,
\eqn\thermScalarDouble{
  b_{T^2_{8,0}}\beta^{-8}=\lambda_{\OO_H\OO_H T^{2}_{8,0}}\Big|_{{\Delta^2_H \over C_T}}.
}

\subsec{Triple-stress tensors with non-minimal twist}

We consider the triple stress tensors with  $\tau=8, s=4$ and $\tau=10,s=2$. The unit-normalized triple stress tensor with $\tau=8$ can be written as
\eqn\tnmt{(T^3)_{\mu\nu\rho\sigma}(x)={1\over \sqrt{3}}\left(:T_{(\mu\nu}T_{\rho |\alpha|}{T^{\alpha}}_{\sigma)}:(x)-({\rm traces})\right),
}
\noindent where $|\alpha|$ denotes that index $\alpha$ is excluded from the symmetrization.  The thermal one-point function follows from large-$N$ factorization
\eqn\tnmtth{\eqalign{\langle (T^3)_{\mu\nu\rho\sigma} \rangle_{\beta}&={1\over \sqrt{3}}\left( \langle T_{(\mu\nu}\rangle_{\beta} \langle T_{\rho |\alpha|}\rangle_{\beta} \langle{T^{\alpha}}_{\sigma)}\rangle_{\beta} -({\rm traces})\right)  \cr
&=  {1\over 2\sqrt{3}}{b_{T^1_{2,2}}^{3}\over \beta^{12}}\left(e_{\mu}e_{\nu}e_{\rho}e_{\sigma}-({\rm traces}) \right)\cr
&= -{{4 \pi ^{12} N^3}\over{30375 \beta ^{12}}}\left(e_{\mu}e_{\nu}e_{\rho}e_{\sigma}-({\rm traces}) \right) ,
}}
\noindent therefore,
\eqn\btt{b_{T^3_{8,4}}=-{{4 \pi ^{12} N^3}\over{30375}}.
}

The OPE coefficient of the operator with same quantum numbers ($\Delta=12$, $s=4$)
is calculated holographically and  is given by (D.1) in \KarlssonGHX . In the large-$\Delta$ limit it can be written as
\eqn\opett{P^{(HH,LL)}_{T^3_{8,4}}=\left(-{1\over 2}\right)^{4}\lambda_{\OO_{\Delta} \OO_{\Delta} T^3_{8,4}}\lambda_{\OO_{H}\OO_{H} T^3_{8,4}}\Bigg|_{\left({\DH\over C_T}\right)^{3}}={64\over 2187}{{\Delta_H^3 \Delta^3}\over{C_{T}^3}}+\OO(\Delta^2).
}

\noindent Now, one can easily read-off $\lambda_{\OO_{\Delta} \OO_{\Delta} T^3_{8,4}}$ in the large-$\Delta$ limit
\eqn\ltt{\lambda_{\OO_{\Delta} \OO_{\Delta} T^3_{8,4}}=-{{32 \Delta ^3}\over{27 \sqrt{3} {C_T}^{3/2}}}+\OO(\Delta^2)=-{{4 \Delta ^3}\over{9 N^3}}+\OO(\Delta^2),
}
\noindent where we use the relation between central charge $C_{T}$ and $N$ given by \ct .  Using \muu\ one can obtain
\eqn\termtt{{b_{T^3_{8,4}} \beta^{-12}}=\lambda_{\OO_{H} \OO_{H} T^3_{8,4}}\Big|_{\Delta^{3}_H\over C_T^{3/2}}.
}

We also consider the triple stress tensors with quantum numbers $\Delta=12$ and $s=2$. There are two linearly independent such operators that schematically can be written as $:T_{\alpha\beta}T^{\alpha\beta}T_{\mu\nu}:$ and $:T_{\mu\alpha}T^{\alpha\beta}T_{\beta\nu}:$. We write the following linear combinations of these operators
\eqn\lkone{(T^{3})_{\mu\nu}(x) ={1\over {10\sqrt{2}}} \left(:T_{\alpha\beta}T^{\alpha\beta}T_{\mu\nu}:(x) + 4:T_{\mu\alpha}T^{\alpha\beta}T_{\beta\nu}:(x) - ({\rm trace})\right),
}
\eqn\lkonetwo{(\tilde{T}^{3})_{\mu\nu}(x)={7\over 20}\left(:T_{\alpha\beta}T^{\alpha\beta}T_{\mu\nu}:(x) -{12\over 7}:T_{\mu\alpha}T^{\alpha\beta}T_{\beta\nu}:(x) - ({\rm trace})\right).
}

\noindent Both $(T^{3})_{\mu\nu}$ and $(\tilde{T}^{3})_{\mu\nu}$ are unit-normalized and their overlap vanishes in the large-$N$ limit
\eqn\ovelapp{\langle (T^{3})_{\mu\nu}(x) (\tilde{T}^{3})^{\rho\sigma}(y) \rangle = \OO(1/N^2).
}

\noindent The thermal one-point functions of these operators, obtained by large-$N$ factorization, in the large-$N$ limit are given by
\eqn\thertt{\eqalign{\langle(T^{3})_{\mu\nu}\rangle_{\beta}& = -\sqrt{{2\over 3}}{{N^3 \pi^{12}}\over {10125 \beta^{12}}}\left(e_{\mu}e_{\nu}-({\rm trace})\right),\cr
\langle(\tilde{T}^{3})_{\mu\nu}\rangle_{\beta} &= \OO(N),
}}
\noindent therefore, 
\eqn\bbs{\eqalign{b_{T^{3}_{10,2}}&=-\sqrt{{2\over 3}}{{N^3 \pi^{12}}\over {10125}},\cr
b_{\tilde{T}^{3}_{10,2}}&=0.
}}

The holographic OPE coefficient of the operator with the same quantum numbers ($\Delta=12$, $s=2$), with external scalar operators is given by (5.57) in \KarlssonGHX. In the large-$\Delta$ limit it can be written as
\eqn\hope{P^{(HH,LL)}_{T^3_{10,2}}=\left(-{1\over 2}\right)^2\lambda_{\OO_{\Delta} \OO_{\Delta} T^3_{10,2}}\lambda_{\OO_{H}\OO_{H} T^3_{10,2}}\Bigg|_{\left({\DH\over C_T}\right)^{3}}={32\over 729}{{\Delta_{H}^{3}\Delta^{3}}\over{C_T^{3}}} + \OO(\Delta^{2}).
}
\noindent We can read-off $\lambda_{\OO_{\Delta} \OO_{\Delta} T^3_{10,2}}$:
\eqn\ltwoo{\lambda_{\OO_{\Delta} \OO_{\Delta} T^3_{10,2}}=-{{8\sqrt{2}}\over{27}}{\Delta^{3}\over C_{T}^{3/2}}+\OO(\Delta^2) = -{\sqrt{2}\over{3\sqrt{3}}}{{\Delta^{3}}\over N^3} + \OO(\Delta^2).
}

\noindent Again, using  \muu, one can confirm that this operator thermalizes
\eqn\termtt{{b_{T^3_{10,2}} \beta^{-12}}= \lambda_{\OO_{H} \OO_{H} T^3_{10,2}}\Big|_{\Delta^{3}_{H}\over C_T^{3/2}}.
}

%
%

\newsec{Thermal two-point function and block decomposition}
In this section we study the thermal two-point function $\langle \OO_\Delta\OO_\Delta\rangle_\beta$ and decompose it in thermal blocks. 
We determine the contributions of a few low-lying operators, including the stress tensor $T^{1}_{2,2}$ and the double stress tensor $T^{2}_{4,4}$.
They exactly match the corresponding OPE coefficients and thermal expectation values computed in  previous sections.
Due to the presence of multiple operators with equal scaling dimension and spin, there is a  mixing problem which we solve explicitly in a few cases. 
Related  appendices include  Appendix F, where we review the statement that the thermal one-point functions of multi-trace operators with derivatives vanish on $S^1\times {\bf R}^{d-1}$
and Appendix G, where we consider  two-dimensional thermal two-point functions.
In Appendix H we do a similar analysis for the vector model in four dimensions.

\subsec{Thermal two-point function of a single trace scalar operator}
The correlator at finite temperature $\beta^{-1}$ in the free theory can be calculated by Wick contractions
using the propagators on $S^1\times {\bf R}^{3}$. Explicitly, the two-point function at finite temperature is given by\foot{Here and below we assume that $\Delta>4$. We further drop the disconnected term $\langle \OO_\Delta\rangle_\beta^2\sim N^2$.}
\eqn\thermaltwopt{\eqalign{
	\langle \OO_{\Delta}(x)\OO_\Delta(0)\rangle_\beta= \tilde g(x^0_E,|{\bf x}|)^\Delta+{\pi^4\Delta(\Delta-2)\over 9\beta^4} \tilde g(x^0_E,|{\bf x}|)^{\Delta-2}+\ldots,
}}
where 
\eqn\ggfunc{\eqalign{
	\tilde g(x^0_E,|{\bf x}|) &= \sum_{m=-\infty}^{\infty}{1\over (x^0_E+m\beta)^2+{\bf x}^2}\cr
	&={\pi\over 2\beta|{\bf x}|}\Big[ {\rm Coth}\Big({\pi\over\beta}(|{\bf x}|-ix^0_E)\Big)+{\rm Coth}\Big({\pi\over\beta}(|{\bf x}|+ix^0_E)\Big)\Big].
}}
The dots in \thermaltwopt\ contain contributions due to further self-contractions which will not be important below\foot{These terms will be proportional to $\beta^{-2a} \tilde g(x^0_E,|{\bf x}|)^{\Delta-a}$, with $a\geq 4$. When decomposed into thermal blocks, these will not affect the operators with dimension $\Delta<8$ or $\Delta=8$ with non-zero spin $s\neq0$.}.
Taking the $\beta\to\infty$ limit of \thermaltwopt\ we can read off the decomposition of the two-point function in terms of thermal conformal blocks on $S^1\times {\bf R}^3$ with coordinates $x=(x^{0}_{E},{\bf x})$. 

Following \IliesiuFAO, if $|x|=\sqrt{(x^0_{E})^2+{\bf x}^2}\leq \beta$ the two-point function can be evaluated using the OPE: 
\eqn\twoptOPE{
	\langle \OO_{\Delta}(x)\OO_\Delta(0)\rangle_\beta = \sum_{\OO}\lambda_{\OO_\Delta\OO_\Delta\OO}|x|^{\tau-2\Delta} x_{\mu_1}\cdots x_{\mu_{s_\OO}}\langle \OO^{\mu_1\cdots\mu_{s_\OO}}\rangle_\beta,
}
where $\lambda_{\OO_\Delta\OO_\Delta\OO}$ is the OPE coefficient, $\tau$ and $s_\OO$ is the twist and spin of $\OO$, respectively. Using \onept\ together with \twoptOPE, the two-point function on $S^1\times {\bf R}^3$ can be organized in the following way \IliesiuFAO:
\eqn\decomp{
	\langle \OO_{\Delta}(x)\OO_\Delta(0)\rangle_\beta = \sum_{\OO_{\tau,s}\in\OO_\Delta\times\OO_\Delta}{a_{\OO_{\tau,s}}\over\beta^{\Delta_\OO}} 
	     {1\over |x|^{2\Delta-\tau +s}}C^{(1)}_{s}\left({x^{0}_E\over |x|}\right),
} 
where we sum over primary operators $\OO_{\tau, s}$, with twist $\tau$ and spin $s$, appearing in the OPE $\OO_\Delta\times\OO_\Delta\sim\OO_{\tau,s}+\ldots$. 
In \decomp\ $C^{(1)}_{s}(x^{0}_E/ |x|)$ is a Gegenbauer polynomial
which, together with a factor of $|x|^{-2\Delta+\tau -s}$, forms a thermal conformal block in $d=4$ dimensions
and
the coefficients $a_{\OO_{\tau,s}}$ are given by 
\eqn\block{ a_{\OO_{\tau,s}} = \left({1\over 2}\right)^{s}\lambda_{\OO_{\Delta}\OO_{\Delta}\OO_{\tau,s}}b_{\OO_{\tau,s}} .
}

Expanding \thermaltwopt\ for $\beta\to\infty$ one finds: 
\eqn\twoptexp{\eqalign{
	\langle \OO_{\Delta}(x)\OO_\Delta(0)\rangle_\beta =& {1\over |x|^{2\Delta}}\Big[1+{\pi^2\Delta\over 3\beta^2}|x|^2\cr
	&+{\pi^4\Delta\over 90\beta^4}|x|^2(3{\bf x}^2(5\Delta-9)+(15\Delta-19) (x^0_E)^2)+\OO(\beta^{-6})\Big].
}
} 
From the expansion \twoptexp, we can read off the coefficients $a_{\tau',s'}:=\sum_{\OO_{\tau',s'}}a_{\OO_{\tau',s'}}$ where we sum over all operators with twist $\tau'$ and spin $s'$:
\eqn\acoeff{\eqalign{
	a_{2,0} &= {\pi^2\Delta\over 3},\cr
	a_{4,0} &= {\pi^4\Delta(3\Delta-5)\over 18},\cr
	a_{2,2} &= {\pi^4\Delta\over 45}.
}}
For future reference, expanding \thermaltwopt\ to $\OO({1\over\beta^8})$ one finds 
\eqn\adimsixandeight{\eqalign{
	a_{2,4} &= {2\pi^6\Delta\over 945},\cr
	a_{4,4} &= {\pi^8\Delta(\Delta-1)\over 1050}.
}}
Note that due to the mixing of operators with the same twist and spin, $a_{\tau,s}$ generically contains the contribution from multiple operators. In the following section we calculate the OPE coefficients and thermal one-point functions of operators which are not multi stress tensors but contribute to \acoeff\ and \adimsixandeight. 

\subsec{CFT data of scalar operators with dimensions two and four}
We explicitly calculate the thermal one-point functions $\langle \OO\rangle_\beta=b_\OO \beta^{-\Delta_\OO}$ and OPE coefficients $\lambda_{\OO_\Delta\OO_\Delta\OO}$ for scalar operators $\OO$ with twist $\tau'=2$ and $\tau'=4$ using Wick contractions. 
This is done to find which operators contribute to the thermal two-point function and to resolve a mixing problem. 

For $\tau'=2$ there is only one such operator, the single trace operator $\OO_2(x)={1\over \sqrt{2}N}:Tr(\phi^2):(x)$ given in \singlop. The OPE coefficient is found by considering the three-point correlator
\eqn\opeCoeffdimtwo{
  \langle \OO_\Delta(x_1)\OO_\Delta(x_2)\OO_2(x_3)\rangle = {\lambda_{\OO_\Delta\OO_\Delta\OO_2}\over |x_{12}|^{2\Delta-2}|x_{13}|^{2}|x_{23}|^2}. 
}
The three-point function is calculated in Appendix A, in the large-$N$ limit, and it is given by
\eqn\opeCoeffdimtwoRes{
  \langle \OO_\Delta(x_1)\OO_\Delta(x_2)\OO_2(x_3)\rangle = {\sqrt{2}\Delta\over N}{1\over |x_{12}|^{2\Delta-2}|x_{13}|^{2}|x_{23}|^2},
}
and therefore $\lambda_{\OO_\Delta\OO_\Delta\OO_2}={\sqrt{2}\Delta\over N}$ to leading order in $1/N$. To calculate the thermal one-point function $\propto\langle Tr(\phi^2)\rangle_\beta$, we include self-contractions, i.e.\ contractions of fundamental fields within the same composite operator separated by a distance $m\beta$ along the thermal circle for $m\neq0$ and integer. Explicitly, the one-point function of $\OO_{2}$ is given by
\eqn\dimtwoonept{\eqalign{
  \langle \OO_2(x)\rangle_\beta&= {1\over\sqrt{2}N}\sum_{m\neq 0}{N^2\over (m\beta)^2}= {\pi^2 N\over 3\sqrt{2}\beta^2},
}}
\noindent therefore,
\eqn\bbtwo{b_{\OO_2}={\pi^2 N\over 3\sqrt{2}}.
}
The contribution to the thermal two-point function $a_{\OO_2}$ is found using \opeCoeffdimtwoRes\ and \bbtwo
\eqn\atwo{
  a_{2,0}=b_{\OO_{2}}\lambda_{\OO_\Delta\OO_\Delta\OO_2}={\pi^2\Delta\over3}.
}
This agrees with \acoeff\  which was obtained from the thermal two-point function. 

We now continue with scalar operators of twist four. There are two such linearly independent  operators appearing in the $\OO_{\Delta}\times\OO_\Delta$ OPE. In order to construct an orthonormal basis, consider the following single and double trace operators:
\eqn\dimfourops{\eqalign{
  \OO_{4}(x) &= {1\over 2N^2}:Tr(\phi^4):(x),\cr
  \OO_{4,{\rm DT}}(x) &= {1\over 2\sqrt{2}N^2}:Tr(\phi^2)Tr(\phi^2):(x).
}}
We further construct the operator $\tilde{\OO}_4$ that has vanishing overlap with $\OO_{4,{\rm DT}}(x)$ as follows:
\eqn\tildeOfour{
  \tilde{\OO}_4=\NN\Big[\OO_{4}-c_{\OO_{4}\OO_{4,{\rm DT}}}\OO_{4,{\rm DT}}\Big],
}
with $\NN$ a normalization constant and $c_{\OO_{4}\OO_{4,{\rm DT}}}$ is the overlap defined by 
\eqn\ccoef{
  \langle \OO_{4}(x)\OO_{4,{\rm DT}}(y)\rangle = {c_{\OO_{4}\OO_{4,{\rm DT}}}\over |x-y|^8}. 
}
Explicit calculation gives $c_{\OO_{4}\OO_{4,{\rm DT}}}={2\sqrt{2}\over N}$ and $\NN={1\over\sqrt{2}}$ in the large-$N$ limit, and the scalar dimension four operator orthogonal to the double trace operator $\OO_{4,{\rm DT}}$ is therefore
\eqn\tildeOfourfinal{
  \tilde{\OO}_4 = {1\over\sqrt{2}}\Big[\OO_{4}-{2\sqrt{2}\over N}\OO_{4,{\rm DT}}\Big]. 
}
Note that even though the second term in \tildeOfour\ is suppressed by $1/N$, it can still contribute to the thermal two-point function due to the  scaling of OPE coefficients and one-point function of a $k$-trace operator $\OO^{(k)}$:
\eqn\scaling{\eqalign{
  b_{\OO^{(k)}} &\sim N^k,\cr
  \lambda_{\OO_\Delta\OO_\Delta\OO^{(k)}}&\sim {1\over N^k},
}}
in the limit $N\to\infty$.

The one-point function and the OPE coefficient for $\OO_4$ is found analogously to that of $\OO_2$ in the large-$N$ limit
\eqn\oSTfourdata{\eqalign{
  &b_{\OO_4} = {\pi^4N\over 9},\cr
  &\lambda_{\OO_\Delta\OO_\Delta\OO_4} = {2\Delta\over N}.
}}

Consider now the double trace operator given in \dimfourops. The one-point function factorizes in the large-$N$ limit:
\eqn\oneptDoubletrace{\eqalign{
  \langle \OO_{4,{\rm DT}}(x)\rangle_\beta &= {1\over\sqrt{2}}(\langle\OO_2(x)\rangle_\beta)^2\cr
  &={\pi^4N^2\over 18\sqrt{2}\beta^4}.
}}
Likewise, the OPE coefficient can be computed in the large-$N$ limit (see Appendix A)
\eqn\opeDoubletracedimfour{
  \lambda_{\OO_\Delta\OO_\Delta\OO_{4,{\rm DT}}} = {\sqrt{2}\Delta(3\Delta-5)\over N^2}. 
}
Consider now the thermal one-point function of $\tilde{\OO}_4$ in \tildeOfourfinal
\eqn\tildefouronept{\eqalign{
  \langle \tilde{\OO}_4\rangle_{\beta}=& {1\over\sqrt{2}\beta^4}\Big[b_{\OO_4}-{2\sqrt{2}\over N}b_{\OO_{4,{\rm DT}}}\Big]\cr
  &=\OO(N^{-1}),
}}
where we have used \oSTfourdata\ and \oneptDoubletrace. Since the corresponding OPE coefficient is suppressed by $N^{-1}$, it follows that the only scalar operator with dimension four contributing to the thermal two-point function is the double trace operator $\OO_{4,{\rm DT}}$. From the OPE coefficient and thermal one-point function of this double trace operator, using \oneptDoubletrace\ and \opeDoubletracedimfour, we find the following contribution to the thermal two-point function 
\eqn\afourscalar{
  a_{4,0} = {\pi^4 \Delta(3\Delta-5)\over 18},
}
which agrees with \acoeff.

\subsec{CFT data of single-trace operator with twist two and spin four}
The primary single trace operator $\Xi =\OO_{2,4}$ with twist $\tau=2$ and spin $s=4$ is given by 
\eqn\operful{\eqalign{\Xi_{\mu\nu\rho\sigma}(x) = {1\over {96 \sqrt{35}N}} :Tr\big(&\phi (\pa_{\mu}\pa_{\nu}\pa_{\rho}\pa_{\sigma}\phi)-16(\pa_{(\mu}\phi)(\pa_{\nu}\pa_{\rho}\pa_{\sigma)}\phi)\cr
&+18(\pa_{(\mu}\pa_{\nu}\phi)(\pa_{\rho}\pa_{\sigma)}\phi)-({\rm traces})\big):(x).
}}
The relative coefficients follow from requiring that the operator is a primary, see Appendix E for details. 

The thermal one-point function of this operator is found from Wick contractions in the large-$N$ limit to be
\eqn\thopf{\langle \Xi_{\mu\nu\rho\sigma} \rangle_{\beta}={{8 \pi^6 N}\over{27 \sqrt{35}\beta^{6}}}\left(e_{\mu}e_{\nu}e_{\rho}e_{\sigma} - ({\rm traces})\right).
}
Moreover, the OPE coefficient in the $\OO_{\Delta}\times\OO_{\Delta}$ OPE can again be calculated using Wick contractions similarly to how it was done for $T^2_{4,4}$ in Appendix A. By explicit calculation one finds
\eqn\tpf{\langle\OO_{\Delta}(x_1)\OO_{\Delta}(x_2)\Xi_{\mu\nu\rho\sigma}(x_3) \rangle = {{4 \Delta }\over{\sqrt{35} N}}{{Z_{\mu}Z_{\nu}Z_{\rho}Z_{\sigma} - ({\rm traces})}\over{|x_{12}|^{2\Delta -2} |x_{13}|^2 |x_{23}|^2}},
}
and therefore the OPE coefficient $\lambda_{\OO_{\Delta}\OO_{\Delta}\OO_{2,4}}$ is given by
\eqn\flambda{\lambda_{\OO_{\Delta}\OO_{\Delta}\OO_{2,4}}={{4 \Delta }\over{\sqrt{35} N}}.
}
\noindent Now, it is easy to check that
\eqn\finrel{{1\over 2^4}\lambda_{\OO_{\Delta}\OO_{\Delta}\OO_{2,4}}b_{\OO_{2,4}}={{2 \pi^6 \Delta }\over{945}},
}
which agrees with $a_{2,4}$ in \adimsixandeight.

\subsec{CFT data of double-trace operators with twist and spin equal to four}
To find the full contribution to the thermal two-point function from the operators with $\tau=4$ and $s=4$ 
we need to take into account the contribution of all operators with these quantum numbers and solve a mixing problem. 
In addition to  the double-stress tensor operator with these quantum numbers,
the other double trace primary operator which contributes  is given by 
\eqn\dtodimeight{\eqalign{
  \OO^{\rm DT}_{\mu\nu\rho\sigma}(x) = {1\over 96\sqrt{70}N^2}:Tr(\phi^2)\Big(Tr(\phi\pa_\mu\pa_\nu\pa_\rho\pa_\sigma\phi)-16 Tr(\pa_{(\mu}\phi\pa_\nu\pa_\rho\pa_{\sigma)}\phi)\cr
 +18Tr(\pa_{(\mu}\pa_\nu\phi\pa_\rho\pa_{\sigma)}\phi)(x) -({\rm traces})\Big):(x),
}}
where the operator is unit-normalized. Notice that this is the double trace operator obtained by taking the normal ordered product of two single trace operators, the scalar operator with dimension $2$ and the single trace spin-$4$ operator with dimension $6$. There are more double trace operators with these quantum numbers which are, however, not simply products of single trace operators. These do not contribute to the thermal two-point function to leading order in ${1\over N^2}$
(see appendix F).

Note that it follows from large-$N$ factorization that the overlap of this operator with $(T^2)_{\mu\nu\rho\sigma}$ is suppressed by powers of ${1\over N}$; since both of these are double trace operators and obey the  scaling \scaling, to study the thermal two-point function to leading order in $N^2$, one can therefore neglect this overlap. 

The thermal one-point function of $\OO^{\rm DT}_{\mu\nu\rho\sigma}$ follows from the large-$N$ factorization and we find that
\eqn\oneptOO{
  b_{\OO^{\rm DT}_{4,4}}=\sqrt{{{2}\over{35}}}{{4 \pi^8 N^2}\over81},
}
where we used the thermal one-point functions for each single trace operator given by \dimtwoonept\ and \thopf. The OPE coefficient is calculated in Appendix A,
\eqn\opeseconddt{
  \lambda_{\OO_{\Delta}\OO_{\Delta}\OO^{\rm DT}_{4,4}}=\sqrt{{{2}\over{35}}}{{4\Delta (\Delta -1) }\over{N^2}}.
}
Using the thermal one point function and the OPE coefficient in \oneptOO\ and \opeseconddt\ respectively, it is found that it the operator $\OO^{\rm DT}_{\mu\nu\rho\sigma}$ gives the following contribution to the thermal two point function:
\eqn\aseconddt{
  a_{\OO^{\rm DT}_{4,4}} = \left({1\over 2}\right)^4b_{\OO^{\rm DT}_{4,4}}\lambda_{\OO_{\Delta}\OO_{\Delta}\OO^{\rm DT}_{4,4}} = {2\pi^8\Delta(\Delta-1)\over 2835}.
}

The total contribution from $T^2_{4,4}$ together with that of $\OO^{\rm DT}_{4,4}$, using \opeTSq, \tsqonept\ and \aseconddt, is  
\eqn\fullatdimeight{
 a_{4,4}= (a_{T^2_{4,4}}+a_{\OO^{\rm DT}_{4,4}}) = {\pi^8\Delta(\Delta-1)\over 1050}.
}
This agrees with $a_{4,4}$ in \adimsixandeight. 

%
%

\newsec{Comparison with the eigenstate thermalization hypothesis}

In this section we discuss the relation of our results to the eigenstate thermalization hypothesis (ETH).
We argue that the stress tensor sector of the free $SU(N)$ adjoint scalar theory in $d=4$ satisfies the ETH to leading order in $C_T\sim N^2\gg1$. 
We explain the equivalence of the micro-canonical and canonical ensemble when $\DH\sim C_T$ in large-$C_T$ theories. In this regime, the diagonal part of the ETH is (up to exponentially suppressed terms which we do not consider), equivalent to thermalization. 
Note that in two dimensions the Virasoro descendants of the identity
satisfy the ETH (see e.g. \BasuKZO\ for a recent discussion).

We begin by showing the equivalence between the micro-canonical and the canonical ensemble on $S^1_\beta\times S^{d-1}$ when $\DH\sim C_T\gg1$. See \refs{\PappadopuloJK,\LashkariVGJ,\LashkariHWQ,\GobeilFZY,\DymarskyNTG} for a similar discussion at infinite volume as well as \FaulknerHLL\ in the two-dimensional case. The expectation value in the micro-canonical ensemble of an operator $\OO$, which we take to be a scalar for simplicity, at energy $E=\DH/R$ is given by 
\eqn\microDef{
  \langle \OO\rangle_{E}^{\rm (micro)} = {1\over N(E)}\sum_{\tilde{\OO}} \langle\tilde{\OO}|\OO|\tilde{\OO}\rangle,
}
where we sum over states $|\tilde{\OO}\rangle$ with energy $(E,E+\delta E)$ and $N(E)$ is the number of states in this interval. On the other hand, consider the partition function at inverse temperature $\beta$ given by 
\eqn\partFunctionDef{
  Z(\beta) = \sum_{\tilde{\OO}} e^{-{\beta\tilde{\Delta}\over R}}= \int d\tilde{\Delta} \rho(\tilde{\Delta})e^{-{\beta \tilde{\Delta}\over R}},
}
where we sum over all states in the theory. In the second line in \partFunctionDef\ we have approximated the sum of delta-functions by a continuous function $\rho(\tilde{\Delta})$. Expectation values in the canonical ensemble is then computed by\foot{It was argued in \GobeilFZY\ that the existence of the thermodynamic limit implies that we only need to sum over operators with low spin.}
\eqn\canEnsDef{
  \langle \OO\rangle_\beta = Z(\beta)^{-1}\int d\tilde{\Delta} \rho(\tilde{\Delta})\langle \OO\rangle_{E}^{\rm (micro)}e^{-{\beta\tilde{\Delta}\over R}}.
}

Consider the partition function in \partFunctionDef\ with a free energy $F=-\beta^{-1}\log Z(\beta)$. By an inverse Laplace transform of \partFunctionDef\ we find the density of states
\eqn\densOfStates{
  \rho(\DH) = {1\over 2\pi i R}\int d\beta' e^{\beta'({\DH\over R}-F(\beta'))}.
}
For $\DH\sim C_T$ and a large free energy\foot{We consider a CFT in a high temperature phase.} $F\sim C_T$, we can evaluate \densOfStates\ using a saddlepoint approximation with the saddle at $\beta$ given by 
\eqn\saddle{
  {\DH\over R} = \pa_{\beta'} (\beta' F)|_{\beta}. 
}
Consider now the thermal expectation value in \canEnsDef, multiplying both sides by $Z(\beta)$ and doing an inverse Laplace transform evaluated at $\DH\sim C_T$ we find 
\eqn\thermres{
  \rho(\DH)\langle \OO\rangle_{\DH/R}^{\rm (micro)} = {1\over 2\pi i R}\int d\beta' \langle \OO\rangle_{\beta'} e^{\beta'({\DH\over R}-F(\beta'))}.
}
For $F\sim C_T\gg1$ we again use a saddlepoint approximation to evaluate \thermres\ with the saddle at $\beta$ determined by \saddle, assuming $\langle \OO\rangle_{\beta'}$ does not grow exponentially with $C_T$. The RHS of \thermres\ is therefore the thermal expectation value $\langle \OO\rangle_{\beta}$ multiplied by the saddlepoint approximation of the density of states in \densOfStates. It then follows that
\eqn\eqvEns{
  \langle \OO\rangle_{\DH/R}^{\rm (micro)} \approx \langle \OO\rangle_{\beta},
}
with $\beta$ determined by \saddle. In particular, in the infinite volume limit $R\to\infty$, the free energy is given by\foot{Here we denote the canonically normalized stress tensor by $T^{\rm (can)}_{\mu\nu}$, whose two-point function is given by $\langle {T^{\rm (can)}}^{\mu\nu}(x)T_{\rho\sigma}^{\rm (can)}(y)\rangle = {C_T\over S_d^2}({I^{\mu}}_{(\rho}{I^{\nu}}_{\sigma)}-({\rm trace}))$.} 
\eqn\freeEnergy{
  F = {b_{T^{\rm (can)}_{\mu\nu}}S_{d}R^{d-1}\over d\beta^{d}},
}
where $S_{d}=Vol(S^{d-1})=2\pi^{d\over 2}/\Gamma({d\over2})$. Inserting \freeEnergy\ in \saddle\ we find \PappadopuloJK
\eqn\DeltaHSaddle{
  {\beta \over R} = \left({-(d-1)b_{T^{\rm (can)}_{\mu\nu}}S_d\over d\DH}\right)^{1\over d}.
}

We can use \eqvEns\ to see the thermalization of the stress tensor. The free energy is related to the expectation value of the stress tensor $T^{\rm (can)}_{\mu\nu}$ \SimmonsDuffinGJK
\eqn\stressFreeEnergy{
  \langle T_{00}^{\rm (can)}\rangle_\beta  = {1\over S_d R^{d-1}}\pa_\beta(-\beta F(\beta)).
}
On the other hand, the expectation value of $T_{00}^{\rm (can)}$ in a heavy state $|\OH\rangle$ is fixed by the Ward identity to be
\eqn\heavyStateFree{
  \langle \OH|T_{00}^{\rm (can)}|\OH\rangle = -{\DH\over S_d R^d}. 
}
Multiplying \saddle\ with $(S_d R^{d-1})^{-1}$ and comparing with \stressFreeEnergy-\heavyStateFree\ we find that
\eqn\StressThermP{
  \langle \OH|T_{00}^{\rm (can)}|\OH\rangle=\langle T_{00}^{\rm (can)}\rangle_\beta. 
}
This shows the thermalization of the stress tensor in heavy states where $F\sim \DH\sim C_T$ in large-$C_T$ theories. Note that this follows from \eqvEns\ since we can replace the micro-canonical expectation value at $E=\DH/R$, on the LHS, with the expectation value in any single heavy state with dimension $\DH$ due to the Ward identity, independent of the  heavy state. Put differently, the stress tensor satisfies the ETH as we will review below. 

We now consider the eigenstate thermalization hypothesis for CFTs at finite temperature on the sphere $S^{d-1}$ of radius $R$. 
The diagonal part of the  ETH is given by
\eqn\eth{\langle \OO_H |\OO_{\tau,s} |\OO_H \rangle = \langle \OO_{\tau,s}\rangle_{E}^{\rm (micro)} + \OO\left(e^{-S(E)}\right),
}
\noindent where $\OO_{H}$ and $\OO_{\tau,s}$ are local primary operators and $\langle \OO_{\tau,s}\rangle_{E}^{\rm (micro)}$ is the expectation value of $\OO_{\tau,s}$ in the micro-canonical ensemble at energy $E={\DH\over R}$. Here we assume that the operator $\OH$ is a heavy scalar operator with large conformal dimension 
$\DH\propto C_T \gg 1 $. The operator $\OO_{\tau,s}$ on the other hand can have non-zero spin.\foot{The tensor structure in \eth\ is suppressed.}. In \eth, $e^{S(E)}$ is the density of states at energy $E=\DH/R$. As shown in \eqvEns, in the limit $\DH\sim C_T\gg1$, the micro-canonical ensemble is equivalent to the canonical ensemble at inverse temperature $\beta$ determined by \saddle. It then follows from \eth\ that the diagonal part of the ETH can written in terms of OPE coefficients and thermal one-point functions:
\eqn\ethcft{{\lambda_{\OH\OH\OO_{\tau,s}}\over R^{\tau+s}}={{b_{\OO_{\tau,s}}f_{\OO_{\tau,s}}\left(\beta/R\right)}\over \beta^{\tau+s}}+\OO\left(e^{-S(E)}\right),
}
where $f_{\OO_{\tau,s}}$  also appears in \deff. This is equivalent to the statement of thermalization discussed in the rest of the paper.

In this paper we observed that the multi stress tensor operators satisfy \ethcft.
One can also ask if \ethcft\ holds for any operator in the specific heavy state we considered. By comparing eqs.~\opeCoeffdimtwoRes\ and \dimtwoonept\ using \muu, one can check that operator $\OO_{2}={1\over{\sqrt{2}N}}:Tr(\phi^2):$ does not satisfy \ethcft. Since this is a free theory, it is not a surprise that the ETH is not satisfied by all operators in the spectrum which is seen explicitly in this case.

%
%
%

\newsec{Discussion}
In this paper we argued that multi stress tensor operators $T^k_{\tau,s}$ in CFTs with a large central charge $C_T$
thermalize: their expectation values in  heavy states are the same as their thermal expectation values.
This is equivalent  to the statement that multi stress tensor operators in higher-dimensional CFTs satisfy the diagonal part of the ETH in the thermodynamic limit.
The analogous statement in the $d=2$ case is that  the Virasoro descendants of the identity  satisfy the ETH condition in the large-$C_T$ limit. 

We  observed that the operator $\OO_{2}={1\over {\sqrt{2}N}}:Tr(\phi^2):$ does not satisfy the ETH. 
This is seen by comparing eqs.~\opeCoeffdimtwoRes\ and \dimtwoonept\ using \muu. 
While this operator does not thermalize in the heavy states we considered, the OPE coefficient averaged over all operators with $\DH\sim C_T$ is expected to be proportional to the thermal one-point function. 
The averaged OPE coefficients should therefore scale like $\sim\sqrt{\DH}$ compared to $\lambda_{\OH\OH \OO_2}\sim\DH/\sqrt{C_T}$ for the heavy states we considered. 
It would be interesting to exhibit  heavy operators that produce the former scaling.

We provided a bootstrap argument in favor of the thermalization of multi stress tensor operators.
One should be able to refine it to give an explicit form for
leading behavior of the multi stress tensor OPE coefficients -- we leave it for future work.
The holographic/bootstrap OPE coefficients for the leading twist double stress tensor operators can be found in e.g.
 \KulaxiziTKD\ -- they are nontrivial functions of the spin.
As explained in \refs{\KulaxiziTKD,\KarlssonDBD}, the leading $\Delta$ behavior of the minimal-twist double- and triple-stress tensor OPE coefficients is consistent with the exponentiation of the  
near lightcone stress tensor conformal block.
One can go beyond the leading twist multi stress tensors.
In holographic HHLL correlators  each term of the type $(\Delta \mu)^k \sim (\Delta \DH /C_T)^k$
comes from the exponentiation of the  stress-tensor block -- this follows from the Wilson line calculation of the correlator
in the AdS-Schwarzschild background \refs{\MaxfieldRKN,\KulaxiziTKD,\ParnachevFNA}.

In this paper we argue that this behavior is universal, and is not just confined to holographic theories.
 Hence, one can formulate another statement equivalent to the 
thermalization of  multi stress tensor operators.
Namely, scalar correlators of pairwise identical operators of dimensions $\Delta_{1,2}$
in large-$C_T$ theories in the limit $\Delta_{1,2} \gg 1$, $\Delta_1 \Delta_2 /C_T$ fixed are given by the exponentials
of the stress tensor conformal block\foot{See \FitzpatrickZHA\ for previous work on the eikonalization of the multi stress tensor
OPE coefficients at large spin.}. 
This is similar to what happens in two-dimensional CFTs. 

Note that the universality of  the OPE coefficients is naively in tension with the results of \FitzpatrickYJB, where
 finite gap ($\Delta_{\rm gap}$) corrections to the multi stress tensor OPE coefficients were considered.
In particular, for double stress tensors, such corrections behave like $\Delta^3/\Delta_{\rm gap}$ which is clearly
at odds with the universality statement.
Of course, the results of \FitzpatrickYJB\ are obtained in the limit $\Delta \ll \Delta_{\rm gap}$, while 
in this paper we consider 
the opposite regime $\Delta \gg \Delta_{\rm gap}$.

One may also wonder what happens with the universality of the OPE coefficients beyond  leading order in $\Delta$.
In particular, in \ParnachevFNA , it was shown that the bootstrap result for the HHLL correlator  exactly matches the holographic Wilson 
line calculation (in the double scaling limit where only the minimal twist multi stress tensor operators contribute).
This corresponds to including terms beyond the exponential of the stress tensor block --
one needs to compute the HHLL correlator, take a logarithm of the  result, divide by $\Delta$, and then
take the large-$\Delta$ limit.
The result is sensitive to  terms subleading in the large-$\Delta$ limit of the multi stress tensor OPE coefficients.
In four spacetime dimensions the result in \ParnachevFNA\ is given by an elliptic integral -- is it applicable beyond holography?

 In \KulaxiziTKD\ terms subleading in $\Delta$ were shown to be important for the computation of the phase shift. 
The simplest nontrivial case in two spacetime dimensions is the operator $\Lambda_4$ which is a level four Virasoro descendant of the identity (see e.g.\PerlmutterIYA).
  One could also get it by 
  using the CFT normal ordering and imposing the quasi-primary condition \DiFrancescoNK .
Consider now the case of minimal twist (twist four) operators in four dimensions.
How do we determine 
the analog of $\Lambda_4$?
There is no Virasoro algebra now.

Presumably, one can   reconstruct the analog of $\Lambda_4$  in four spacetime dimensions by considering a CFT normal ordered product 
of stress tensors, and adding a single trace term to ensure that the resulting operator is a primary and
is orthogonal to the stress tensor itself.
Note that the CFT normal ordering differs from the oscillator normal ordering in a free theory by the addition of a single trace
operator, as reviewed in Appendix G.
This procedure can then be generalized to other multi-trace operators.
We leave it for future work.

It is also helpful to imagine what happens in a theory like $\NN=4$ Super Yang-Mills, where there is a marginal line
connecting the weak and the strong coupling (the latter admits a holographic description).
Presumably, as the coupling is turned on, only one operator remains light (with dimension eight and spin four), while  others get
 anomalous dimensions.
It would be interesting to see this explicitly even to the leading nontrivial order in the 't Hooft coupling.
It would also be interesting to see how the corresponding OPE coefficient interpolates between its free 
and strong coupling values.

Using crossing symmetry, we argued that the universality of multi stress tensor OPE coefficients is related to the OPE coefficients $\lambda_{\OH T_{\mu\nu}\OO'}$, with $\OO'\neq\OH$ being either heavy or light, present in the cross-channel expansion. Such OPE coefficients with at least one operator being heavy were recently studied in \refs{\DelacretazNIT,\BelinHEA}. It would be interesting to further study the connection of our results to this  work. 

Another interesting question concerns the fate of the double trace operators  of the schematic form $[\OO_{\Delta}\OO_{\Delta}]_{n,l}$. 
Consider the $d=4$ case in the large volume limit and $n,l=0$, for simplicity.
We expect that the corresponding OPE coefficients in the free theory behave like $\lambda_{\OO_{H}\OO_{H}[\OO_{\Delta}\OO_{\Delta}]_{0,0}}\propto \Delta_H^2/C_T \propto C_T \mu^2$,\foot{This scaling is obtained by computing
the OPE coefficient $\lambda_{\OO_{H}\OO_{H}[\OO_{\Delta}\OO_{\Delta}]_{0,0}}$
for $1 \ll \DH \ll C_T$ and extrapolating it to the  $ \DH \sim C_T$ regime. } 
while their thermal one-point functions behave like $\la[\OO_{\Delta}\OO_{\Delta}]_{0,0} \ra_{\beta}\propto C_T \beta^{-2\Delta}$. 
Comparing the two results with the help of  \muu\ one observes that such operators  do not thermalize in the free theory for generic $\Delta$.
The situation is more nontrivial in holography where we do not know the large $\mu$ behavior of the OPE coefficients\foot{
Note that the large-$N$ scaling in holography is different. Both the OPE coefficients and the thermal expectation values behave like $C_T^0$ as opposed to $C_T \sim N^2$.}.
As pointed out in \FitzpatrickZQZ,  the contribution of double-trace operators to thermal two-point functions is different
from that of multi stress tensors.
The latter is only sensitive to the behavior of the metric near the boundary, but the former knows about the full black hole metric.
This seems to indicate that the thermalization of the double trace operators in holographic theories is also unlikely\foot{A
simple way to decouple such operators is to take the large-$\Delta$ limit.}.

It is a natural question how generic are the heavy states for which the stress tensor sector thermalizes.
The results of our paper seem to suggest that such thermalization is more 
generic than the thermalization of other light operators\foot{
A closely related question of finding ``typical" states where the stress tensor sector thermalizes in the large volume limit in $d=2$ was recently discussed in \DattaJEO.
There it was observed  that such states are Virasoro descendants when the central charge is finite.}.
Other interesting questions include generalizations to the case of finite but large central charge and
to non-conformal quantum field theories.

\bigskip
\bigskip
\bigskip

 \bigskip
 \bigskip
 
\noindent {\bf Acknowledgments}: 
We thank Aleksandar Bukva, Ilija Buri\' c, Sa\v so Grozdanov, Manuela Kulaxizi, Eric Perlmutter and Larry Yaffe for discussions, correspondence and comments on the draft.
The work of R.K. and A.P. is supported in part by an Irish Research Council Laureate Award. The work of P.T. is supported in part by an Ussher Fellowship Award. 
 \bigskip
 \bigskip

\appendix{A}{OPE coefficients from Wick contractions}
In this appendix we go through the calculations needed for finding the OPE coefficients of various operators using Wick contractions. This mainly amounts to counting the number of contractions leading to a planar diagram. For simplicity, the figures are shown for external operators with $\Delta=4$ while we write down the result for general $\Delta$ as this is needed for the main body of the paper. 

To begin with, since we consider a large-$N$ matrix theory, it is convenient to use the double-line notation for fundamental field propagators. In Fig.~1 the two-point function $\langle :Tr(\phi^4)::Tr(\phi^4):\rangle$ is visualised. 
\ifig\LABEL{The two-point function $\langle :Tr(\phi^4)::Tr(\phi^4):\rangle$ before any contractions.}{\epsfxsize4in\epsfbox{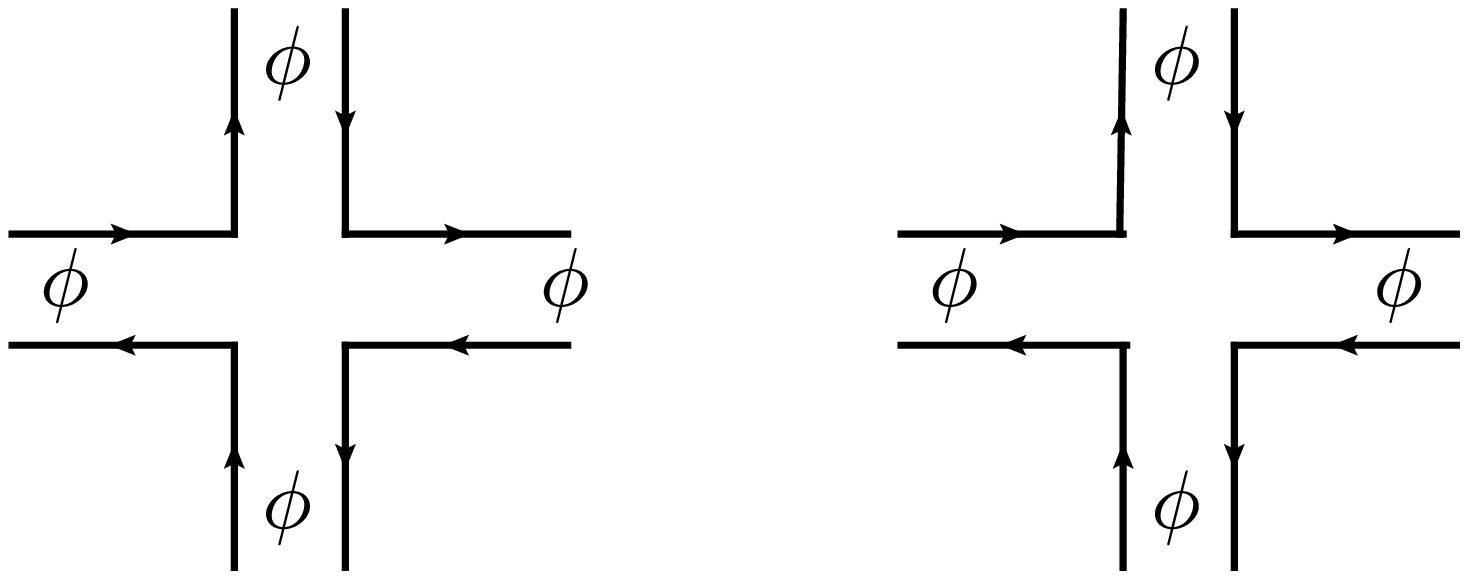}}
In Fig.~2, the planar diagram is shown for $\Delta=4$ and there are $\Delta$ number of such contractions giving a planar diagram 
\eqn\normm{
	P_{\langle :Tr(\phi^\Delta)::Tr(\phi^\Delta):\rangle}=\Delta,
}
where the $P_{\langle ...\rangle}$ denotes the number of planar diagrams for $\langle ...\rangle$. 
\ifig\LABEL{The two-point function $\langle:Tr(\phi^4)::Tr(\phi^4):\rangle$ completely contracted.}{\epsfxsize4in\epsfbox{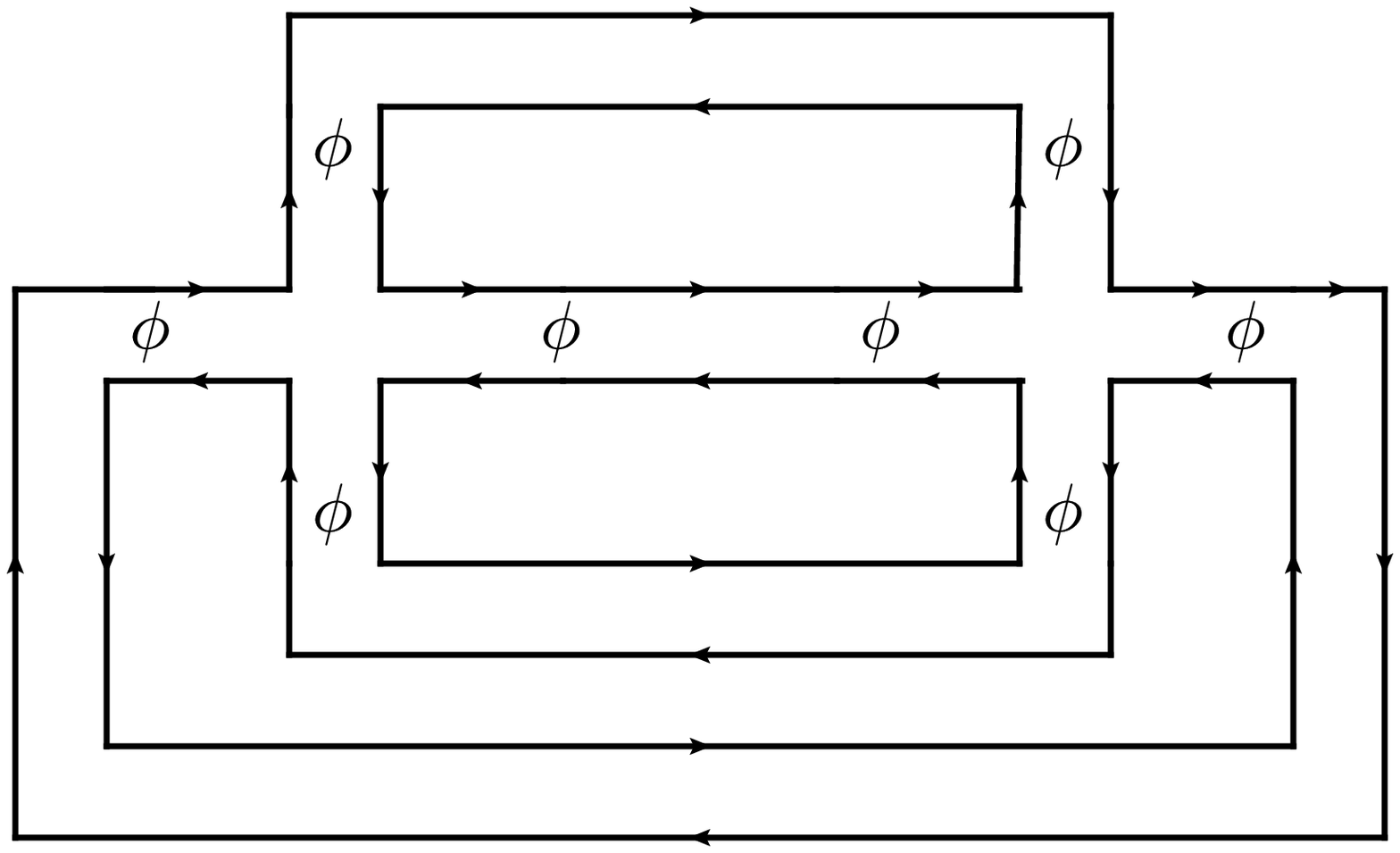}}

We further need the OPE coefficient $\lambda_{\OO_\Delta\OO_\Delta\OO_2}$. This is shown in Fig.~3 for $\Delta=4$ and there are $2\Delta$ possibilities for step (1), $\Delta$ number of possibilites for step (2) after which everything is fixed assuming that the diagram is planar. This gives
\eqn\phiDphiDphitwo{
	P_{\langle :Tr(\phi^\Delta)::Tr(\phi^\Delta)::Tr(\phi^2):\rangle} = 2\Delta^2.
} 
\ifig\LABEL{The three-point function $\langle:Tr(\phi^4)::Tr(\phi^4)::Tr(\phi^2):\rangle$ completely contracted. }{\epsfxsize4in\epsfbox{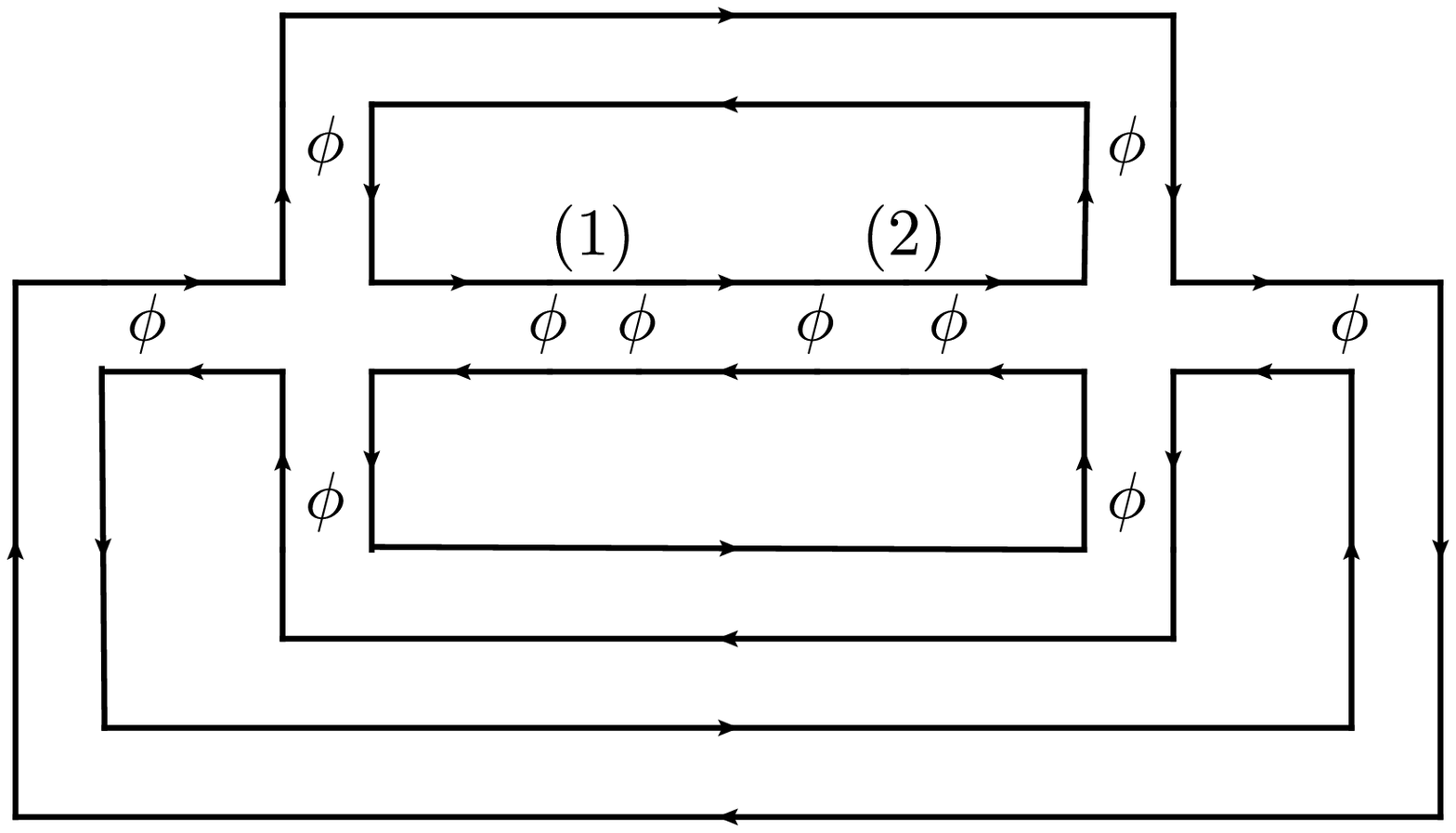}}

In Fig.~4 the three-point function $\langle :Tr(\phi^\Delta)::Tr(\phi^\Delta)::Tr(\phi^4):\rangle$ for $\Delta=4$ is shown. For the first contraction (1) there are $2\Delta$ possibilites, for the second contraction there are $\Delta$ and for step (3) there are two possibilites. This gives overall
\eqn\pphifourcub{
  P_{\langle:Tr(\phi^\Delta)::Tr(\phi^\Delta)::Tr(\phi^4):\rangle} = 4\Delta^2.
} 
\ifig\LABEL{The three-point function $\langle:Tr(\phi^4)::Tr(\phi^4)::Tr(\phi^4):\rangle$ completely contracted. }{\epsfxsize4in\epsfbox{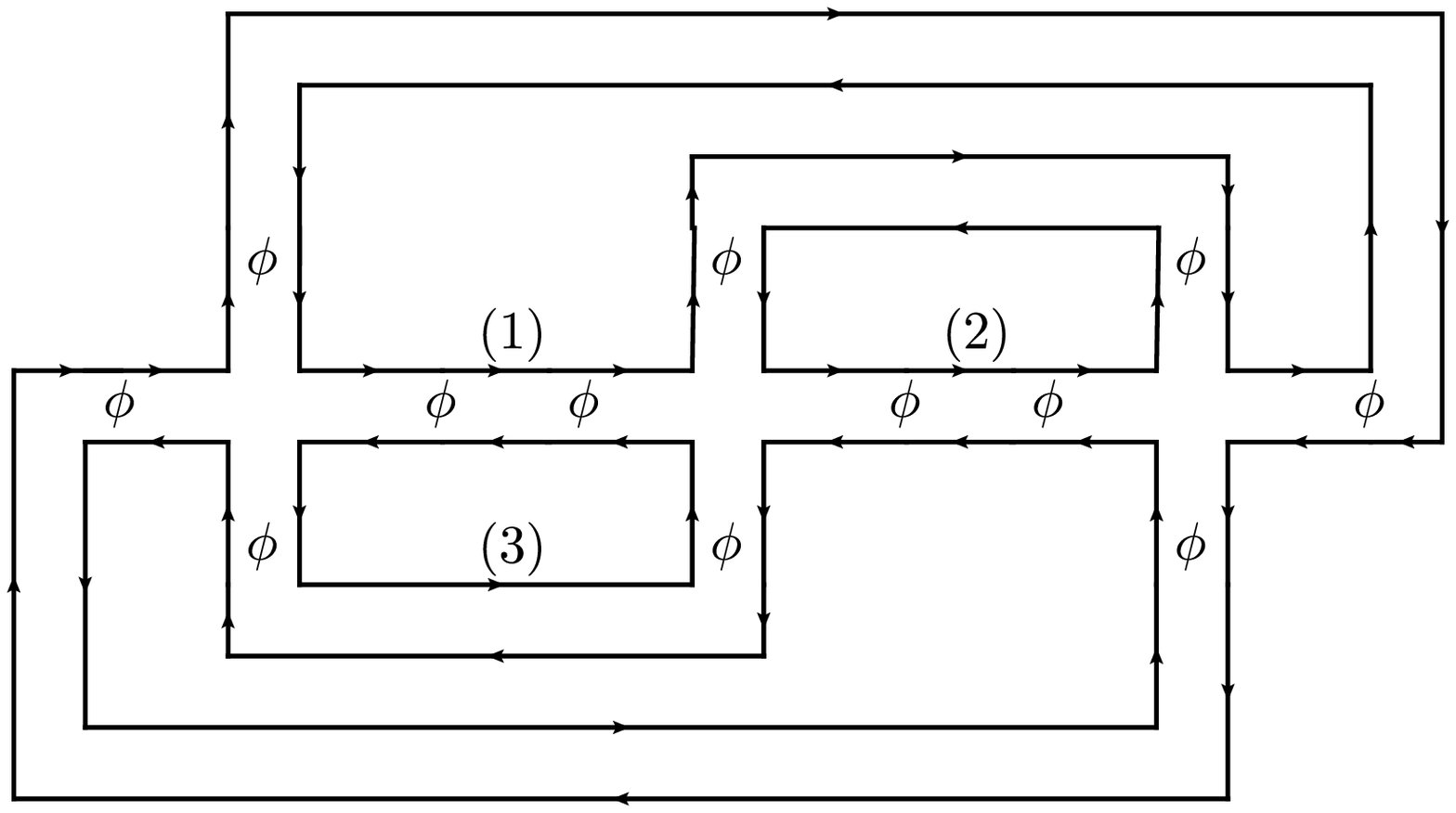}}

In Fig.~5 and Fig.~6, the three-point function $\langle :Tr(\phi^4)::Tr(\phi^4):Tr(\phi^2)Tr(\phi^2):\rangle$ is shown. The reason for there being two different types of diagrams is because each trace term in the double trace operator $:Tr(\phi^2)Tr(\phi^2):$ can either be contracted with the same $:Tr(\phi^4):$ (Fig.~5, type B), or to both (Fig.~6, type A). 

Consider first the type of diagrams in Fig.~5.  For the first contraction there are $2\Delta$ such terms and the second contraction gives another factor of $2$. Contraction (3) and (4) contributes factors of $\Delta$ and $2$ respectively. What remains is equivalent to the two-point function $\langle :Tr(\phi^{\Delta-2})::Tr(\phi^{\Delta-2}):\rangle$ which further give a factor of $(\Delta-2)$ and therefore there are $8\Delta^2(\Delta-2)$ contractions of type B in Fig.~5. 

Continuing with Fig.~6, the first contraction gives a factor of $2\Delta$, the second contraction $\Delta$ and the third one a factor of $2(\Delta-1)$. What remains is then fixed by imposing that the diagram is planar. The type A diagrams in Fig.~6 therefore further contributes $4\Delta^2(\Delta-1)$ planar diagrams to $\langle :Tr(\phi^\Delta)::Tr(\phi^\Delta):Tr(\phi^2)Tr(\phi^2):\rangle$. It is therefore found that 
\eqn\phifourDTO{
  P_{\langle :Tr(\phi^\Delta)::Tr(\phi^\Delta):Tr(\phi^2)Tr(\phi^2):\rangle} = 4\Delta^2(3\Delta-5).
}

\ifig\LABEL{The three-point function $\langle:Tr(\phi^4)::Tr(\phi^4)::Tr(\phi^2)Tr(\phi^2):\rangle$. There are two such types of contractions that give planar diagrams, here it shown when each $:Tr(\phi^2):$ connect to a separate $:Tr(\phi^4):$.}{\epsfxsize4in\epsfbox{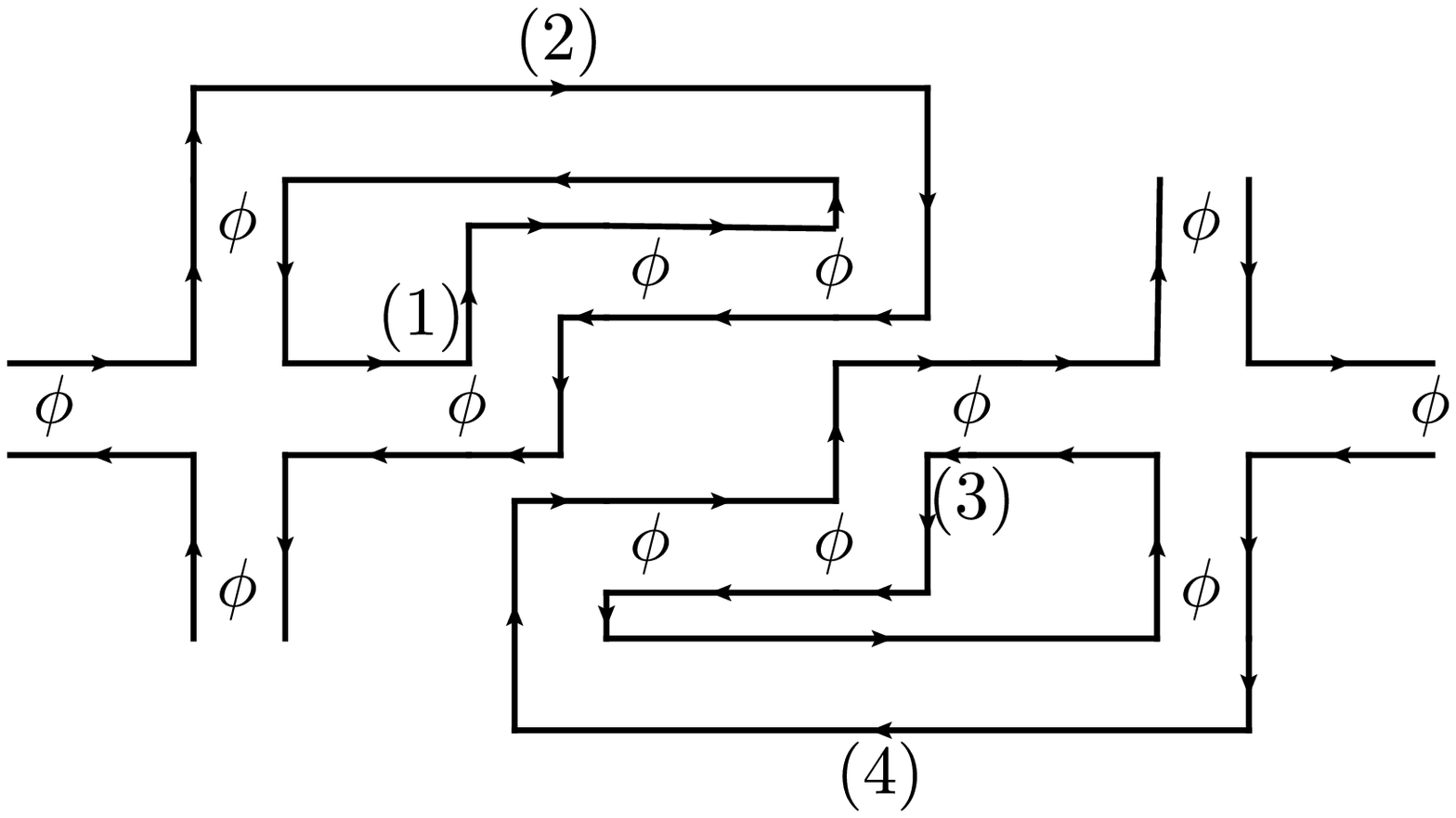}}

\ifig\LABEL{The three-point function $\langle:Tr(\phi^4)::Tr(\phi^4)::Tr(\phi^2)Tr(\phi^2):\rangle$. There are two such types of contractions that give planar diagrams, here it shown when each $:Tr(\phi^2):$ connect to both $:Tr(\phi^4):$ operators.}{\epsfxsize4in\epsfbox{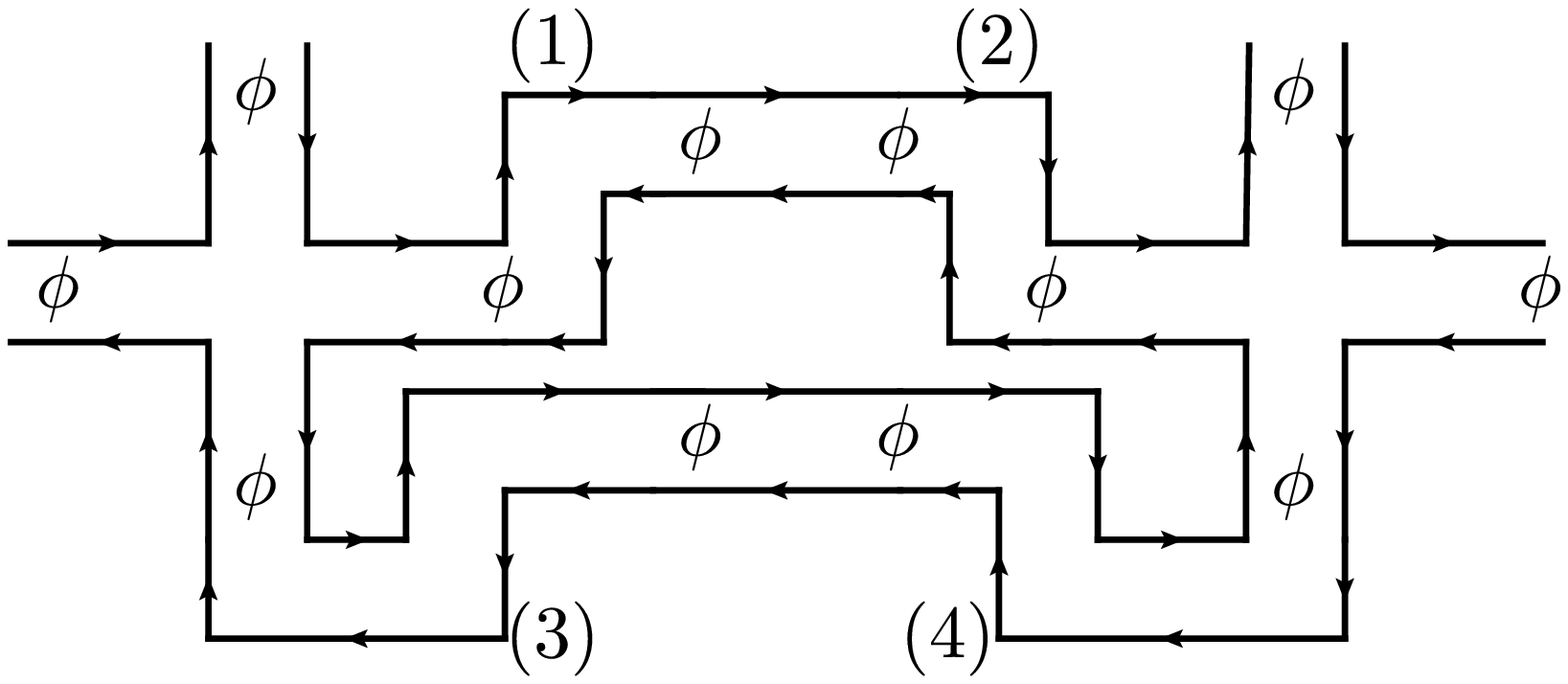}}

Consider now the stress tensor OPE coffiecient $\lambda_{\OO_\Delta\OO_\Delta T_{\mu\nu}}$ where 
\eqn\stresssecond{
  T_{\mu\nu}(x) = {1\over 2\sqrt{3}N}:Tr\left(\pa_\mu\phi\pa_\nu\phi-{1\over 2}\phi\pa_\mu\pa_\nu\phi -({\rm trace})\right):(x)
}
and the three-point function $\langle \OO_\Delta\OO_\Delta T_{\mu\nu}\rangle$:
\eqn\threeptstree{
  \langle \OO_\Delta(x_1)\OO_\Delta(x_2)T_{\mu\nu}(x_3)\rangle = \lambda_{\OO_\Delta\OO_\Delta T_{\mu\nu}} {Z_{\mu}Z_\nu-{\rm traces}\over |x_{12}|^{2\Delta-2}|x_{23}|^{2}|x_{13}|^2},
}
where $Z_\mu={{x_{13}}_\mu\over |x_{13}|^2}-{{x_{12}}_\mu\over |x_{12}|^2}$. From the definition of $T_{\mu\nu}$ in \stresssecond\ it is clear that the only term that contributes to term ${x_{13}}_\mu {x_{13}}_\nu$ comes from the second term in \stresssecond\ that is of the form $\propto Tr(\phi\pa_\mu\pa_\nu\phi)$. Up to the derivatives, the diagram will look like those visualised in Fig.~3. The number of diagrams is half of that given in \phiDphiDphitwo\ since we restrict to terms proportional to ${x_{13}}_\mu {x_{13}}_\nu$:
\eqn\ttt{
  P_{\langle \OO_\Delta\OO_\Delta T_{\mu\nu}\rangle|_{{x_{13}}_\mu {x_{13}}_\nu}} = \Delta^2,
}
from which we reproduce \TOPE. 

Now we want to find the OPE coefficient $\lambda_{\OO_\Delta\OO_\Delta T^2_{4,4}}$ for the double-stress tensor $T^2_{4,4}$. This is done similarly to the way the stress tensor OPE coefficient was found. First, the operator $(T^2)_{\mu\nu\rho\sigma}$ was given in \doublestress\ to be 
\eqn\tsqsec{
  (T^2)_{\mu\nu\rho\sigma}(x) = {1\over\sqrt{2}}:T_{(\mu\nu}T_{\rho\sigma)}:(x)-({\rm traces})
}
and the three-point function $\langle \OO_\Delta\OO_\Delta (T^2)_{\mu\nu\rho\sigma}\rangle$ is fixed by conformal symmetry to be 
\eqn\ootsq{
  \langle\OO_\Delta(x_1)\OO_{\Delta}(x_2)(T^{2})_{\mu\nu\rho\sigma}(x_3)\rangle = {\lambda_{\OO_\Delta\OO_\Delta T^2_{4,4}}\over |x_{12}|^{2\Delta-4}|x_{13}|^{4}|x_{23}|^{4}}\left(Z_\mu Z_\nu Z_\rho Z_\sigma-({\rm traces})\right).
}
Consider the term in \ootsq\ proportional to ${x_{13}}_\mu {x_{13}}_\nu{x_{13}}_\rho {x_{13}}_\sigma$. This will be due to the term in $(T^2)_{\mu \nu\rho\sigma}$ of the form $Tr(\phi\pa_{(\mu}\pa_\nu\phi)Tr(\phi\pa_\rho\pa_{\sigma)}\phi)$. Using this we find that 
\eqn\OOTTSq{\eqalign{
  \langle\OO_\Delta(x_1)\OO_{\Delta}(x_2)(T^{2})_{\mu\nu\rho\sigma}(x_3)\rangle&|_{{x_{13}}_\mu {x_{13}}_\nu{x_{13}}_\rho {x_{13}}_\sigma} = {1\over \Delta N^\Delta}{1\over \sqrt{2}}\left({-1\over 4\sqrt{3}N}\right)^28^2N^{\Delta}\cr
  &\times{P_{\langle \OO_\Delta\OO_\Delta T^2_{4,4}\rangle|_{{x_{13}}_\mu {x_{13}}_\nu{x_{13}}_\rho {x_{13}}_\sigma}}\over |x_{12}|^{2(\Delta-2)}|x_{23}|^{4}|x_{13}|^{12}}.
}}
The number of contractions giving a planar diagram, $P_{\langle \OO_\Delta\OO_\Delta T^2_{4,4}\rangle|_{{x_{13}}_\mu {x_{13}}_\nu{x_{13}}_\rho {x_{13}}_\sigma}}$, come from diagrams of the form given in Fig.~6. Since we are considering the term proportional ${x_{13}}_\mu {x_{13}}_\nu{x_{13}}_\rho {x_{13}}_\sigma$, the number of such diagrams are reduced compared to scalar double trace operator. Instead the first contraction, (1) in Fig.~6, give a factor of $\Delta$, the second contraction, (2), a factor of $(\Delta-1)$, the third contraction (3) gives a further factor $\Delta$ after which everything is fixed by imposing that the diagram is planar. We therefore find that 
\eqn\numbPTT{
  P_{{\langle \OO_\Delta\OO_\Delta T^2_{4,4}\rangle}|_{{x_{13}}_\mu {x_{13}}_\nu{x_{13}}_\rho {x_{13}}_\sigma}}= \Delta^2(\Delta-1),
}
and inserting this in \OOTTSq\ gives
\eqn\opeExpl{
  \lambda_{\OO_\Delta\OO_\Delta T^2_{4,4}} = {2\sqrt{2}\Delta(\Delta-1)\over 3N^2},
}
and therefore reproduces \opeTSq.

Similar to the double-stress tensor, consider the dimension-eight spin-four double trace operator 
\eqn\dtoApp{\eqalign{
  \OO^{\rm DT}_{\mu\nu\rho\sigma}(x) = {1\over 96\sqrt{70}N^2}:Tr(\phi^2)\Big(Tr(\phi\pa_\mu\pa_\nu\pa_\rho\pa_\sigma\phi)-16 Tr(\pa_{(\mu}\phi\pa_\nu\pa_\rho\pa_{\sigma)}\phi)\cr
 +18Tr(\pa_{(\mu}\pa_\nu\phi\pa_\rho\pa_{\sigma)}\phi)(x) -({\rm traces})\Big):(x).
}}
The three-point function $\langle\OO_\Delta(x_1)\OO_{\Delta}(x_2)\OO^{\rm DT}_{\mu\nu\rho\sigma}(x_3)\rangle$ is given by
\eqn\threesecdDTO{
  \langle\OO_\Delta(x_1)\OO_{\Delta}(x_2)\OO^{\rm DT}_{\mu\nu\rho\sigma}(x_3)\rangle = {\lambda_{\OO_\Delta\OO_\Delta \OO^{\rm DT}_{\mu\nu\rho\sigma}}\over |x_{12}|^{2\Delta-4}|x_{13}|^{4}|x_{23}|^{4}}\left(Z_\mu Z_\nu Z_\rho Z_\sigma-({\rm traces})\right).
}
By again considering terms in \threesecdDTO\ proportional to ${x_{13}}_\mu {x_{13}}_\nu{x_{13}}_\rho {x_{13}}_\sigma$ we find that each term in \dtoApp\ will contribute planar diagram of the type in Fig.~5, while only the term $\sim Tr(\phi\pa^4\phi)$ also give a contribution of the type in Fig.~6. Considering first the terms coming from the diagram in Fig.~5, one finds that this contribution vanishes. The remaining contribution to the term \threesecdDTO\ proportional to ${x_{13}}_\mu {x_{13}}_\nu{x_{13}}_\rho {x_{13}}_\sigma$ comes from the first term in \dtoApp\ and the planar diagram pictured in Fig.~6; there are $2\Delta^2(\Delta-1)$ contractions giving such a planar diagram leading to 
\eqn\threesecdDTOres{\eqalign{
  \langle\OO_\Delta(x_1)\OO_{\Delta}(x_2)\OO^{\rm DT}_{\mu\nu\rho\sigma}(x_3)\rangle&|_{{x_{13}}_\mu {x_{13}}_\nu{x_{13}}_\rho {x_{13}}_\sigma} = {1\over \Delta N^\Delta}{384\over 96\sqrt{70}N^2}N^{\Delta}\cr
  &\times{2\Delta^2(\Delta-1)\over |x_{12}|^{2(\Delta-2)}|x_{23}|^{4}|x_{13}|^{12}},
}} 
where the $384$ in the numerator come from the derivatives.
This gives the OPE coefficient:
\eqn\opeSecDT{
  \lambda_{\OO_{\Delta}\OO_{\Delta}\OO^{\rm DT}_{\mu\nu\rho\sigma}}=\sqrt{{{2}\over{35}}}{{4\Delta (\Delta -1) }\over{N^2}}+\OO(N^{-4}).
}


\appendix{B}{Subleading twist double-stress tensors}
In this Appendix we study the subleading twist double-stress tensors, both with dimension $8$ and spin $s=0,2$ denoted $(T^2)$ and $(T^2)^{\mu\nu}$ respectively. The calculations needed to find the OPE coefficient in the $\OO_\Delta\times\OO_\Delta$ OPE are reviewed as well as the normalization of $(T^2)^{\mu\nu}$.

The $(T^{2})^{\mu\nu}$ was defined in \defSub\ which we repeat here:
\eqn\defSubApp{
  (T^{2})^{\mu\nu}(x) = {1\over \sqrt{2}}:{T^{\mu}}_\alpha T^{\alpha\nu}:(x)-{\delta^{\mu\nu}\over 4\sqrt{2}}:{T^{\beta}}_\alpha {T^{\alpha}}_{\beta}:(x).
}
The operator $(T^2)^{\mu\nu}$ can be seen to be unit-normalized to leading order in $N$:
\eqn\normSub{\eqalign{
  \langle (T^{2})^{\mu\nu}(x_1)(T^{2})_{\rho\sigma}(x_2)\rangle =& {1\over\sqrt{2}}\langle T^{\mu\alpha}(x_1)T_{\rho\beta}(x_2)\rangle\langle {T^{\nu}}_{\alpha}(x_1){T^{\beta}}_{\sigma}\rangle\cr
  &+(\rho\longleftrightarrow\sigma)-({\rm traces})+\OO(N^{-2}). 
}}
Using the two-point function of the stress tensor in \stressnorm\ and ${I^{\mu}}_\alpha {I^\alpha}_\rho={\delta^\mu}_\rho$ one finds
\eqn\normsubSec{
  \langle (T^{2})^{\mu\nu}(x_1)(T^{2})_{\rho\sigma}(x_2)\rangle = {1\over |x|^{16}}\left({I^{(\mu}}_\rho{I^{\nu)}}_\sigma-({\rm traces})\right),
}
from which it is seen that $(T^2)^{\mu\nu}$ is unit-normalised. 

We now want to find the OPE coefficient of $(T^2)^{\mu\nu}$ in the $\OO_\Delta\times\OO_\Delta$ OPE. It can be found from the basic objects $I^{(1)}_{\mu\nu\rho\sigma}$, $I^{(2)}_{\mu\nu\rho\sigma}$ and $I^{(3)}_{\mu\nu\rho\sigma}$ which we calculate below. 

We first consider a similar quantity ${J^{(1)}}^{\mu\nu\rho\sigma}$:
\eqn\eqone{\eqalign{
  {J^{(1)}}^{\mu\nu\rho\sigma} =&\langle :Tr(\phi^\Delta):(x_1):Tr(\phi^\Delta):(x_2)::Tr(\pa_\mu\phi\pa_\nu\phi)Tr(\pa_\rho\phi\pa_\sigma\phi):(x_3)\rangle\cr
  &= {2^4N^\Delta\over |x_{13}|^8|x_{23}|^8|x_{12}|^{2\Delta-4}}\times\Big[(2\Delta)^2(\Delta-2)(x_{13}^\mu x_{13}^\nu x_{23}^\rho x_{23}^\sigma+x_{23}^\mu x_{23}^\nu x_{13}^\rho x_{13}^\sigma)+\cr
  &\Delta^2(\Delta-1)(x_{13}^\mu x_{23}^\nu (x_{13}^\rho x_{23}^\sigma+x_{23}^\rho x_{13}^\sigma)+x_{23}^\mu x_{13}^\nu (x_{13}^\rho x_{23}^\sigma+x_{23}^\rho x_{13}^\sigma))\Big].
}}
Definining $X_{13}^{\mu\nu}={1\over |x_{13}|^4}(-\delta^{\mu\nu}+4{x_{13}^\mu x_{13}^\nu\over |x_{13}|^2})$ we then study ${{J}^{(2)}}^{\mu\nu\rho\sigma}$:
\eqn\eqtwo{\eqalign{
{{J}^{(2)}}^{\mu\nu\rho\sigma} =&\langle :Tr(\phi^\Delta):(x_1):Tr(\phi^\Delta):(x_2)::Tr(\phi\pa_\mu\pa_\nu\phi)Tr(\phi\pa_\rho\pa_\sigma\phi):(x_3)\rangle\cr
=&{N^{\Delta}\over |x_{12}|^{2\Delta-4}}\Big[\Delta^2(\Delta-1)2^2\Big(X_{13}^{\mu\nu}{1\over |x_{23}|^2}X_{13}^{\rho\sigma}{1\over |x_{23}|^2}+X_{13}^{\mu\nu}{1\over |x_{23}|^2}X_{23}^{\rho\sigma}{1\over |x_{13}|^2}\Big)\cr 
+&((2\Delta)^2(\Delta-2))2^2 X_{13}^{\mu\nu}{1\over |x_{13}|^2}X_{23}^{\rho\sigma}{1\over |x_{23}|^2}\cr
+&(13)\longleftrightarrow(23)\Big].
}}
And lastly ${{J}^{(3)}}^{\mu\nu\rho\sigma}$:
\eqn\eqthree{\eqalign{
  {{J}^{(3)}}^{\mu\nu\rho\sigma} =&\langle :Tr(\phi^\Delta):(x_1):Tr(\phi^\Delta):(x_2)::Tr(\phi\pa_\mu\pa_\nu\phi)Tr(\pa_\rho\phi\pa_\sigma\phi):(x_3)\rangle\cr
  &={N^\Delta\over |x_{12}|^{2\Delta-4}}\Big[((2\Delta)^2(\Delta-2))2^3 X_{13}^{\mu\nu}{1\over |x_{13}|^2}{x_{23}^\rho x _{23}^\sigma\over |x_{23}|^8}+\cr
  &+\Delta^2(\Delta-1)2^3X_{13}^{\mu\nu}{1\over |x_{23}|^2}{x_{13}^\rho x_{23}^\sigma+x_{23}^\rho x_{13}^\sigma\over |x_{13}|^4 |x_{23}|^4}\cr
  &+(13)\longleftrightarrow(23)\Big].
}}
We further need to make \eqone-\eqthree\ traceless in the pairs $(\mu,\nu)$ and $(\rho,\sigma)$ and therefore define ${I^{(i)}}^{\mu\nu\rho\sigma}$ as 
\eqn\Ifinal{
  {I^{(i)}}^{\mu\nu\rho\sigma} = {J^{(i)}}^{\mu\nu\rho\sigma}-{\delta^{\mu\nu}\over 4}{{{J^{(i)}}^{\alpha}}_{\alpha}}^{\rho\sigma}-{\delta^{\rho\sigma}\over 4}{{J^{(i)}}^{\mu\nu\alpha}}_{\alpha}+{\delta^{\mu\nu}\delta^{\rho\sigma}\over 16}{{{{J^{(i)}}^{\alpha}}_{\alpha}}^{\gamma}}_{\gamma}.
}

From \eqone-\eqthree, the three-point function $\langle\OO_\Delta(x_1)\OO_\Delta(x_2)(T^2)^{\mu\nu}(x_3)\rangle$ is given by 
\eqn\threeSub{
  \langle\OO_\Delta(x_1)\OO_\Delta(x_2)(T^2)^{\mu\nu}\rangle = {1\over 12\sqrt{2}\Delta N^{\Delta+2}}({{{I^{(1)}}^{(\mu|\alpha}}_{\alpha}}^{|\nu)}-{{{I^{(3)}}^{(\mu|\alpha}}_{\alpha}}^{|\nu)}+{1\over 4}{{{{I^{(2)}}^{(\mu |\alpha}}_{\alpha}}^{|\nu)}-({\rm trace})).
}}
Explicitly we find that 
\eqn\threeSubResult{
  \langle\OO_\Delta(x_1)\OO_\Delta(x_2)(T^2)^{\mu\nu}(x_3)\rangle = {\sqrt{2}\Delta(\Delta-1)\over3N^2}{Z^\mu Z^\nu-({\rm trace})\over |x_{12}|^{2\Delta-6}|x_{13}|^{6}|x_{23}|^{6}}+\OO(N^{-4}).
}

Consider now the scalar operator $(T^2)$ defined by 
\eqn\scaldoubleDef{
	(T^2)(x) = {1\over 36\sqrt{2}N^2} :T_{\mu\nu}T^{\mu\nu}:(x). 
}
The three-point function $\langle \OO_{\Delta}(x_1)\OO_\Delta(x_2)(T^2)(x_3)\rangle$ can be found using $I^{(i)}$ defined in \Ifinal\ as follows
\eqn\threeptScalarDoubleExpl{\eqalign{
	\langle \OO_{\Delta}(x_1)\OO_\Delta(x_2)(T^2)(x_3)\rangle &={1\over 36\sqrt{2}\Delta N^{2+\Delta}}({{I^{(1)}}^{\mu\nu}}_{\mu\nu}-{{I^{(3)}}^{\mu\nu}}_{\mu\nu}+{1\over 4}{{I^{(2)}}^{\mu\nu}}_{\mu\nu})+\OO(N^{-4})\cr
	&= {\Delta(\Delta-1)\over 3\sqrt{2}N^2}{1\over |x_{12}|^{2\Delta-8}|x_{13}|^{8}|x_{23}|^{8}}+\OO(N^{-4}).
}}


\appendix{C}{Single trace operator with dimension $\Delta\sim C_T$}
In this appendix we study the single trace scalar operator $\OO_{\Delta_H}$ given by 
\eqn\heavySingleTrace{
  \OH(x)={1\over \sqrt{{\cal N}_\DH}}:Tr(\phi^{\DH}):(x),
}
with $\DH\sim C_T$ and ${\cal N}_\DH$ a normalization constant\foot{Mixing with other operators with $\Delta\sim C_T$ is not important for this discussion.}. When calculating the normalization constant ${\cal N}_\DH$ as well as the three-point functions $\langle \OH(x_1)\OH(x_2)\OO(x_3)\rangle$, non-planar diagrams generically gets enhanced by powers of $\DH$ and therefore invalidates the naive planar expansion. The goal of this appendix is to show that 
\eqn\nonplanarCorr{
  \langle \OH(x_1)\OH(x_2)\hat{\OO}(x_3)\rangle = \langle \OO_\Delta(x_1)\OO_\Delta(x_2)\hat{\OO}(x_3)\rangle|_{\Delta=\DH},
}
where $\hat{\OO}$ is either $:Tr(\phi^2):$ or, more importantly, minimal-twist multi stress tensors with any spin. Moreover, note that the LHS in \nonplanarCorr\ is in principle exact in $C_T\sim N^2$ while the RHS is obtained by keeping only planar diagrams with $\Delta\ll C_T$ and then setting $\Delta=\DH$ in the end. 

The propagator for the field $\phi$ was given in \fundprop\ by
\eqn\fundpropApp{
  \langle {\phi^i}_j(x){\phi^k}_l(y)\rangle = \left({\delta^{i}}_l{\delta^{k}}_j-{1\over N}{\delta^{i}}_j{\delta^{k}}_l\right){1\over|x-y|^2}. 
}
Consider now the three-point function $\langle:Tr(\phi^{\DH}):(x_1):Tr(\phi^{\DH}):(x_2):Tr(\phi^{2}):(x_3)\rangle$. Due to the normal ordering, one $\phi$ field in $:Tr(\phi^2):(x_3)$ need to be contracted with $:Tr(\phi^{\DH}):(x_1):$ and the other one with $:Tr(\phi^{\DH}):(x_2):$. Note that for this contraction the second term in \fundpropApp\ give a contribution proportional to $Tr(\phi(x_3))=0$. It is therefore seen that 
\eqn\heavyheavyphisq{
  \langle:Tr(\phi_1^\DH)::Tr(\phi_2^\DH)::Tr(\phi_3^{2}):\rangle = 2\DH \langle :Tr(\phi_3\phi_1^{\DH-1})::Tr(\phi_2^\DH):\rangle,
}
where we introduced the notation $\phi_i=\phi(x_i)$ and dropped the $|x_{ij}|^{-2}$ coming from \fundpropApp. The position dependence is easily restored in the end. Now it is seen that the RHS of \heavyheavyphisq\ is proportional to the two-point function\foot{Up to the position dependence.} of $\OO_H$ and we therefore find that 
\eqn\heavyheavyphisqCal{
  \langle:Tr(\phi_1^\DH)::Tr(\phi_2^\DH)::Tr(\phi_3^{2}):\rangle = 2\DH {\cal N}_\DH,
}
which is exact to all orders in $C_T$. Including the normalization factor of $\OH$ in \heavySingleTrace\ and $\OO_2$ from \singlop\ we find that 
\eqn\heavyheavyphisqexact{
  \langle\OH(x_1)\OH(x_2)\OO_2(x_3)\rangle = {\sqrt{2}\DH\over N}{1\over |x_{12}|^{2\DH-2}|x_{13}|^2|x_{23}|^{2}}+\OO(N^{-3}).
}
By comparing \heavyheavyphisqexact\ with \opeCoeffdimtwoRes\ we find that 
\eqn\comphisq{
  \lambda_{\OH\OH \OO_2} = \lambda_{\OO_\Delta\OO_\Delta\OO_2}|_{\Delta=\DH}.
}
Note that in \heavyheavyphisqexact\ the normalization of $\OH$ cancels the contribution from non-planar diagrams in limit $\Delta_H\sim C_T$. For $\Delta=2$ in \singlop, it is trivial to compute the normalization exact in $N$ to get the correction to $\lambda_{\OO_\Delta\OO_\Delta\OO_2}$ in \heavyheavyphisqexact.

Consider now the stress tensor operator defined in \stress\ and the three-point function $\langle\OH(x_1)\OH(x_2)T_{\mu\nu}(x_3)\rangle$. This is fixed by the Ward identity but is an instructive example before considering more general multi stress tensors. In the same way as the OPE coefficient was found in the $\OO_\Delta\times\OO_\Delta$ OPE, due to the tensor structure being fixed by conformal symmetry, we consider the term proportional to $x_{13}^\mu x_{13}^\nu$ in the three-point function. This comes from the $-{1\over 6\sqrt{C_T}}Tr(\phi\pa_\mu\pa_\nu\phi)$ term in the stress tensor when $\pa_\mu\pa_\nu\phi$ is contracted with one of the $\DH$ number of $\phi(x_1)$ fields. Doing this contraction we therefore see that 
\eqn\stressTensOPE{
  \langle:Tr(\phi_1^\DH)::Tr(\phi_2^\DH)::Tr(\phi_3\pa_\mu\pa_\nu\phi_3):\rangle|_{x_{13}^\mu x_{13}^\nu}= 8\DH\langle :Tr(\phi_3\phi_1^{\DH-1})::Tr(\phi_2^\DH):\rangle,
}
where the factor $8$ comes from the derivatives and we again suppress the spacetime dependence. The RHS of \stressTensOPE\ is also proportional to the normalization constant of $\OH$. Including the normalization factor of the stress tensor in \stress\ and that of $\OH$ in \heavySingleTrace, the three-point function $\langle\OH\OH T_{\mu\nu}\rangle$ can be obtained from \stressTensOPE\ from which we read off the OPE coefficient 
\eqn\stressHeavy{
  \lambda_{\OH\OH T_{\mu\nu}} = -{4\DH\over 3\sqrt{C_T}}.
}
This agrees with \TOPE . 

We now want to show that is true for minimal-twist multi stress tensors with any spin. For simplicity, consider the double-stress tensor with spin $4$ defined in \doublestress\
\eqn\doublestressApp{
  (T^2)_{\mu\nu\rho\sigma}(x) = {1\over\sqrt{2}}:T_{(\mu\nu}T_{\rho\sigma)}:(x)-({\rm traces}). 
}
Similarly to the calculation of the three-point function with the stress tensor, we can obtain the three-point function $\langle \OH(x_1)\OH(x_2)(T^2)_{\mu\nu\rho\sigma}(x_3)\rangle$ by considering the term proportional to $x_{13}^\mu x_{13}^\nu x_{13}^\rho x_{13}^\sigma$. This will be due to the term ${1\over \sqrt{2}6^2C_T}Tr(\phi\pa_\mu\pa_\nu\phi)Tr(\phi\pa_\rho\pa_\sigma\phi)$ when contracting $\pa_\mu\pa_\nu\phi$ with some $\phi(x_1)$ and likewise contracting $\pa_\rho\pa_\sigma\phi$ with some other $\phi(x_1)$. The number of such contractions is given by $\DH(\DH-1)$ and we find that 
\eqn\DoublestressTensOPE{\eqalign{
  \langle:Tr(\phi_1^\DH)::Tr(\phi_2^\DH):&:Tr(\phi_3\pa_\mu\pa_\nu\phi_3)Tr(\phi_3\pa_\rho\pa_\sigma\phi_3):\rangle|_{x_{13}^\mu x_{13}^\nu x_{13}^\rho x_{13}^\sigma} \cr
  &= 8^2\DH(\DH-1)\langle :Tr(\phi_3^2\phi_{1}^{\DH-2})::Tr(\phi_2^\DH):\rangle, 
}}
where the factor of $8^2$ again is due to acting with the derivatives and note that the position of the $\phi_3$ fields in in the last line is not important. It is again seen that the RHS of \DoublestressTensOPE\ is proportional to the normalization constant of $\OH$. Including the normalization in \doublestress\ and \heavySingleTrace\ we find the three-point function $\langle \OH\OH (T^2)_{\mu\nu\rho\sigma}\rangle$ and read off the OPE coefficient:
\eqn\tSqOPEcoeffHeavy{
  \lambda_{\OH\OH T^2_{4,4}} = {8\sqrt{2}\DH(\DH-1)\over 9C_T}+\OO(C_T^{-3/2}). 
}
which is seen to agree with $\opeTSq$ when setting $\DH=\Delta$. Note that the corrections in \tSqOPEcoeffHeavy\ are solely due to corrections in the normalization of $T^2_{4,4}$ and therefore $\lambda_{\OH\OH T^2_{4,4}}=\lambda_{\OO_\Delta\OO_\Delta T^2_{4,4}}$ to all orders in $C_T$. These arguments generalize straightforwardly to minimal-twist multi stress tensor with any spin such that the results are the same as those obtained in the planar limit for $\Delta\ll C_T^2$ in Section 3 by setting $\DH=\Delta$. The only correction in $C_T$ is then due to the normalization of the multi stress tensor.

The same argument applies to any scalar primary multi-trace operator $\OO_\Delta$, without any derivatives, with OPE coefficients given by \heavyheavyphisqexact, \stressHeavy\ and \tSqOPEcoeffHeavy.


\appendix{D}{Stress tensor thermal one-point function}
In order to calculate thermal one-point functions in the free adjoint scalar model we use the fact that the thermal correlation function is related to the zero-temperature case by summing over images. Consider now the thermal one-point function of the stress tensor. Generally, the one-point function of a spin-$s$ symmetric traceless operator with dimension $\Delta_\OO$ on $S^1\times {\bf R}^{d-1}$ is given by \IliesiuFAO
\eqn\STTOnePt{
  \langle \OO^{\mu_1\ldots\mu_s}(x)\rangle_{\beta} = {b_\OO\over \beta^{\Delta_\OO}}(e^{\mu_1}\ldots e^{\mu_s}-({\rm traces})),
}
where $e^{\mu_1}$ is a unit-vector along the thermal circle. Consider first the canonically normalized stress tensor given by $T^{\rm (can)}_{\mu\nu}={1\over 3S_d}(Tr(\pa_\mu\phi\pa_\mu\phi)-{1\over 2}Tr(\phi\pa_\mu\pa_\nu\phi)-({\rm traces}))$. In order to find the one-point function, use the following: 
\eqn\derOnepT{
  \langle Tr(\pa_\mu^{(x)}\phi(x)\pa_\nu^{(y)}\phi(y))\rangle  = {2(N^2-1)\over|x-y|^4}(\delta_{\mu\nu}-4(y-x)_\mu(y-x)_\nu{1\over |x-y|^2})
}
and 
\eqn\derOnepTTwo{
  \langle Tr(\pa_\mu^{(x)}\pa_\nu^{(x)}\phi(x)\phi(y))\rangle = {2(N^2-1) \over|x-y|^4}(-\delta_{\mu\nu}+4(y-x)_\mu(y-x)_\nu {1\over |x-y|^2}).
}
To get the thermal correlator, we use \derOnepT\ and \derOnepTTwo\ with $x,y$ along the thermal circle separated by a distance $m\beta$, with $m$ integer, and sum over $m\neq0$. The relevant terms for calculating the one-point functions in terms of fundamental fields are therefore 
\eqn\fundField{\eqalign{
  \langle Tr(\pa_\mu\phi\pa_\nu\phi)\rangle_{\beta,m} &=  -{8(N^2-1)\over (m\beta)^4}e^{\mu}e^{\nu}+{2(N^2-1)\over (m\beta)^4}\delta_{\mu\nu},\cr
  \langle Tr(\pa_\mu\pa_\nu\phi\phi)\rangle_{\beta,m} &=  {8(N^2-1)\over (m\beta)^4}e^{\mu}e^{\nu}-{2(N^2-1)\over (m\beta)^4}\delta_{\mu\nu},
}}
where we note that only the first term in each equation in \fundField\ contribute to the non-trace term in \STTOnePt.

We therefore find for the stress tensor one-point function: 
\eqn\stressOnePtEx{\eqalign{
  \langle T^{\rm (can)}_{\mu\nu}\rangle_{\beta} &= {1\over 3S_d}(\langle Tr(\pa_\mu\phi\pa_\nu\phi)\rangle_{\beta}-{1\over 2}\langle Tr(\pa_\mu\pa_\nu\phi\phi)\rangle_{\beta}- {\rm trace})\cr
  &= {-12(N^2-1)\over 3S_d} {2\zeta(4)\over \beta^4}(e_{\mu}e_{\nu}-({\rm trace})),
}}
where the $2\zeta(4)$ comes from summing over images and we therefore have 
\eqn\bt{
  b_{T^{\rm (can)}_{\mu\nu}} = -{4(N^2-1)\over S_d} 2\zeta(4) = -{4\pi^4\over 45 S_d}(N^2-1).
} 
This agrees with $f={b_{T^{\rm (can)}_{\mu\nu}}\over d}$ in eq. (2.17) in \IliesiuFAO\ for $(N^2-1)$ free scalar fields. This also agrees with $a_{2,2}={\pi^4\Delta\over 45}$ found from the two-point thermal correlator using: 
\eqn\twoone{
  a_{2,2}= {\pi^4\Delta\over 45} = \left({1\over 2}\right)^2 {\lambda_{\OO_\Delta\OO_\Delta T^{\rm (can)}}b_{T^{\rm (can)}_{\mu\nu}}\over {C_T\over S_d^2}},
}
using $\lambda_{\OO_\Delta\OO_\Delta T^{\rm (can)}}=-{4\Delta\over 3S_d}$ in this normalization and $C_T = {4\over 3}(N^2-1)$. This is simply related to the one-point function for the unit-normalized stress tensor by (to leading order in $N$)
\eqn\btnorm{\eqalign{
	b_{T_{\mu\nu}} &= {b_{T^{\rm (can)}_{\mu\nu}}\over{\sqrt{C_T}\over S_d}}\cr
	&\approx-{2\pi^4N\over 15\sqrt{3}}.
}}

Let us now consider the thermalization of the stress tensor, keeping all the index structures.
To compare the thermal two-point function with the heavy-heavy-light-light correlator, we want to relate the dimension of the heavy operator, $\Delta_H$, to the inverse temperature $\beta$. Consider the expectation value of the stress tensor in a heavy state created by $\OH$ on the cylinder ${\bf R}\times S^{3}$ 
\eqn\three{\langle \OO_{H}|T^{\mu\nu}(x_{E,2}^0,\hat{n})|\OO_{H}\rangle_{\rm cyl} =\lim_{x_3\to\infty}|x_3|^{2\DH} |x_2|^{4}\lambda_{\OH\OH T_{\mu\nu}}{{Z^{\mu}Z^{\nu}-{1\over 4}\delta^{\mu\nu}Z^{\rho}Z_{\rho}}\over{|x_{13}|^{2\Delta_{H}-2}|x_{23}|^{2}|x_{12}|^{2}}},
}
\noindent where the RHS is found by a conformal transformation to the plane with $Z^{\mu}=\left({{x_{12}^{\mu}}\over{|x_{12}|^2}}+{{x_{23}^{\mu}}\over{|x_{23}|^2}}\right)$. When $x_1=0$ and $x_3\to\infty$, it is seen that $Z^\mu= -{x_2^\mu\over |x_2|^2}$ and \three\ only depends on $\hat{x}^\mu = {x_{21}^\mu\over |x_{21}|} =\hat{r}$, where $\hat{r}$ is a radial unit vector. In radial quantization it follows that
\eqn\threee{\langle \OO_{H}|T^{\mu\nu}(x_{E,2}^0,\hat{n})|\OO_{H}\rangle_{\rm cyl}={\lambda_{\OH\OH T_{\mu\nu}}\over R^{4}}(\hat{e}_\mu \hat{e}_\nu-{1\over 4}\delta_{\mu\nu})}
where we reintroduced the radius of the sphere $R$, $\lambda_{\OH\OH T_{\mu\nu}}$ is the OPE coefficient of $T_{\mu\nu}$ in the $\OO_H\times\OO_H$ OPE and $\hat{e}_{\mu}=(1,0,0,0)$.  

The thermal one-point function of an operator $\OO_{\tau,s}$, with twist $\tau$ and spin $s$, on $S^1\times S^3$ is fixed by conformal symmetry \IliesiuFAO\
\eqn\thermSphere{
	\langle \OO_{\tau,s}(x)\rangle_\beta = {b_{\OO_{\tau,s}}f_{\OO_{\tau,s}}({\beta\over R})\over \beta^{\tau+s}}(e^{\mu_1}\cdots e^{\mu_s}-({\rm traces})),
}
where $f_{\OO_{\tau,s}}(0)=1$ and $e^{\mu}=(1,0,0,0)$.

We assume thermalization of the stress tensor in the heavy state: 
\eqn\therm{\eqalign{
  \langle \OO_{H}|T_{\mu\nu}(x)|\OO_{H}\rangle&=\langle T_{\mu\nu}(x)\rangle_{\beta}
}}
\noindent where $\langle T^{\mu\nu}(x)\rangle_{\beta}$ is the thermal one-point function at inverse temperature $\beta$ evaluated on $S^{1}\times S^{3}$, with $R$ being the radius of $S^{3}$. Using \threee-\therm\ we find 
\eqn\thermstressFin{
	{\lambda_{\OH\OH T_{\mu\nu}}\over R^{4}} = {b_{T_{\mu\nu}}f_{T_{\mu\nu}}({\beta\over R})\over \beta^{4}}.
}

Using \thermstressFin\ for $R\to\infty$ in the free adjoint scalar theory, together with the one-point function $b_{T_{\mu\nu}}=-{2\pi^4N\over 15\sqrt{3}}$ and the OPE coefficient $\lambda_{\OH\OH T_{\mu\nu}}=-{4\DH\over 3\sqrt{C_T}}$, one finds the following relation between $\mu={160\Delta_H\over 3C_T}$ and the inverse temperature $\beta$:
\eqn\muuApp{
  \mu={8\over3}\left({\pi R\over \beta}\right)^4.
}
This agrees with \muu.


\appendix{E}{Dimension-six spin-four single trace operator}
We want to calculate the contribution of the single trace operator with $\tau=2$ and $s=4$. The unit-normalised $\OO_{2,4}$ operator is given by\foot{We denote this operator either as $\OO_{2,4}$ or  $\Xi_{\mu\nu\rho\sigma}$ depending whether we want to explicitly list the indices or not.}
\eqn\operfulll{\eqalign{\Xi_{\mu\nu\rho\sigma}(x) = {1\over {96 \sqrt{35}N}} :Tr\big(&\phi (\pa_{\mu}\pa_{\nu}\pa_{\rho}\pa_{\sigma}\phi)-16(\pa_{(\mu}\phi)(\pa_{\nu}\pa_{\rho}\pa_{\sigma)}\phi)\cr
&+18(\pa_{(\mu}\pa_{\nu}\phi)(\pa_{\rho}\pa_{\sigma)}\phi)-({\rm traces})\big):(x).
}}
The relative coefficients are fixed by demanding that it is a primary operator $[K_\alpha,\Xi_{\mu\nu\rho\sigma}]=0$. Explictily, this is done using the conformal algebra
\eqn\alg{\eqalign{
[K_{\mu},P_{\nu}]&=2i(\eta_{\mu\nu}D- M_{\mu\nu}),\cr
[M_{\mu\nu},P_{\rho}]&=-i(\eta_{\rho\mu}P_{\nu}-\eta_{\rho\nu}P_{\mu}),
}}
and the action on the fundamental field $\phi$
\eqn\mom{\eqalign{
P_{\mu}\phi(0) &=-i\partial_{\mu}\phi(0),\cr
D\phi(0) &= i\phi(0).
}}
The relevant commutators in order to fix $\Xi_{\mu\nu\rho\sigma}$ are
\eqn\coms{\eqalign{[K_{\alpha},P_{\mu}\phi]=&-2\eta_{\alpha\mu}\phi,\cr
[K_{\alpha},P_{\mu}P_{\nu}\phi]=&-4\eta_{\alpha\mu}P_{\nu}\phi -4\eta_{\alpha\nu}P_{\mu}\phi+ 2\eta_{\mu\nu}P_{\alpha}\phi,\cr
[K_{\alpha},P_{\mu}P_{\nu}P_{\rho}\phi]=&-6\eta_{\alpha\mu}P_{\nu}P_{\rho}\phi -6\eta_{\alpha\nu}P_{\mu}P_{\rho}\phi -6\eta_{\alpha\rho}P_{\nu}P_{\mu}\phi\cr
&+ 2\eta_{\mu\nu}P_{\rho}P_{\alpha}\phi + 2\eta_{\rho\nu}P_{\mu}P_{\alpha}\phi + 2\eta_{\mu\rho}P_{\nu}P_{\alpha}\phi,\cr
[K_{\alpha},P_{\mu}P_{\nu}P_{\rho}P_{\sigma}\phi]=&-8\eta_{\alpha\mu}P_{\nu}P_{\rho}P_{\sigma}\phi -8\eta_{\alpha\nu}P_{\mu}P_{\rho}P_{\sigma}\phi -8\eta_{\alpha\rho}P_{\nu}P_{\mu}P_{\sigma}\phi -8\eta_{\alpha\sigma}P_{\nu}P_{\rho}P_{\mu}\phi\cr
&+ 2\eta_{\mu\nu}P_{\rho}P_{\sigma}P_{\alpha}\phi + 2\eta_{\mu\rho}P_{\nu}P_{\sigma}P_{\alpha}\phi + 2\eta_{\mu\sigma}P_{\rho}P_{\nu}P_{\alpha}\phi + 2\eta_{\nu\rho}P_{\mu}P_{\sigma}P_{\alpha}\phi \cr 
&+ 2\eta_{\nu\sigma}P_{\mu}P_{\rho}P_{\alpha}\phi + 2\eta_{\rho\sigma}P_{\mu}P_{\nu}P_{\alpha}\phi,
}}
which can also be found in e.g.\ Appendix F in \PenedonesUE.

The thermal one-point function of this operator is found from Wick contractions to be
\eqn\thopfapp{\langle \Xi_{\mu\nu\rho\sigma} \rangle_{\beta}={{8 (\pi T)^6 N}\over{27 \sqrt{35}}}\left(e_{\mu}e_{\nu}e_{\rho}e_{\sigma} - ({\rm traces})\right).
}
Moreover, the three-point function with operators $\OO_{\Delta}(x)={1\over \sqrt{\Delta N^{\Delta}}}: Tr\left(\phi^{\Delta}\right):(x)$ can again be calculated using Wick contractions similarly to how it was done for $T^2_{\mu\nu\rho\sigma}$ in Appendix A. By explicit calculation one finds
\eqn\tpfapp{\langle\OO_{\Delta}(x_1)\OO_{\Delta}(x_2)\Xi_{\mu\nu\rho\sigma}(x_3) \rangle = {{4 \Delta }\over{\sqrt{35} N}}{{Z_{\mu}Z_{\nu}Z_{\rho}Z_{\sigma} - ({\rm traces})}\over{|x_{12}|^{2\Delta -2} |x_{13}|^2 |x_{23}|^2}},
}
and therefore the OPE coefficient $\lambda_{\OO_{\Delta}\OO_{\Delta}\OO_{2,4}}$ is given by
\eqn\flambdaapp{\lambda_{\OO_{\Delta}\OO_{\Delta}\OO_{2,4}}={{4 \Delta }\over{\sqrt{35} N}}.
}
\noindent Now, it is easy to check that
\eqn\finrel{{1\over 2^4}\lambda_{\OO_{\Delta}\OO_{\Delta}\OO_{2,4}}b_{\OO_{2,4}}={{2 \pi ^6 \Delta }\over{945}},
}
which agrees with $a_{2,4}$ in \adimsixandeight.


\appendix{F}{Thermal one-point functions of multi-trace operators in the large-$N$ limit}
In \fullatdimeight, it was shown that $a_{4,4}$ was due to double trace operators which were normal ordered products of single trace operators without any derivatives. There are, however, other double trace operators that have the same quantum numbers and are schematically represented as $[\OO_a \OO_b]_{n,l}$. Concretely, the double trace operators with twist and spin four besides $(T^2)_{\mu\nu\rho\sigma}$ and $(\OO^{\rm DT})_{\mu\nu\rho\sigma}$ are $[\OO_{2}\OO_{2}]_{0,4}$ and $[\OO_{2}T_{\mu\nu}]_{0,2}$. We argue that the thermal one-point functions of these operators are subleading in the large-$N$ limit when evaluated on the plane. 

Consider the thermal one-point function of a double trace operator $[\OO_a\OO_b]_{n,l}=\OO_{a}\pa^{2n}\pa^{l}\OO_{b} + \ldots$, where $\OO_{a}$ and $\OO_{b}$ are single trace primary operators and dots represent terms where derivatives acts on $\OO_{a}$ as well, in order to make $[\OO_a\OO_b]_{n,l}$ a primary operator.  The term in the thermal one-point function that behaves as $N^{k}$ ($N^{2}$ for double trace operators) comes from contracting the fundamental field within each trace separately. Therefore we have 
\eqn\toff{\langle \OO_{a}\pa^{2n}\pa^{l}\OO_{b} \rangle_{\beta} \approx \langle \OO_{a} \rangle_{\beta} \langle\pa^{2n}\pa^{l}\OO_{b} \rangle_{\beta}+\OO(1),
}
which is simply due to large-$N$ factorization.
\noindent As $\pa^{2n}\pa^{l}\OO_{b}$ is a descendant of $\OO_{b}$, it is easy to explicitly show that $\langle\pa^{2n}\pa^{l}\OO_{b} \rangle_{\beta} = 0$ for $n\neq 0$ or $l\neq 0$, from which it follows that 
\eqn\toff{\langle \OO_{a}\pa^{2n}\pa^{l}\OO_{b} \rangle_{\beta} = \OO(1).
}
\noindent Similar reasoning holds for all terms in $[\OO_a\OO_b]_{n,l}$, so we conclude for $n\neq 0$ or $l\neq 0$ that
\eqn\tofff{\langle [\OO_{a}\OO_{b}]_{n,l} \rangle_{\beta} = \OO(1).
}
It is easy to generalise ($n$ and/or $l$ non-zero)
\eqn\toff{\langle [\OO_{a_1}\ldots\OO_{a_k}]_{n,l} \rangle_{\beta} = \OO(N^{k-2}).
}
\noindent Using the canonical scaling for the OPE coefficients \scaling\ it is found that these multi-trace operators give a suppressed contribution to the thermal two point function in the large-$N$ limit: 
\eqn\ff{\lambda_{\OO_{\Delta}\OO_{\Delta}[\OO_{a_1}\ldots \OO_{a_k}]_{n,l}}\langle [\OO_{a_1}\ldots\OO_{a_k}]_{n,l} \rangle_{\beta}= \OO\Big({1\over N^2}\Big).
}

The conclusion is that these operators with $n\neq0$ or $l\neq0$ do not contribute to the thermal two-point functions to leading order in $N$. Note that for $n=l=0$, the operator is just $:\OO_{a_1}\OO_{a_2}\ldots\OO_{a_k}:$ and it does contribute to the thermal 2pt function since
\eqn\ffnonzero{
  \lambda_{\OO_{\Delta}\OO_{\Delta}[\OO_{a_1}\ldots\OO_{a_k}]_{n=0,l=0}}\langle [\OO_{a_1}\ldots\OO_{a_k}]_{n=0,l=0} \rangle_{\beta}=\OO(1).
}

From \ff\ it is seen that multi stress tensor operators of the schematic form $[T^k]_{n,l}$ with either $n$ or $l$, or both, being non-zero will not contribute to the thermal correlator to leading order in $N$ on the plane.

\appendix{G}{Free boson in two dimensions}

In this appendix we discuss free scalars in two dimensions.
We first consider a single scalar and then the case of the $SU(N)$  adjoint scalar. 
We compute two-point functions of a particular class of quasi-primary operators at finite temperature $1/\beta$. These two-point functions are not determined by the conformal symmetry, because the quasi-primary operators do not transform covariantly from the plane to the cylinder. They transform covariantly only with respect to the global conformal transformations. The only operators that have the non-zero thermal one-point functions are the Virasoro descendants of the vacuum and therefore, only these operators contribute to the thermal two-point function of the quasi-primary operators\foot{We check this explicitly up to the $\OO(1/\beta^4)$.}. Virasoro descendants of the vacuum have different OPE coefficients with external quasi-primary operators compared with the case when primary external operators are considered.\foot{Deviation from the Virasoro vacuum block in the Regge limit of four-point HHLL correlator is observed in \GiustoMUP\ as well.}

\subsec{Review free boson in two dimensions}

We consider single free boson $\phi(z)$ in two dimensions. The stress tensor can be written in terms of Virasoro modes as
\eqn\stssvm{T(z)=\sqrt{2}\sum_{n}z^{-n-2}L_{n}.
}

\noindent This stress tensor is unit-normalized 
\eqn\un{\langle T(z)T(w) \rangle = {1\over (z-w)^4}.
}

The fundamental field can be expressed as Laurent series  
\eqn\fdfex{\pa \phi(z)=\sum_{n=-\infty}^{+\infty}z^{-n-1}\alpha_{n},
} 
\noindent where oscillators $\alpha_{n}$ obey the following algebra
\eqn\alg{[\alpha_{n},\alpha_{m}]=n\delta_{n+m,0}.
}
\noindent They act on the vacuum as
\eqn\va{\alpha_{n}\vac = 0, \qquad n\geq 0.
}
\noindent The two-point function of the fundamental fields is given by
\eqn\tpfff{\langle\pa\phi(z)\pa\phi(w)\rangle ={1\over {(z-w)^2}}.
}

The unit-normalized stress tensor can be expressed in terms of the fundamental field as 
\eqn\stsff{T(z)={1\over \sqrt{2}}:\pa \phi \pa \phi:(z)={1\over \sqrt{2}}\sum_{m,n}z^{-m-n-2}:\alpha_{m}\alpha_{n}:,
}
\noindent where $:ab:$ denotes product of operators $a$ and $b$ with the corresponding free theory oscillators being normally ordered such that the operators annihilating the vacuum are put at the rightmost position. Then, it follows
\eqn\lalp{L_{n}={1\over 2}\sum_{m}:\alpha_{n-m}\alpha_{m}:={1\over 2}\left(\sum_{m\geq 0}\alpha_{n-m}\alpha_{m}+\sum_{m<0}\alpha_{m}\alpha_{n-m}\right).
}



\subsec{Thermal two-point function of quasi-primary operator}

We are interested in computing the thermal two-point function of quasi-primary operators at temperature $1/\beta$. Quasi-primary operators $\OO(z)$ are defined as $[L_1,\OO(z)]=0$, or equivalently, in therms of their asymptotic in-states $\OO(0)\vac =|\OO\rangle$, as $L_{1}|\OO\rangle = 0$. We denote the quantum numbers of quasi-primary operators that correspond to eigenvalues of $L_{0}$ and $\bar{L}_{0}$ by $(h,\bar{h})$. We consider the following unit-normalized quasi-primary operator with quantum numbers $(h,0)$
\eqn\exst{\OO_{h}(z)={1\over \sqrt{h!}}:(\pa \phi)^h:(z)={1\over \sqrt{h!}} \sum_{m_{1},m_{2},\ldots ,m_{h}}z^{-\sum_{i=1}^{h}m_{i}-h}:\alpha_{m_{1}}\ldots\alpha_{m_{h}}:,
}
\noindent which is properly defined when $h$ is a positive integer. Its asymptotic in-state is given by
\eqn\exvac{|\OO_{h}\rangle = \OO_{h}(0)\vac = {1\over \sqrt{h!}}(\alpha_{-1})^{h}\vac.
}
\noindent One can check that this operator is a quasi-primary but not a Virasoro primary. 

The thermal two-point function of this operator for even $h$ is given by  
\eqn\thermtwopt{\eqalign{
  \langle\OO_h(z)\OO_h(0)\rangle_{\beta} =& \sum_{n=0}^{{1\over 2}(h-2)}{h!\over {4^n(h-2n)!}}\Big({2\zeta(2)\over\beta^{2}}\Big)^{2n}\left(\sum_{m=-\infty}^\infty {1\over (z+m\beta)^2}\right)^{h-2n}\cr
  &+{{2^h \pi}\over{\Gamma \left({{1}\over{2}}-{{h}\over{2}}\right)^2 \Gamma (h+1)}}\Big({2\zeta(2)\over\beta^2}\Big)^{h}.
}}
\noindent This expression is obtained by writing all possible Wick contractions between fundamental fields $\pa \phi$, including those that belong to same operator $\OO_{h}$, that we call self-contractions. Fundamental fields are separated along the thermal circle in all Wick contractions. Factors $\left({2\zeta(2)\over\beta^2}\right)$ are due to the self-contractions, 
\eqn\fact{\sum_{m=-\infty, m\neq 0}^{\infty}{1\over \beta^{2}m^2}=\left({2\zeta(2)\over\beta^2}\right).
}
\noindent The sum over $n$ comes from doing $n$ self-contractions within each of the external operators. Term ${h!\over {4^n(h-2n)!}}$ counts the number of Wick contractions with $n$ self-contractions for each external operator, including $1/\sqrt{h!}$ normalization factors. The term in the second line of \thermtwopt\ is due to the case when we take $n=h/2$ self-contractions in both external operators, i.e. it represents the disconnected contribution. 

Since the state $\OO_{h}$ is quasi-primary, it transforms properly only with respect to the global conformal transformation. These are just the M\"obius transformations in two-dimensional spacetime $z \to {{az + b}\over {cz+d}}$, with $ad-bc=1$. On the other hand, the usual way to calculate the thermal two-point function of primary operators in two dimensions is to do a conformal transformation from the plane to the cylinder with radius $\beta$, $z \to {\beta \over 2\pi}\log(z)$. This transformation is clearly not one of the M\"obius transformations and that is why we can not use this method to compute the thermal two-point functions of quasi-primary operators. 

Expanding \thermtwopt\ for $T={1\over \beta}\to0$ one finds 
\eqn\exptwoptther{
  z^{2h}\langle\OO_h(z)\OO_h(0)\rangle_{\beta}=1+{h\over 3}{(\pi z)^2\over \beta^2}+{h(h-{1\over 5})\over 12}{(\pi z)^4\over \beta^{4}}+\OO\left({1\over \beta^6}\right).
}

\subsec{Quasi-primaries, OPE coefficients, and thermal one-point functions}

In expansion \exptwoptther , terms $\OO(z^{h_{1}})$ are due to the quasi-primary operator with quantum numbers $(h_{1},0)$ in the operator product expansion $\OO_{h}\times\OO_{h}$. Identity in the expansion is due to the identity operator. We show that the second term on the RHS is due to the stress tensor. The quantum numbers of stress tensor $T(z)$ are $(2,0)$. First, we evaluate the thermal one-point function of the stress tensor
\eqn\opfst{\langle T\rangle_{\beta}={1\over \sqrt{2}}\sum_{m=-\infty, m\neq 0}^{\infty}{1\over {\beta^2 m^2}}={{\pi^2}\over{3\sqrt{2}\beta^{2}}}.
}
\noindent This is obtained by the Wick contractions of fundamental fields in the stress tensor, that are separated along the thermal circle. The same result can be obtained by the transform of the stress tensor from the plane to the cylinder using the Schwarzian derivative.

We define the OPE coefficient of unit-normalized operator $\OO$, with quantum numbers $(h_{\OO},0)$, with two $\OO_{h}$ operators as
\eqn\opedef{\langle\OO_{h}(z_{1})\OO_{h}(z_{2})\OO(z_{3}) \rangle ={\lambda_{\OO_{h}\OO_{h}\OO}\over{(z_1-z_3)^{h_{\OO}} (z_2-z_3)^{h_{\OO}}(z_1-z_2)^{2h-h_{\OO}}}}.
}

Next, we evaluate its OPE coefficient of the stress tensor with $\OO_{h}$ by doing the Wick contractions between fundamental fields
\eqn\opest{\langle\OO_{h}(z_{1})\OO_{h}(z_{2})T(z_{3}) \rangle =\sqrt{2}h{1\over{(z_1-z_3)^2 (z_2-z_3)^2(z_1-z_2)^{2(h-1)}}},
}
\noindent therefore $\lambda_{\OO_{h}\OO_{h}T}=\sqrt{2}h$. This OPE coefficient is fixed by the Ward identity. Now, it follows
\eqn\stf{z^{2}\lambda_{\OO_{h}\OO_{h}T}\langle T \rangle_{\beta}={h\over 3}{(\pi z)^2\over \beta^2},
}
\noindent which reproduces the second term on the RHS of \exptwoptther .

We are now interested in the contributions of quasi-primary operators with quantum numbers $(4,0)$. There are only two linearly independent operators with these quantum numbers given by\foot{Both of them are unit-normalized.}
\eqn\dst{:TT:(z)={1\over \sqrt{24}}:(\pa \phi)^{4}:(z)={1\over \sqrt{24}}\sum_{a,b,c,d}z^{-a-b-c-d-4}:\alpha_{a}\alpha_{b}\alpha_{c}\alpha_{d}:,
}
\eqn\lbfdef{\Lambda_{4}(z)=\sqrt{{10\over 27}}\left(\sum_{m,n=-\infty}^\infty z^{-m-n-4} *L_mL_n*-{3\over 10}\sum_{m=-\infty}^\infty z^{-m-4}(m+2)(m+3)L_m\right),
}
\noindent where $*ab*$ denotes the product where the relevant Virasoro generators are normally ordered. It should be noted that the operator $\Lambda_{4}(z)$ is Virasoro descendant of unity, while $:TT:(z)$ is not. The relevant asymptotic in-states are given by
\eqn\dstvac{\eqalign{&|:TT:\rangle =:TT:(0)|0\rangle = {1\over \sqrt{24}}(\alpha_{-1})^{4}|0\rangle, \cr
&|\Lambda_{4}\rangle =\Lambda_{4}(0)\vac=\sqrt{{10\over 27}}\left(L_{-2}^{2}-{3\over 5}L_{-4}\right)\vac.
}}

\noindent In terms of oscillators, $|\Lambda_{4}\rangle$ state can be represented as
\eqn\lbal{|\Lambda_{4}\rangle=\sqrt{{10\over 27}}\left({1\over 4}(\alpha_{-1})^4+{2\over 5}\alpha_{-1}\alpha_{-3}-{3\over 10}(\alpha_{-2})^2\right)\vac.
}
\noindent From eqs.~\dstvac\ and \lbal\ one can see that $|:TT:\rangle$ and $|\Lambda_{4}\rangle$ are the only quasi-primary states with quantum numbers $(4,0)$. Namely, all such states have to be linear combinations of the following states
\eqn\basis{\alpha_{-4}\vac,\qquad \alpha_{-3}\alpha_{-1}\vac, \qquad \alpha_{-2}^2\vac, \qquad \alpha_{-2}\alpha_{-1}^{2}\vac, \qquad \alpha_{-1}^{4}\vac,} 
\noindent because 
\eqn\lzero{L_{0}\left(\prod_{i=1}^{N}\alpha_{-k_{i}}\right)\vac=\left(\sum_{i=1}^{N}k_{i}\right)\left(\prod_{i=1}^{N}\alpha_{-k_{i}}\right)\vac,
}
\noindent where $k_{i}>0$. It is straightforward to check 
\eqn\states{\eqalign{L_{1}\alpha_{-4}\vac &= 4\alpha_{-3}\vac,\cr
L_{1}\alpha_{-3}\alpha_{-1}\vac &= 3\alpha_{-2}\alpha_{-1}\vac,\cr
L_{1}\alpha_{-2}^2\vac &= 4\alpha_{-2}\alpha_{-1}\vac,\cr
L_{1}\alpha_{-2}\alpha_{-1}^2\vac &= 2\alpha_{-1}^{3}\vac,\cr
L_{1}\alpha_{-1}^{4}\vac &= 0.\cr
}}
\noindent It follows that $\alpha_{-1}^{4}\vac$ is already quasi-primary and one can make only one more as $\alpha_{-3}\alpha_{-1}\vac -{3\over 4} \alpha_{-2}\alpha_{-2}\vac$.\foot{These states are not unit-normalized.} $|:TT:\rangle$ and $|\Lambda_{4}\rangle$ are just the linear combination of these two states with overall normalization.

Now, one can calculate the overlap of $|:TT:\rangle$ and $|\Lambda_{4}\rangle$ states as
\eqn\overlap{\langle 0|\Lambda_{4}(0):TT:(0)\vac = {\sqrt{5}\over 3}.
}

\noindent The state orthogonal to $|\Lambda_{4}\rangle$ can be written as 
\eqn\ortlb{|\tilde{\Lambda}_{4}\rangle = {3\over 2}\left(:TT:(0)- {\sqrt{5}\over 3} \Lambda_{4}(0)\right)\vac.
}
\noindent Using \dstvac\ and \lbal , it can be written in terms of free theory oscillators.

We compute the OPE coefficients of $:TT:$ and $\Lambda_{4}$ with two $\OO_{h}$ operators. We express all states in terms of free theory oscillators and use algebra \alg\ to find
\eqn\ttope{\lambda_{\OO_{h}\OO_{h}:TT:} = \langle \OO_{h} | \OO_{h}(1)|:TT:\rangle = {{\sqrt{6}}\over 2}h(h-1),
}
\eqn\lfope{\lambda_{\OO_{h}\OO_{h}\Lambda_4} = \langle \OO_{h} | \OO_{h}(1)|\Lambda_4\rangle = \sqrt{{5 \over 6}}h\left(h-{1\over 5}\right),
}
\eqn\lfope{\lambda_{\OO_{h}\OO_{h}\tilde{\Lambda}_4} = \langle \OO_{h} | \OO_{h}(1)|\tilde{\Lambda}_4\rangle = {2\over \sqrt{6}}h\left(h-2\right).
}

Now, we evaluate the thermal one-point functions of $\Lambda_{4}$ and $\tilde{\Lambda}_{4}$. From (3.4) in \DattaJEO\ we have 
\eqn\kraus{\langle *T^2* \rangle_{\beta}={{3 \pi ^4}\over{20 \beta ^4}},
}
\noindent which is the thermal one-point function of the first term on the RHS of \lbfdef .  The second term can be written as $-{3\over 10}\sum_{m=-\infty}^\infty z^{-m-4}(m+2)(m+3)L_m = -{3\over {10\sqrt{2}}}\pa^{2}T(z).$ It is clear that it will not affect the thermal one-point function of $\Lambda_{4}(z)$, as $\langle\pa^2 T\rangle_{\beta}=0$.

Therefore, from \lbfdef , we have 
\eqn\lbdth{\langle\Lambda_{4}\rangle_{\beta}=\sqrt{{10\over 27}}\langle *T^2* \rangle_{\beta} = {{\pi ^4}\over{2 \sqrt{30} \beta ^4}}.
}
\noindent Now, it follows
\eqn\dst{z^{4}\langle\Lambda_{4}\rangle_{\beta}\lambda_{\OO_{h}\OO_{h}\Lambda_{4}}={{\pi ^4 z^4}\over{12 \beta ^4}}h \left(h-{{1}\over{5}}\right),
}
\noindent which is the third therm at the RHS of \exptwoptther . On the other hand, we can evaluate the thermal one-point function of $:TT:(z)$ operator by Wick contractions of fundamental fields separated along the thermal circle
\eqn\ttopf{\langle :TT: \rangle_{\beta}={{\pi ^4}\over{6 \sqrt{6} \beta ^4}}.
}
\noindent Using \ortlb , it is straightforward to confirm that $\langle \tilde{\Lambda}_{4}\rangle_{\beta} = 0$. Therefore, as we expected, operator $\tilde{\Lambda}_{4}$ does not contribute to the thermal two-point function of $\OO_{h}$ operators, even thought it is present in the operator product expansion $\OO_{h}\times\OO_{h}$. 

This is a general property of two-dimensional CFTs, that only the operators in the Virasoro vacuum module have non-zero expectation value on the cylinder.


\subsec{Free adjoint scalar model in two dimensions}
In this subsection we study a large-$c$ theory.
Consider the free adjoint $SU(N)$ scalar in 2d with 
\eqn\phiosc{
  {\pa\phi(z)^a}_b = \sum_m z^{-m-1}{{(\alpha_m)}^a}_b
}
with 
\eqn\commtwod{
  [{{(\alpha_m)}^a}_b,{{(\alpha_n)}^c}_d] = m \delta_{m+n}\Big({\delta^a}_d{\delta^c}_b-{1\over N}{\delta^a}_b{\delta^c}_d\Big).
}
The thermal two point of the quasi-primary operator $\OO_h={1\over \sqrt{h N^h}}:Tr((\pa \phi)^h):$ follows immediately from the result in four dimensions upon replacing the propagator of fundamental fields. We find that
\eqn\thermaltwopttwod{\eqalign{
  \langle \OO_{h}(z)\OO_h(0)\rangle_\beta=g_{2d}(z)^h +{\pi^4 h(h-2)\over 9\beta^4}g_{2d}(z)^{h-2}+\ldots,
}}
where 
\eqn\ggfunctwod{\eqalign{
  g_{2d}(z) &= \sum_{m=-\infty}^{\infty}{1\over (z+m\beta)^2}\cr
  &=\Big({\pi\over \beta \sin(\pi z/\beta)}\Big)^2.
}}
Expanding \thermaltwopttwod\ for $\beta\to\infty $ we find 
\eqn\exptwod{
  \langle \OO_{h}(z)\OO_{h}(0)\rangle_\beta= z^{-2h}\Big[1+{{\pi^2 h} \over {3\beta^2}}z^2+{\pi^4 h(15h -19)\over 90\beta^4}z^4+\OO(\beta^{-6})\Big].
}

Consider first the normalized stress tensor which is given by 
\eqn\stresstensrtwod{
  T={1\over \sqrt{2}N}:Tr(\pa \phi\pa\phi):,
}
with $c=N^2$ so that $\langle T(z)T(0)\rangle={1\over z^4}$. By calculating the OPE coefficient with $\OO_h$ and the thermal one-point function of $T$, one finds that these are the same as those for the scalar $Tr(\phi^2)$ operator in four dimensions so that $\la T\ra_{\beta}={\pi^2N\over 3\sqrt{2}\beta^2}$ and $\lambda_{\OO_h\OO_h T}={\sqrt{2}h\over N}$, and the product reproduces the weight two term in \exptwod:
\eqn\prodttwod{
	\la T\ra_{\beta} \lambda_{\OO_h\OO_h T} = {\pi^2 h \over 3\beta^2}.
}

Consider now $*TT*$ defined by 
\eqn\confOrdTSQ{
  *TT*(0) = \lim_{z\to 0}T(z)T(0)-({\rm sing.\ terms}).
} 
The OPE of the stress tensor in \stresstensrtwod\ can be found in the free theory by first performing Wick contractions
\eqn\ttOPE{\eqalign{
  T(z)T(0) &= {1\over 2N^2}:Tr(\pa \phi(z)\pa\phi(z))::Tr(\pa \phi(0)\pa\phi(0)):\cr
  &=:TT:(0)+\ldots +{2\over N^2 z^2}:Tr(\pa \phi(z)\pa\phi(0)):+ {1\over z^4},
}
}
and expanding the second term in \ttOPE\ for $z\to0$ we find 
\eqn\ttOPEcont{\eqalign{
  T(z)T(0) &= :TT:(0)+\ldots +{2\over N^2 z^2}:Tr(\pa \phi(0)\pa\phi(0)):\cr
  &+{2\over N^2 z}:Tr(\pa^2 \phi(0)\pa\phi(0)):+{1\over N^2}:Tr(\pa^3 \phi(0)\pa\phi(0)):+\ldots\cr
  &+ {1\over z^4},
}}
where the dots refer to higher order terms in $z$. Inserting the OPE \ttOPEcont\ in \confOrdTSQ\ we find that 
\eqn\conf{
  *TT*(0) = :TT:(0)+{1\over N^2}:Tr(\pa^3 \phi(0)\pa\phi(0)):.
}
Consider the state $*TT*(0)|0\rangle$, which is given in terms of oscillator modes by  
\eqn\statestarT{
  *TT*(0)|0\rangle = {1\over 2N^2}Tr(\alpha_{-1}^2)Tr(\alpha_{-1}^2)|0\rangle + 2{1\over N^2}Tr(\alpha_{-3}\alpha_{-1})|0\rangle.
}
Now $Tr(\alpha_{-1}^m)|0\rangle$ is a quasi-primary while $Tr(\alpha_{-3}\alpha_{-1})|0\rangle$ is not. One way to make it a quasi-primary is to simply remove the second term in \statestarT\ and then we get a quasi-primary state which is just $:TT:|0\rangle$. Another option is to remove a descendant of the stress tensor to construct $|\Lambda_4\rangle$. To do the latter we need to remove the descendant of the stress tensor with weight $4$ given by $\pa^2 T$ 
\eqn\pasqT{
  \pa^2 T = {\sqrt{2}\over N}:Tr(\pa^3\phi \pa \phi):+{\sqrt{2}\over N}:Tr(\pa^2\phi\pa^2\phi):.
}
Acting on the vacuum we find
\eqn\pasqTState{
  \pa^2 T(0)|0\rangle = {2\sqrt{2}\over N}Tr(\alpha_{-3}\alpha_{-1})|0\rangle+{\sqrt{2}\over N}Tr(\alpha_{-2}^2)|0\rangle.
}
Consider now $L_1={\sqrt{2}\over N}(Tr(\alpha_{-1}\alpha_{2})+Tr(\alpha_{-2}\alpha_{3}+\ldots))$ which acts as $L_1Tr(\alpha_{-2}^2)|0\rangle={4\sqrt{2}\over N}Tr(\alpha_{-1}\alpha_{-2})|0\rangle$ and as $L_1Tr(\alpha_{-3}\alpha_{-1})|0\rangle={3\sqrt{2}\over N}Tr(\alpha_{-1}\alpha_{-2})|0\rangle$. We can therefore construct a quasi-primary state annihilated by $L_1$: $Tr(\alpha_{-3}\alpha_{-1})|0\rangle-{3\over 4}Tr(\alpha_{-2}^2)|0\rangle$. The quasi-primary $|\Lambda_4\rangle$ is then given by:
\eqn\lambdafour{\eqalign{
  |\Lambda_4\rangle &= {1\over\sqrt{2}}\Big[*TT*(0)|0\rangle -{3\over 5\sqrt{2}N}\pa^2T(0)|0\rangle\Big]\cr.
    &={1\over2\sqrt{2}N^2}\Big[Tr(\alpha_{-1}^2)Tr(\alpha_{-1}^2)|0\rangle-{6\over 5}Tr(\alpha_{-2}^2)|0\rangle+{8\over 5}Tr(\alpha_{-1}\alpha_{-3})|0\rangle\Big] 
}}

There are two more weight $4$ single trace quasi-primary operators given by
\eqn\ohfour{\eqalign{
  \OO^{(1)} &= {1\over 2N^2}Tr((\pa\phi)^4)\cr
  \OO^{(2)} &= {n_{\OO^{(2)}}\over N}(Tr(\pa^3\phi\pa\phi)-{3\over 2}Tr(\pa^2\phi\pa^2\phi)),\cr
        &= {n_{\OO^{(2)}}\over N}({1\over 2}\pa^2 Tr(\pa\phi\pa\phi)-{5\over 2}Tr(\pa^2\phi\pa^2\phi)),
}}
where $n_{\OO^{(2)}}$ is some $N$-independent normalization constant. The state $|\Lambda_4\rangle$ can be written in terms of $:TT:(0)|0\rangle+a \OO_2(0)|0\rangle$ in the following way\
\eqn\lambdafourtt{
  |\Lambda_4\rangle = {1\over\sqrt{2}}\Big[:TT:(0)|0\rangle+{2\over 5N n_{\OO^{(2)}}}\OO^{(2)}|0\rangle\Big].
}
The OPE coefficient for $:TT:$ is up to a normalization the same as the scalar dimension $4$ double trace operator in 4d and is given by 
\eqn\ttOPEtwod{\eqalign{
  \langle \OO_h \OO_h :TT:\rangle&={1\over h N^h}{1\over 2N^2}4 h^2(3h-5)N^h{1\over z_{13}^4z_{23}^4z_{12}^{2h-4}}\cr
    &= {1\over N^2}2h(3h-5){1\over z_{13}^4z_{23}^4z_{12}^{2h-4}},
}
}
where $4h^2(3h-5)$ come from the number of contractions giving planar diagrams. Consider now the OPE coefficient for $\OO^{(2)}$. One finds
\eqn\otwoOPEcoeff{\eqalign{
  \langle \OO_h\OO_h \OO^{(2)}\rangle &= {n_{\OO^{(2)}} N^h \over h N^{h+1} z_{13}^4z_{23}^4z_{12}^{2h-2}}\Big[(-2)(-3)h^2(z_{13}^2+z_{23}^2)-{3\over 2}2 h^2(-2)^2)z_{13}z_{23}\Big]\cr
  &= {6h n_{\OO^{(2)}}\over N z_{13}^4z_{23}^4z_{12}^{2h-4}}.
}}
Using \ttOPEtwod, \otwoOPEcoeff\ and \lambdafourtt\ we find the OPE coefficient for $|\Lambda_4\rangle$ 
\eqn\opeLambdafourfin{
  \langle \OO_h \OO_h \Lambda_4\rangle = {\sqrt{2}h(15h-19)\over 5N^2}.
}
Note that the $h$ dependence matches that of the weight $4$ term in the two-point function \exptwod . Additionally, the OPE coefficient given by \opeLambdafourfin\ can not be extrapolated to the limit when $h \sim C_T$, as in this limit the planar expansion used for calculating \opeLambdafourfin\ breaks down. For this reason, we can not test the thermalization of $\Lambda_4$ in heavy state $\OO_{h_H}$. Let us consider the thermal one-point function which is given by 
\eqn\lambdafourOnepT{
  \langle\Lambda_4\rangle_{\beta} = \Big[{1\over\sqrt{2}}b_T^2+\OO(1)\Big] = {\pi^4N^2\over 18\sqrt{2}\beta^4},
}
where the term $\propto{1\over N}\langle\OO^{(2)}\rangle_\beta$ is subleading since it is single trace. We find that
\eqn\opetimesthermal{
  \langle\Lambda_4\rangle_{\beta}\lambda_{\OO_h\OO_h \Lambda_4} = {\pi^4 h(15h-19)\over 90\beta^4},
}
which agrees with the weight $4$ term in \exptwod. 

Note that it is explicitly seen that one can write $\Lambda_4$ either as $*TT*+({\rm desc.\, of\, T})$ or as $:TT:+{1\over N}\OO_{ST}$ with $\OO_{ST}$ a quasi-primary single trace operator. In this case the single trace operator which one needs to add to $:TT:$ to get $\Lambda_4$ can be written as a sum of descendants $\OO^{(2)}\propto \pa^2 T-{5\over\sqrt{2}}Tr(\pa^2\phi\pa^2\phi)$. Explicitly, we have
\eqn\Lambdafourtwoways{\eqalign{
  |\Lambda_4\rangle &= {1\over\sqrt{2}}\Big[*TT*(0) -{3\over 5\sqrt{2}N}\pa^2T(0)\Big]|0\rangle \cr
  &={1\over \sqrt{2}}\Big[:TT:(0)+{2\over 5N n_{\OO^{(2)}}}\OO^{(2)}\Big]|0\rangle.
}}
As we saw above, using the second line in \Lambdafourtwoways\ it is straightforward to calculate correlation functions using Wick contractions to see that $\Lambda_4$ gives the full weight four contributions to the thermal two-point function for large-$N$ theories.

Now, we consider the following quasi-primary operator 
\eqn\stdel{
\OO_{\Delta}(z,\zbar) = {{\sqrt{2}}\over{\sqrt{\Delta}N^{\Delta/2}}} :Tr\left((\pa \phi \bar{\pa}\bar{\phi})^{\Delta\over 2} \right):(z,\zbar),
}
\noindent where we denote the anti-holomorphic part of the free field by $\bar{\phi}=\bar{\phi}(\zbar)$. The thermal two-point function of this operator, up to the terms subleading in large-$N$ expansion, is given by
\eqn\ttwpfno{\eqalign{
&\la \OO_{\Delta}(z,\zbar) \OO_{\Delta}(0,0) \ra_\beta = {{\pi ^{2\Delta}}\over{\beta ^{2\Delta} \sin ^\Delta \left({{\pi  z}\over{\beta }}\right) \sin ^\Delta \left({{\pi  \zbar}\over{\beta }}\right)}}\cr
& ={1\over{(z \bar{z})^{\Delta}}}\left(1+{{\pi^2\Delta(z^2+\bar{z}^2)}\over{6\beta^2} } + {{\pi ^4 \Delta  (5 \Delta +2)}\over{360 \beta ^4}}(z^4+\bar{z}^4 )+{{\pi ^4 \Delta ^2}\over{36 \beta ^4}}z^2\bar{z}^2  +\OO\left({1\over\beta^6}\right) \right).
}} 
One can easily check that the OPE coefficients of stress tensor $T$ and its anti-holomorphic partner $\bar{T}$ with $\OO_{\Delta}$ are given by
\eqn\opetns{\lambda_{\OO_{\Delta}\OO_{\Delta}T}=\lambda_{\OO_{\Delta}\OO_{\Delta}\bar{T}}={{\Delta}\over{\sqrt{2}N}},
}
while their thermal one-point function are given by
\eqn\thtns{\la T\ra_{\beta}=\la \bar{T}\ra_{\beta}={{\pi^2 N}\over{3\sqrt{2}}\beta^{2}}.
}
It is easy to check that terms proportional to $\beta^{-2}$ in \ttwpfno\ are contributions of $T$ and $\bar{T}$ operators 
\eqn\tcontns{\la T\ra_{\beta}\lambda_{\OO_{\Delta}\OO_{\Delta}T}z^2+\la \bar{T}\ra_{\beta}\lambda_{\OO_{\Delta}\OO_{\Delta}\bar{T}}\bar{z}^2={{\pi^2\Delta(z^2+\bar{z}^2)}\over{6\beta^2}}. 
}
We compute the OPE coefficient of operators $\Lambda_{4}$, defined by \lambdafour , and its anti-holomorphic partner $\bar{\Lambda}_{4}$ with $\OO_{\Delta}$ and obtain
\eqn\opettns{\lambda_{\OO_{\Delta}\OO_{\Delta}\Lambda_4}=\lambda_{\OO_{\Delta}\OO_{\Delta}\bar{\Lambda}_{4}}={{\Delta  (5 \Delta +2)}\over{10\sqrt{2}N^2}},
}
\noindent which agrees with (C.26) in \KulaxiziTKD . Its thermal one-point function (which is the same as $\la \bar{\Lambda}_4 \ra_{\beta}$) is given by \lambdafourOnepT . Another operator that contributes to thermal two-point function \ttwpfno\ is $:T\bar{T}:$. Its OPE coefficient with $\OO_{\Delta}$ and thermal one-point function are given by
\eqn\ttbns{\eqalign{\lambda_{\OO_{\Delta}\OO_{\Delta}:T\bar{T}:}&={\Delta^2\over{2N^2}}\cr
\la :T\bar{T}:\ra_{\beta}&={{\pi^4 N^2}\over{18\beta^{4}}}.
}}
Again, it is easy to check
\eqn\conttwns{\eqalign{\la \Lambda_4 \ra_{\beta} \lambda_{\OO_{\Delta}\OO_{\Delta}\Lambda_{4}}z^4+  \la \bar{\Lambda}_{4}\ra_{\beta}\lambda_{\OO_{\Delta}\OO_{\Delta}\bar{\Lambda}_{4}}\bar{z}^4 &+\la :T\bar{T}:\ra_{\beta}\lambda_{\OO_{\Delta}\OO_{\Delta}:T\bar{T}:}z^2\bar{z}^2 =\cr
&= {{\pi ^4 \Delta  (5 \Delta +2)}\over{360 \beta ^4}}(z^4+\bar{z}^4 )+{{\pi ^4 \Delta ^2}\over{36 \beta ^4}}z^2\bar{z}^2,
}}
which matches with the corresponding terms in \ttwpfno . 

The OPE coefficients $\lambda_{\OO_{\Delta}\OO_{\Delta}\Lambda_{4}}$, $\lambda_{\OO_{\Delta}\OO_{\Delta}\bar{\Lambda}_{4}}$, and $\lambda_{\OO_{\Delta}\OO_{\Delta}:T\bar{T}:}$ can be extrapolated to the limit $\Delta \sim N^2$, by the same logic as in Appendix C. Then, we can explicitly check the thermalization property of $\Lambda_{4}$, $\bar{\Lambda}_{4}$, and $:T\bar{T}:$. To establish a relation between the inverse temperature $\beta$ and the conformal dimension $\Delta_{H}$ of heavy state $\OO_{H}=\OO_{\Delta\sim N^2}$, we assume the thermalization of stress tensor
\eqn\thttdns{\la T \ra_{\beta} = \lambda_{\OO_{H}\OO_{H}T}, 
}
which implies
\eqn\reltdns{{\Delta_{H}\over{N^2}}={\pi^2\over{3\beta^2}}.
}
Using this relation, it is easy to show
\eqn\thermaltdns{\eqalign{\la \Lambda_{4}\ra_{\beta}&=\lambda_{\OO_{H}\OO_{H}\Lambda_{4}}\Big|_{\Delta_{H}^2\over N^2},\cr
\la \bar{\Lambda}_{4}\ra_{\beta}&=\lambda_{\OO_{H}\OO_{H}\bar{\Lambda}_{4}}\Big|_{\Delta_{H}^2\over N^2},\cr
\la :T\bar{T}:\ra_{\beta}&=\lambda_{\OO_{H}\OO_{H}:T\bar{T}:}\Big|_{\Delta_{H}^2\over N^2}.
}}
This means that operators $\Lambda_{4}$, $\bar{\Lambda}_{4}$, and $:T\bar{T}:$ thermalize in the quasi-primary state $\OO_{H}$ similarly to the thermalization in a Virasoro primary states in large-$c$ theory, that was analyzed in \BasuKZO .

\appendix{H}{Vector model}
In this section we study the free scalar vector model at large-$N$. Consider the scalar operator 
\eqn\scalVec{
  \OO_\Delta = {1\over \sqrt{\NN(\Delta)}}:(\varphi^i \varphi^i)^{\Delta\over 2}:(x),
}
where $\NN(\Delta)$ is a normalization constant which to leading order in $N$ is given by
\eqn\normV{
  \NN(\Delta) \approx (\Delta)!! N^{\Delta\over 2}. 
}
The thermal two-point function is given by 
\eqn\thermaltwoptVec{\eqalign{
  \langle \OO_{\Delta}(x)\OO_\Delta(0)\rangle_\beta=\tilde{g}(x_{E}^{0},|{\bf x}|)^\Delta+\Big({\Delta\over 2}\Big)^2 {1\over \Delta}\tilde{g}(x_{E}^{0},|{\bf x}|)^{\Delta-2}+\ldots,
}}
where 
\eqn\ggfunc{\eqalign{
  \tilde{g}(x_{E}^{0},|{\bf x}|) &= \sum_{m=-\infty}^{\infty}{1\over (x_{E}^{0}+m\beta)^2+{\bf x}^2}\cr
  &={\pi\over 2\beta|{\bf x}|}\Big[ {\rm Coth}\Big({\pi\over\beta}(|{\bf x}|-i x_{E}^{0})\Big)+{\rm Coth}\Big({\pi\over\beta}(|{\bf x}|+i x_{E}^{0})\Big)\Big].
}}
The thermal $a_{\tau,J}$ coefficients $a_{2,2}$ and $a_{4,4}$ are the same as in the adjoint model (this is so since the second term in \thermaltwoptVec\ does not affect these):
\eqn\aV{\eqalign{
  a_{2,2} &= {\pi^4\Delta\over 45},\cr
  a_{4,4} &= {\pi^8\Delta(\Delta-1)\over 1050}.
}
}

The unit-normalized stress tensor is given by 
\eqn\stressVec{
  T_{\mu\nu}(x) = {1\over 3\sqrt{C_T}}:\left(\pa_\mu \varphi^i\pa_\nu \varphi^i-{1\over 2}\varphi^i\pa_\mu\pa_\nu \varphi^i-({\rm trace})\right):(x),
}
where $C_T={4\over 3}N$. The OPE coefficient of the stress tensor is again found by Wick contractions to be
\eqn\OpeStressVect{
  \lambda_{\OO_\Delta\OO_\Delta T_{\mu\nu}} = -{4\Delta\over 3\sqrt{C_T}},
}
in agreement with the stress tensor Ward identity. The double-stress tensor is given by 
\eqn\duoblstress{
  T^2_{\mu\nu\rho\sigma} = {1\over \sqrt{2}}:T_{(\mu\nu}T_{\rho\sigma)}:-({\rm traces}),
}
and the OPE coefficient is calculated precisely as for the adjoint model and we find
\eqn\doublestressOPEVECT{
  \lambda_{\OO_\Delta\OO_\Delta T^2_{4,4}} = {8\sqrt{2}\over 9C_T}\Delta(\Delta-1).
}
There is another double-trace operator with twist $4$ and spin $4$ and takes the same form $:\OO_2\OO_{2,4}:$ as for the adjoint model 
\eqn\dtodimeightVec{\eqalign{
  \OO^{\rm DT}_{\mu\nu\rho\sigma}(x) = {1\over 96\sqrt{70}N}:\varphi^i\varphi^i\Big(\varphi^j\pa_\mu\pa_\nu\pa_\rho\pa_\sigma\varphi^j-16 \pa_{(\mu}\varphi^j\pa_\nu\pa_\rho\pa_{\sigma)}\varphi^j\cr
 +18\pa_{(\mu}\pa_\nu\varphi^j\pa_\rho\pa_{\sigma)}\varphi^j -({\rm traces})\Big):(x).
}}
The OPE coefficient and the thermal one-point function yields the same result as for the corresponding operator in the adjoint model\foot{Note that this is not true for all operators but is in line with the fact that $a_{4,4}$ is unaffected by the second term in \thermaltwoptVec.}. It then follows that the $a_{4,4}$ extracted from \thermaltwoptVec\ is reproduced by the sum of the double stress tensor and \dtodimeightVec.

\appendix{I}{Factorization of thermal correlators}
In this appendix we argue for the factorization of thermal expectation values of multi-trace operators in large-$C_T$ theories on $S^1\times {\bf R}^{d-1}$. Consider the thermal two-point function of a scalar operator $\OO$ with dimension $\Delta$:
\eqn\odelta{
  \langle \OO(x)\OO(0)\rangle_\beta = \langle \OO\rangle_\beta\langle\OO\rangle_\beta+\langle \OO(x)\OO(0)\rangle_{\beta,c},
}
where the second term consist of the connected part of the correlator. Note that the disconnected term in \odelta\ is independent of the position $x$. On the other hand we can evaluate \odelta\ using the OPE on the plane which takes the form 
\eqn\opeOdelta{
  \OO(x)\OO(0) = {1\over |x|^{2\Delta}}+\sum_{n,l} \lambda_{\OO\OO[\OO\OO]_{n,l}}x^{2n+l}[\OO\OO]_{n,l}+\ldots,
}
when written in terms of primaries and the dots refer to terms surpressed in the large-$C_T$ limit. Note that $\lambda_{\OO\OO[\OO\OO]_{n,l}}$ are the MFT OPE coefficient which are of order $1$. The term in \opeOdelta\ that is independent of $x$ is due to the $n=l=0$ term in \opeOdelta\ and inserting the OPE on the LHS of \opeOdelta, we find that 
\eqn\factscalar{
  \lambda_{\OO\OO[\OO\OO]_{0,0}}\langle[\OO\OO]_{0,0}\rangle_\beta = \langle \OO\rangle_\beta^2. 
}
When $[\OO\OO]_{0,0}$ is unit-normalized the OPE coefficient is given by $\lambda_{\OO\OO[\OO\OO]_{0,0}}=\sqrt{2}$ and it follows that 
\eqn\factscalarfinal{
  \langle[\OO\OO]_{0,0}\rangle_\beta = {1\over\sqrt{2}}\langle \OO\rangle_\beta^2.
}
We therefore see the that the thermal one-point function of the double-trace operator factorizes on the plane. 
We expect a similar argument to hold for multi stress tensors.

\listrefs

\bye